\documentclass[12pt,a4paper,twoside]{article}

\usepackage{adjustbox}
\usepackage{amscd}
\usepackage{amsfonts}
\usepackage{amsmath}
\usepackage{amssymb}
\usepackage{bm}
\usepackage{cancel}
\usepackage{color}
\usepackage{comment}
\usepackage{enumitem}
\usepackage{mathabx}
\usepackage{mathdots}
\usepackage{mathtools}
\usepackage{stmaryrd}
\usepackage{subdepth} 
\usepackage{tcolorbox}
\usepackage{theorem}
\usepackage{tikz-cd}
\newtheorem{thm}{Theorem}
\newtheorem{assumption}[thm]{Assumption}
\newtheorem{definition}[thm]{Definition}
\newtheorem{lemma}[thm]{Lemma}
\newtheorem{proposition}[thm]{Proposition}
\newtheorem{defprop}[thm]{Definition/Proposition}
{\theorembodyfont{\upshape}\newtheorem{remark}[thm]{\it Remark}}
{\theorembodyfont{\upshape}\newtheorem{example}[thm]{Example}}
\newtheorem{corollary}[thm]{Corollary}
\newcommand{\R}{{\mathord{\mathbb R}}}
\newcommand{\Z}{{\mathord{\mathbb Z}}}
\newcommand{\N}{{\mathord{\mathbb N}}}
\newcommand{\C}{{\mathord{\mathbb C}}}
\newcommand{\T}{{\mathord{\mathbb T}}}
\newcommand{\mA}{\mathcal A}
\newcommand{\mB}{\mathcal B}
\newcommand{\mE}{\mathcal E}
\newcommand{\mH}{\mathcal H}
\newcommand{\mL}{\mathcal L}
\newcommand{\mM}{\mathcal M}
\newcommand{\mP}{\mathcal P}
\newcommand{\mQ}{\mathcal Q}
\newcommand{\mS}{\mathcal S}
\newcommand{\e}{{\rm e}}

\newcommand{\pf}{{\rm pf}}
\newcommand{\ran}{{\rm ran\,}}		
\newcommand{\sign}{{\rm sign}}
\newcommand{\sgn}{{\rm sgn}}
\newcommand{\spa}{{\rm span\,}}
\newcommand{\tr}{{\rm tr}}
\def\slim{\mathop{\rm s-lim}}
\newcommand{\sAut}{{{}^\ast\hspace{-0.5mm}\rm Aut}}
\newcommand{\sIso}{{{}^\ast\rm Iso}}
\newcommand{\Hom}{{\rm Hom}}
\newcommand{\sHom}{{{}^\ast\rm Hom}}
\newcommand{\card}{{\rm card}}
\newcommand{\rd}{\hspace{-0.5mm}{\rm d}}
\newcommand{\diag}{{\rm diag}}
\newcommand{\h}{{{\mathfrak h}}}
\newcommand{\fh}{{\wh{\mathfrak h}}}
\newcommand{\hh}{{{\mathfrak h}^{\oplus 2}}}
\newcommand{\fhh}{{\wh{\mathfrak h}^{\oplus 2}}}
\newcommand{\ff}{{{\mathfrak f}}}
\newcommand{\fA}{{\mathfrak A}}
\newcommand{\fB}{{\mathfrak B}}
\newcommand{\Fin}{{\rm Fin}}
\newcommand{\Cs}{$\text{C}^\ast$\hspace{-0.8mm}-} 
\newcommand{\str}{${}^\ast$\hspace{-0.4mm}-} 
\newcommand{\al}{\alpha}
\newcommand{\gm}{\gamma}
\newcommand{\Gm}{\Gamma}
\newcommand{\veps}{{\varepsilon}}
\newcommand{\kp}{\kappa}
\newcommand{\io}{\iota}
\newcommand{\lm}{\lambda}
\newcommand{\Lm}{\Lambda}
\newcommand{\sg}{\sigma}
\newcommand{\dom}{{\rm dom}}
\newcommand{\eig}{{\rm eig}}
\newcommand{\spec}{{\rm spec}}
\newcommand{\supp}{{\rm supp}}
\newcommand{\vi}{\varphi}
\newcommand{\wt}{\widetilde}
\newcommand{\wh}{\widehat}
\newcommand{\wc}{\widecheck}
\newcommand{\bd}{\begin{definition}}
\newcommand{\ed}{\end{definition}\vspace{1mm}}
\newcommand{\bt}{\begin{thm}}
\newcommand{\et}{\end{thm}\vspace{2mm}}
\newcommand{\bc}{\begin{corollary}}
\newcommand{\ec}{\end{corollary}\vspace{2mm}}
\newcommand{\bl}{\begin{lemma}}
\newcommand{\el}{\end{lemma}\vspace{2mm}}
\newcommand{\bdp}{\begin{defprop}}
\newcommand{\edp}{\end{defprop}}
\newcommand{\bp}{\begin{proposition}}
\newcommand{\ep}{\end{proposition}\vspace{2mm}}
\newcommand{\bx}{\begin{example}}
\newcommand{\ex}{\end{example}}
\newcommand{\br}{\begin{remark}}
\newcommand{\er}{\end{remark}}
\newcommand{\bass}{\begin{assumption}}
\newcommand{\eass}{\end{assumption}\vspace{2mm}}
\newcommand{\bprf}{\noindent{\bf Proof.}\hspace{2mm}}
\newcommand{\eprf}{\hfill $\Box$\vspace{5mm}}
\newcommand{\ie}{i.e.}
\def\bas#1\eas{\begin{align*}#1\end{align*}}
\def\ba#1\ea{\begin{align}#1\end{align}}
\newcommand{\bn}{\begin{enumerate}}
\newcommand{\en}{\end{enumerate}}
\newcommand{\Ev}{{\rm Ev}}
\newcommand{\Od}{{\rm Od}}
\newcommand{\Int}{\int_{-\pi}^\pi\frac{\rd k}{2\pi}\hspace{0.5mm}}
\newcommand{\Inti}{\int_{-\pi}^\pi\rd k/(2\pi)\hspace{0.5mm}}
\newcommand{\G}[1]{{\rm #1}}
\newcommand{\M}[1] {\C^{{#1}\times{#1}}}
\newcommand{\nm}[2] {\C^{{#1}\times{#2}}}
\newcommand{\num}[2]{\llbracket{#1},{#2}\rrbracket}
\newcommand{\ii}{{\rm i}}

\newcommand{\ei}{\e^{\ii k}}
\newcommand{\An}[1]{{Assumption #1}}
\newcommand{\tikzcircle}[2][red,fill=red]{\tikz[baseline=-0ex]\draw[#1,radius=#2] (-1,0) circle ;}
\DeclareMathOperator*{\esup}{ess\,sup\,}
\renewcommand{\Re}{{\rm Re}}
\renewcommand{\Im}{{\rm Im}}

\begin{document}
\pagestyle{myheadings}
\title{On the asymptotic scaling of the von Neumann entropy in quasifree fermionic right mover/left mover systems}
\author{Walter H. Aschbacher\footnote{walter.aschbacher@univ-tln.fr}
\\ \\
Universit\'e de Toulon, Aix Marseille Univ, CNRS, CPT, Toulon, France
}
\markboth{Walter H. Aschbacher}{Entropy scaling in quasifree R/L mover systems}

\date{}
\maketitle

\begin{abstract}
For the general class of quasifree fermionic right mover/left mover systems over the infinitely extended two-sided discrete line introduced in \cite{As21} within the algebraic framework of quantum statistical mechanics, we study the von Neumann entropy of a contiguous subsystem of finite length in interaction with its environment. 
In particular, under the assumption of spatial translation invariance, we analyze the  asymptotic behavior of the von Neumann entropy for large subsystem lengths and prove that its leading order density is, in general, nonvanishing and displays the signature of a mixture of the independent thermal species underlying the right mover/left mover system. 
As special cases, the formalism covers so-called nonequilibrium steady states, thermal equilibrium states, and ground states. Moreover, for general Fermi functions, we derive a necessary and sufficient criterion for the von Neumann entropy density to vanish. 
\end{abstract}

\noindent {\it Mathematics Subject Classifications (2010)}\,
46L60, 47B35, 82C10, 82C23.

\noindent {\it Keywords}\,
Open quantum systems, quasifree fermionic chains, right mover/left mover systems, nonequilibrium steady states, von Neumann entropy.

\tableofcontents

\section{Introduction}
\label{sec:intro}

Open quantum systems, \ie, quantum systems which are effectively coupled through interactions of various kinds to a so-called extensive external environment, are ubiquitous in nature.  A rigorous study from first principles is therefore of central importance for a broader understanding of many of their properties and, in particular, of their thermodynamic properties in and out of equilibrium. Since open quantum systems have typically a very large number of degrees of freedom and since the finite accuracy of any feasible experiment does not allow an empirical distinction between an infinite system and a finite system with sufficiently many degrees of freedom, a powerful strategy consists in approximating the actual finite system by an idealized one with infinitely many degrees of freedom. Although it seems to head in the direction opposite to the simplicity which characterizes the phenomenological description at the thermodynamic scales of interest, this idealization has, on the contrary, numerous well-known major analytical and algebraic advantages (see \cite{Pr83} for an extensive discussion). 

One of the most important axiomatic frameworks for the study of such idealized infinite systems is the so-called algebraic approach to quantum mechanics based on operator algebras. After having been heavily used from as early as the 1960s on, in particular for the quantum statistical description of quantum systems in thermal equilibrium (see, for example, \cite{Em72, Se86, BrRo8797}), the benefits of this framework have again started to unfold more recently in the physically much more general situation of open quantum systems out of equilibrium. Although the most interesting phenomena which emerge on the macroscopic level are not restricted to systems in thermal equilibrium but, quite the contrary, often occur out of equilibrium, our general theoretical understanding of nonequilibrium order and phase transitions is substantially less developed. 

Most of the rather scarce mathematically rigorous results have been obtained for the so-called nonequilibrium steady states (NESSs) introduced in \cite{Ru01} by means of scattering theory on the algebra of observables. An important role in the construction of such NESSs is played by the so-called quasifree fermionic systems, and this is true not only because of their mathematical accessibility but also when it comes to real physical applications. Indeed, from a mathematical point of view, these systems allow for a simple and powerful representation independent description since scattering theory on the fermionic algebra of observables boils down to scattering theory on the underlying 1-particle Hilbert space over which the fermionic algebra is constructed. This restriction of the dynamics to the 1-particle Hilbert space opens the way for a rigorous mathematical analysis of many purely quantum mechanical properties which are of fundamental physical interest. Moreover, beyond their importance due to their mathematical accessibility, quasifree fermionic systems effectively describe nature: aside from the various electronic systems in their independent electron approximation, they also play an important role in the rigorous approach to physically realizable quantum spin chains under (Araki's extension of) the Jordan Wigner transformation (see \cite{JoWi28, Ar84} and, for example, \cite{As23}).

The first representatives of such quantum spin models, the so-called (Lenz-) Ising and Heisenberg chains, were respectively introduced in 1920 in \cite{Le20} (and analyzed in 1925 in \cite{Is25}) and in 1928 in \cite{He28} in order to describe magnetic properties of crystalline solids. An important special instance is the so-called XY chain whose Hamiltonian density has the form
\ba
\label{XYDensity}
(1+\gamma)\,\sigma_1^{(x)}\sigma_1^{(x+1)}+(1-\gamma)\,\sigma_2^{(x)}\sigma_2^{(x+1)}
+2\lm \sigma_3^{(x)}
\ea
where $\gamma\in\R$ stands for the anisotropy, $\lm\in\R$ for an external magnetic field, and the superscripts of the Pauli matrices for the sites $x\in\Z$. Without the magnetic field (\ie,  $\lm=0$), the so-called isotropic version of \eqref{XYDensity} (\ie, $\gamma=0$) was studied in 1950 in \cite{Na50} and the more general anisotropic version (\ie, $\gamma\in[-1,1]$) was introduced in 1961 in \cite{LiScMa61} (where also the name ''XY model'' was coined). In 1962 in \cite{Ka62}, \eqref{XYDensity} was supplemented by the external magnetic field $\lm$ (see also \cite{Ni67}). Already in 1969, a first physical realization of the XY chain has been identified (see, for example, \cite{CuScPf69}). The impact of the XY chain on the experimental, numerical, theoretical, and mathematical research activity in the field of low-dimensional magnetic systems is ongoing ever since (see, for example,  \cite{MiKo04}). 

In the present paper, we consider 1-dimensional quantum mechanical systems whose configuration space is the 2-sided infinite discrete line $\Z$ and whose algebra of observables is the so-called canonical anticommutation relations (CAR) algebra over the 1-particle Hilbert space of all complex-valued square-summable functions over $\Z$. The class of states we want to study is the class of so-called right mover/left mover states (R/L movers) introduced in \cite{As21} by means of time dependent scattering theory on the 1-particle Hilbert space. For a given Hamiltonian generating a quasifree dynamics on the 1-particle Hilbert space, an R/L mover is specified by a 2-point operator whose main part consists of a mixture of two independent species stemming from the asymptotic right and left side of $\Z$, carrying the inverse temperatures $\beta_L$ and $\beta_R$, respectively. The prototypical example of such an R/L-mover is the general NESS constructed as the large time limit of the averaged trajectory of a time-evolved initial state which is the decoupled product of three thermal equilibrium states over the configuration spaces of the so-called sample and the reservoirs to the left and right of it. This class contains in particular the so-called XY NESS, \ie, the NESS constructed for the full XY model \eqref{XYDensity} with nonvanishing anisotropy and nonvanishing external magnetic field (see \cite{As21} and references therein). Moreover, as particular cases, this class also contains thermal equilibrium states and ground states. 

For the two-sided infinitely extended XY chain with density \eqref{XYDensity}, the complete ground state phase diagram for the parameters $(\gm,\lm)\in\R^2$ is given in \cite{ArMa85}. Notably, the number of pure ground states is equal to $1$ if $|\lm|\ge 1$ or if $\gm=0$ and $|\lm|< 1$ but it equals $2$ if $\gm\neq 0$ and $|\lm|<1$ (and if the XY model does not reduce to the Ising model; the latter is characterized by $|\gm|=1$ and $\lm=0$ and has an infinite number of pure ground states). For a given total system in a pure ground state and a given contiguous subsystem of the total system, the von Neumann entropy of the reduced density matrix of the subsystem frequently serves as a measure which quantifies the degree of entanglement of the subsystem with its complement with respect to the total system. In order to capture relevant information about this quantum correlation on different length scales, one often studies the scaling behavior of the von Neumann entropy with respect to the length of the subsystem (in the 1-dimensional setting). For the XY chain, for instance, it was shown in \cite{JiKo04} that, on a portion of the critical phase diagram (recall that the critical phase diagram of the XY chain equals the subset of all $(\gm,\lm)\in\R^2$ satisfying $|\lm|=1$ or $\gm=0$ and $|\lm|<1$), the von Neumann entropy diverges logarithmically. In contrast, on a portion of the noncritical phase diagram (being the complement of the critical phase diagram), we know from \cite{ItJiKo05} that the von Neumann entropy saturates to a constant. These findings are prototypical  and representative for the rather clear picture which has emerged since then for the scaling laws in 1-dimensional systems for which the von Neumann entropy is expected (under additional assumptions) to saturate to a constant, \ie, to obey a so-called area law, if the system is noncritical and to diverge logarithmically if it is critical (see, for example, \cite{AmFaOsVe08, EiCrPl10} for extensive reviews).

Since the von Neumann entropy has the potential to detect the criticality of a system in a pure ground state through its entanglement scaling law, one is naturally led to the study of its behavior for more general open system at one (or several) nonvanishing temperatures and of the impact of the underlying criticality (see, for example, \cite{Sa99}). Although the von Neumann entropy does not provide a direct interpretation of the entanglement content of the subsystem in question if the total system is at nonvanishing temperature (since it does not,  in the usual way, tell apart classical and quantum correlations), it still is a fundamental quantity also for non pure states of the total system for which it measures the mixedness of the subsystem (\ie, the disorder or lack of information about the subsystem, see \cite{We78, Pr83}, for example). Uhlmann's theory, for instance, introduces the notion of mixing-enhancement and a so-called purer state in this theory indeed has a lower von Neumann entropy (this notion also allows for the construction of a very general equivalence relation, see  \cite{We78}, for example). In addition, the von Neumann entropy also is a key ingredient for numerous quantities which play an important role in various applications of quantum information theory (see, for example, \cite{AmFaOsVe08, EiCrPl10, GoWi21}) and, on a maybe more fundamental level, in the variational characterization, through the maximal entropy principle, of thermal equilibrium states in quantum spin systems. In the latter, the von Neumann entropy enters the expression of the free energy when restricted to a local subsystem. On the other hand, if one considers an open subsystem, the so-called conditional entropy (or entropy of the open subsystem), defined as the difference between the von Neumann entropy of the total system and the von Neumann entropy of the complement of the subsystem, measures the entropy of the subsystem in interaction with its complement with respect to the total system and characterizes the infinite thermal equilibrium state (the so-called KMS state) through the principle of maximal conditional entropy (see \cite{BrRo8797} for precise statements). For the class of translation invariant states of a quantum spin system, the (infinite limit) mean von Neumann entropy (also called the von Neumann entropy density) exists (if the local configuration space goes to infinity in the sense of van Hove) and has many interesting properties (for 1-dimensional systems, the existence of the density is a direct consequence of the subadditivity of the von Neumann entropy and of Fekete's subadditivity lemma). In particular, the von Neumann entropy density enters the mean free energy whose maximum again characterizes the infinite thermal equilibrium state (for certain classes of quantum spin interactions). Moreover, the conditional von Neumann entropy density also exists and, for thermal equilibrium states, these two densities coincide (if the interaction energy across the boundary grows slower than the volume of the local configuration space, see \cite{En82, BrRo8797}). Last but not least, in 1-dimensional systems and for arbitrary translation invariant states, the von Neumann entropy density can also be interpreted as a 1-sided conditional entropy per finite subsystem length (see \cite{En82}).

It is thus fair to say that the von Neumann entropy and its density figure in many different physically relevant situations. Motivated by this fact, we want to study the asymptotic scaling of the von Neumann entropy  in our general setting of L/R movers comprising, notably, an important class of infinitely extended open systems out of thermal equilibrium. In order to clarify the usual strategy of the analysis, we want to bring out the role played by the 1-particle Hilbert space underlying the CAR algebra. Moreover, for the same reason, we disentangle the assumptions used at various places and keep the setting at a mathematically rather general level (in particular, in view of possible future generalizations to locally perturbed systems). By embedding certain important quantities in broader families, we improve our understanding of the asymptotic scaling for general and not necessarily gauge-invariant fermionic systems which, in particular, cover NESSs,  thermal equilibrium states, and ground states within the same formalism. In the main theorem of the present paper, we prove that, in our general L/R mover systems, the von Neumann entropy density is, in general, nonvanishing and displays the signature of a mixture of two independent thermal species stemming from the left and the right reservoir at their respective temperatures.

\vspace{2mm}

The paper is organized as follows.

\vspace{1mm}

{\it Section \ref{sec:infinite} (Infinite fermionic systems)}\, 
We introduce the algebraic framework for the  systems to be studied, \ie, the (doubled) 1-particle position Hilbert space, the CAR algebra of observables, the 2-point operators, and the so-called quasifree states. Moreover, we define what we call a Fermi family, \ie, a set of localized elements of the 1-particle Hilbert space whose corresponding generators satisfy the CAR. Moreover, in view of Section \ref{sec:RDM}, we introduce the local observable algebra and construct a local family of matrix units based on the selfdual generators and prove their properties needed in the sequel.

{\it Section \ref{sec:RDM} (Reduced density matrix)}\, 
We introduce the main ingredient which enters the definition of the von Neumann entropy in Section \ref{sec:Neu}, \ie, the reduced density matrix which characterizes the restriction of the state under consideration to the local observable algebra. In order to extract the desired information on the spectrum of the reduced density matrix, we follow the strategy of \cite{ViLaRiKi03} albeit in a somewhat different form which stresses the role played by the underlying 1-particle Hilbert space. In particular, we show that, for any given Fermi family, there always exists a so-called Bogoliubov transformation on the 1-particle Hilbert space transforming the initial Fermi family into a Fermi family which possesses the required factorization properties allowing for a precise description of the spectrum of the reduced density matrix. Moreover, we introduce the so-called Majorana correlation matrix and relate the spectrum of its imaginary part to spectrum of the reduced density matrix.

{\it Section \ref{sec:Neu} (Von Neumann entropy)}\, 
In order to be able to define our main object of study, \ie, the von Neumann entropy, we introduce the Shannon entropy function and its associated (translated and dilated) binary entropy. Moreover, we derive the functional equation satisfied by the Shannon entropy function on the spectrum of the reduced density matrix and prove, conversely, that the Shannon entropy function is the unique (measurable) solution of this functional equation. We then introduce the von Neumann entropy and compute its usual expression involving the spectrum of the reduced density matrix, \ie, the spectrum of the imaginary part of the Majorana correlation matrix from Section \ref{sec:RDM}.

{\it Section \ref{sec:trans} (Translation invariance)}\, 
It's only from this section on that we make use of translation invariance. By means of the usual unitary Fourier transform, we switch to (doubled) momentum space. Under the assumption of translation invariance, any 2-point operator becomes a matrix multiplication operator and any Fermi family is reduced to the so-called completely localized standard Fermi family which allows us to resort to general Toeplitz theory. In particular, we show that the imaginary part of the Majorana correlation matrix acts as the finite section of a block Toeplitz operator whose block symbol is computed for a general translation invariant 2-point operator. 

{\it Section \ref{sec:R/L} (R/L mover entropy asymptotics)}\, 
Making use of \cite{As21}, we introduce the class of states of our interest, \ie, the R/L movers. In order to do so, we first specify the (selfdual) Hamiltonian and the so-called R/L mover generator based on the asymptotic projections for the underlying right/left geometry.  Then, for general Fermi functions, the class of R/L mover 2-point operators is defined under simple assumptions on the Hamiltonian and for general so-called initial 2-point operators. Subsequently, we discuss the main examples covered by the formalism, \ie, the NESSs, the thermal equilibrium states, and the ground states. Under the assumption of a Hamiltonian of finite range (whose range is bounded by the length of the sample), we display the multiplication operator form of a general R/L mover 2-point operator. We then arrive at the main theorem of our paper in which we prove that, in general L/R mover systems, the von Neumann entropy density is, in general, nonvanishing and carries the signature of a mixture of two independent thermal species stemming from the left and the right reservoir at their respective temperatures. After a discussion of a number of special cases and examples, we finally derive a necessary and sufficient criterion for the von Neumann entropy density to vanish.

 {\it Section \ref{sec:Proofs} (Proofs)}\,
Except for the proof of the main theorem, this section contains all the proofs from the foregoing sections carried out in the present paper.

{\it Appendix  \ref{app:Toep} (Toeplitz operators)}\, 
We introduce most of the function spaces used in the preceding sections. Moreover, we provide a diagrammatic overview of the definitions of the block Toeplitz operators and their corresponding finite section method on all the function spaces needed.

\section{Infinite fermionic systems}
\label{sec:infinite}

In this section, we introduce the axiomatic framework used for the study of infinite systems discussed in the Introduction. Recall that, in this so-called algebraic approach to quantum statistical mechanics, the three fundamental ingredients of a physical system, \ie, the observables, the time evolution, and the states of the system, have a mathematical representation in the form of a \Cs algebra, a 1-parameter group of \str automorphisms, and a normalized positive linear functional on the observable algebra, respectively (see, for example, \cite{Em72, Pr83, Se86, BrRo8797}). 

In order to make these notions precise, we first introduce the following notations. Let $\N:=\{1,2,\ldots\}$ and $\N_0:=\N\cup\{0\}$. For all $n,m\in\N$ (fixed in the following) and for any set $D$, we denote the set of $n\times m$ matrices with entries from $D$ by $D^{n\times m}$ and the entries of $X\in D^{n\times m}$ by $(X)_{ij}$ for all $i\in\num{1}{n}$ and all $j\in\num{1}{m}$. If $m=1$, we write $D^n:=D^{n\times 1}$ and $(X)_i:=(X)_{i1}$ for all $i\in\num{1}{n}$ and all $X\in D^n$. Moreover, $[X_{ij}]_{i\in\num{1}{n}, j\in\num{1}{m}}\in D^{n\times m}$ stands for the matrix with entries $X_{ij}\in D$ for all $i\in\num{1}{n}$ and all $j\in\num{1}{m}$ (\ie, $([X_{ij}]_{i\in\num{1}{n}, j\in\num{1}{m}})_{kl}=X_{kl}$ for all $k\in\num{1}{n}$ and all $l\in\num{1}{m}$), where, for all $x,y\in\Z$, we set
\ba
\label{Setxy}
\num{x}{y}
:=\begin{cases}
\{x,x+1,\ldots,y\}, & x<y,\\
\hfill\{x\}, & x=y,\\
\hfill\emptyset, &x>y.
\end{cases}
\ea
Moreover, for all $X:=[X_{ij}]_{i\in\num{1}{n}, j\in\num{1}{m}}\in D^{n\times m}$, we denote the transpose of $X$ by $X^T:=[X_{ji}]_{i\in\num{1}{m}, j\in\num{1}{n}}\in D^{m\times n}$. If $D=\C$, we set $\bar X:=[\bar X_{ij}]_{i\in\num{1}{n}, j\in\num{1}{m}}\in\nm{n}{m}$, define the conjugate transpose (adjoint) by $X^\ast:=\bar X^T\in\nm{m}{n}$, and write $\Re(X):=(X+\bar X)/2$ and $\Im(X):=(X-\bar X)/(2\ii)$, where $\bar z$ for all $z\in\C$ stands for the usual complex conjugation on $\C$. For any $D\subseteq\R$ and any $\chi\in\C^D$, the function $\bar\chi\in\C^D$ is defined by $\bar\chi(x):=\overline {\chi(x)}$ for all $x\in D$, where, for any two sets $A,B$, we denote by $B^A$ the set of all functions from $A$ to $B$.

Furthermore, if $\mH$ is any complex Hilbert space (with scalar product $(\cdot, \cdot)_\mH$ and induced norm $\|\cdot\|_\mH$), we denote the (external) direct sum of $\mH$ with itself by $\mH^{\oplus 2}$ (being equal to $\mH^2$ as a set), we write $f_1\oplus f_2:=\begin{bmatrix}f_1\\f_2\end{bmatrix}\in\mH^{\oplus 2}$, and we recall that the scalar product on $\mH^{\oplus 2}$ is given by $(F, G)_{\mH^{\oplus 2}}:=\sum_{i\in\num{1}{2}}((F)_i, (G)_i)_\mH$ for all $F,G\in\mH^{\oplus 2}$.
Moreover, $\mL(\mH)$ stands for the \Cs algebra of all bounded operators on $\mH$ with respect to (the usual addition, complex scalar multiplication, multiplication, involution, and) the operator norm defined by $\|a\|_{\mL(\mH)}:=\sup_{f\in\mH,\hspace{0.5mm}\|f\|_\mH=1}\|af\|_\mH$ for all $a\in\mL(\mH)$. For all $a\in\mL(\mH)$  and all $X:=[X_{ij}]_{i,j\in\num{1}{2}}\in\C^{2\times 2}$, we define $(aX)_{ij}:=X_{ij}a\in\mL(\mH)$ for all $i,j\in\num{1}{2}$ and $aX:=[(aX)_{ij}]_{i,j\in\num{1}{2}}\in\mL(\mH)^{2\times 2}$ and we often identify $\mL(\mH)^{2\times 2}$ with $\mL(\mH^{\oplus 2})$ by means of  the usual natural bijection $I:\mL(\mH)^{2\times 2}\to \mL(\mH^{\oplus 2})$ defined, for all $A\in\mL(\mH)^{2\times 2}$ and all $f_1, f_2\in\mH$, by 
\ba
I(A)f_1\oplus f_2
:=((A)_{11}f_1+(A)_{12}f_2)\oplus ((A)_{21}f_1+(A)_{22}f_2).
\ea
Any $A\in\mL(\mH^{\oplus 2})$ (under $I^{-1}$) has a unique so-called Pauli expansion
\ba
\label{PauliExp}
A
=a_0\sigma_0+a\sigma,
\ea
where $a_0\in\mL(\mH)$, $\sigma_0\in\M{2}$, and $a\sigma:=\sum_{i\in\num{1}{3}}a_i\sigma_i\in\mL(\mH^{\oplus 2})$ with $a:=[a_i]_{i\in\num{1}{3}}\in\mL(\mH)^3$ and $\sigma:=[\sigma_i]_{i\in\num{1}{3}}\in (\C^{2\times 2})^3$, and where $\{\sigma_\alpha\}_{\alpha\in\num{0}{3}}\subseteq\M{2}$ is the Pauli basis of $\C^{2\times 2}$,
\ba
\label{pm} 
\sigma_0
:=\left[\begin{array}{cc}
1 & 0
\\ 0&1
\end{array}\right],\quad
\sigma_1
:=\left[\begin{array}{cc}
0 & 1\\
1& 0
\end{array}\right],\quad
\sigma_2
:=\left[\begin{array}{cc}
0 &-\ii\\ 
\ii & 0
\end{array}\right],\quad
\sigma_3
:=\left[\begin{array}{cc}
1 & 0
\\ 0& -1
\end{array}\right].
\ea
If  $a_0, b_0\in\mL(\mH)$ and $a, b\in\mL(\mH)^3$ and if $A:=a_0\sigma_0+a\sigma\in\mL(\hh)$ and $B:=b_0\sigma_0+b\sigma\in\mL(\hh)$, then, using that $(cX)(dY)=(cd)(XY)$ for all $c,d\in\mL(\mH)$ and all $X,Y\in \M{2}$, we have 
\ba
\label{prod}
AB
=(a_0b_0+ab)\sigma_0 +(a_0b+ab_0+\ii a\wedge b)\sigma,
\ea
where we set $ab:=\sum_{i\in\num{1}{3}}a_i b_i\in\mL(\mH)$, $a_0b:=[a_0b_i]_{i\in\num{1}{3}}\in\mL(\mH)^3$, $ab_0:=[a_ib_0]_{i\in\num{1}{3}}\in\mL(\mH)^3$, and $a\wedge b\in\mL(\mH)^3$ is given by $(a\wedge b)_i:=\sum_{j,k\in\num{1}{3}}\veps_{ijk}\hspace{0.5mm} a_jb_k$ for all $i\in\num{1}{3}$, and $\veps_{ijk}$ with $i,j,k\in\num{1}{3}$ is the usual Levi-Civita symbol. Of course, all the foregoing considerations can be analogously applied to the case of bounded antilinear operators which we denote by $\bar\mL(\mH)$. 

Furthermore, writing  $\card$ for cardinality, the set of all finite subsets of $\Z$ is denoted by
\ba
\label{def:Fin}
\Fin(\Z)
:=\{\Lm\subseteq\Z\,|\, \card(\Lm)\in\N\}, 
\ea
and we equip it with a direction (\ie, with an upward directed partial ordering) defined by set inclusion. Note that, if $f\in\C^\Z$, the net limit $\lim_{\Lm\in\Fin(\Z)}\sum_{x\in\Lm}|f(x)|^2$ exists if and only if the limit $\lim_{n\to\infty}\sum_{x\in\num{-n}{n}}|f(x)|^2$ exists and, if so, both limits are equal and are denoted by $\sum_{x\in\Z}|f(x)|^2$ (and $\sum_{x\in\Z}|f(x)|^2<\infty$ means that those limits exist). We now write the usual separable complex Hilbert space of square-summable complex-valued functions on $\Z$ as 
\ba
\label{l2(Z)}
\ell^2(\Z)
:=\{f\in\C^\Z\,|\, \sum\nolimits_{x\in\Z}|f(x)|^2<\infty\}.
\ea
The scalar product on \eqref{l2(Z)} is defined by $(f,g)_{\ell^2(\Z)}:=\sum_{x\in\Z}\bar f(x) g(x)$ for all $f,g\in\ell^2(\Z)$, where we set $\sum_{x\in\Z}\bar f(x) g(x):=\lim_{\Lm\in\Fin(\Z)}\sum_{x\in\Lm}\bar f(x)g(x)$ (which exists), and we again have $(f,g)_{\ell^2(\Z)}=\lim_{n\to\infty}\sum_{x\in\num{-n}{n}}\bar f(x) g(x)$ (the corresponding norm is denoted by $\|\cdot\|_{\ell^2(\Z)}$ and analogously for any other norm; since many different spaces occur, we stick to the heavier notation with subscripts, see, in particular, Appendix \ref{app:Toep}). 

Finally, for any elements $A$ and $B$ in the various algebras occurring in the sequel, the commutator and the anticommutator of $A$ and $B$ are denoted as usual by $[A,B]:=AB-BA$ and $\{A,B\}:=AB+BA$, respectively. For all $r,s\in\R$, the Kronecker symbol is defined by $\delta_{rs}:=1$ if $r=s$ and $\delta_{rs}:=0$ if $r\neq s$ and, for all $M\subseteq\R$, we denote by $1_M:\R\to\{0,1\}$ the characteristic function of $M$ (and we write $1_\lambda:=1_{\{\lambda\}}$ if $\lambda\in\R$). 

\bd[Observables]
\label{def:obs}
\bn[label=(\alph*), ref={\it (\alph*)}]
\setlength{\itemsep}{0mm}
\item
\label{obs-a}
The configuration space of the physical system is the two-sided infinite discrete line $\Z$ and the 1-particle Hilbert space is given by
\ba
\h
:=\ell^2(\Z).
\ea

\item
\label{obs-b}
The algebra of observables is the unital \Cs algebra $\fA$ given by the CAR algebra over $\h$,
\ba
\label{A-CAR}
\fA
:=\G{CAR}(\h).
\ea
It is generated by the set of elements $\{c_x\,|\, x\in\Z\}$ satisfying the CAR, \ie, for all $x,y\in\Z$, 
\ba
\label{dCAR1}
\{c_x, c_y\}
&=0,\\
\label{dCAR2}
\{c_x, c_y^\ast\}
&=\delta_{xy} 1.
\ea

\item
\label{obs-c}
The map $c\in\fA^\h$ is defined, for all $f\in\h$, by
\ba
\label{c(f)}
c(f)
:=\lim_{n\to\infty}\sum_{x\in\num{-n}{n}}\bar f(x)\hspace{0.2mm}c_x,
\ea
and the map $B\in\fA^{\hh}$, for all $F:=f_1\oplus f_2\in\h^{\oplus 2}$, by
\ba
\label{B(F)}
B(F)
:=c^\ast(f_1)+c(\bar f_2).
\ea
Moreover, let $\zeta\in\bar\mL(\h)$ be given by $\zeta f:=\bar f$ for all $f\in\h$ and define $J\in\bar\mL(\hh)$ by
\ba
\label{J}
J
:=\zeta\sigma_1.
\ea
\en
\ed

\br
\label{rem:AJW}
In order to establish a bridge between the spin picture and the fermionic picture \eqref{A-CAR}, the generalization from \cite{Ar84} of the Jordan-Wigner transformation for 1-dimensional systems whose configuration space extends infinitely in both directions makes use of the so-called crossed product of the algebra of the Pauli spins over $\Z$ (a Glimm or UHF [\ie, uniformly hyperfinite] algebra as is $\fA$) by the involutive automorphism describing the rotation around the 3-axis by an angle of $\pi$ of the observables on the nonpositive sites (and, thereby, makes it mathematically rigorous for the Jordan-Wigner transformation to be anchored at minus infinity, see \cite{As23} for a detailed and rigorous construction). Using this bridge, \eqref{XYDensity} can be expressed in the fermionic picture and becomes (up to a global prefactor)
\ba
\label{FDensity}
a_x^\ast a_{x+1} + a_{x+1}^\ast a_x
+\gamma (a_x^\ast a_{x+1}^\ast +a_{x+1}a_x).
\ea
In order to treat the anisotropic case $\gamma\neq 0$, i.e., the case in which there is an asymmetry between the first and the second term in \eqref{XYDensity}, the so-called selfdual setting (developed in \cite{Ar68, Ar71, Ar87}) is most natural since gauge invariance is broken in \eqref{FDensity}. Hence, due to the presence of the $\gamma$-term, the Hamiltonian density acquires non-diagonal entries with respect to $\hh$ (see Example \ref{ex:XYNESS} in Section \ref{sec:R/L}). In many respects, the truly anisotropic XY model is substantially more complicated than the isotropic one.
\er

\br
Let $C\in\fA$ satisfy $C^2=0$ and $\{C,C^\ast\}=1$. Then, we have $(C^\ast C)^2=C^\ast(1-C^\ast C)C=C^\ast C$ and, using (three times) the so-called \Cs property 
(\ie, the property that $\|A^\ast A\|=\|A\|^2$ for any $A$ in any \Cs algebra), we get $\|C\|^4=\|C^\ast C\|^2=\|(C^\ast C)^2\|=\|C^\ast C\|=\|C\|^2$ (where $\|\cdot\|$ stands for the \Cs norm of $\fA$), \ie, $\|C\|=1$. As for \eqref{c(f)}, we first set $S_n:=\sum_{x\in\num{-n}{n}}\bar f(x)\hspace{0.2mm}c_x\in\fA$ for all $f\in\h$ (with $f\neq 0$) and all $n\in\N$. Then, defining the element $C_{n,m}\in\fA$ for all $n, m\in\N$ with $n>m$ (sufficiently large) by
\ba
C_{n,m}
:=\frac{S_n-S_m}{\|1_{\{x\in\Z\,|\,m+1\le|x|\le n\}}f\|_\h},
\ea
we compute that $C_{n,m}^2=0$ and  $\{C_{n,m}, C_{n,m}^\ast\}=1$ (due to \eqref{dCAR1}-\eqref{dCAR2}). Hence, we can write  $\|S_n-S_m\|=\|1_{\{x\in\Z\,|\,m+1\le|x|\le n\}}f\|_\h$ for all $n, m\in\N$ with $n>m$ and the limit in \eqref{c(f)} indeed exists. Moreover, for all $f,g\in\h$,  \eqref{dCAR1}-\eqref{dCAR2} also imply
\ba
\label{CAR1}
\{c(f), c(g)\}
&=0,\\
\label{CAR2}
\{c(f), c^\ast(g)\}
&=(f,g)_\h1,
\ea
and, using \eqref{B(F)} and \eqref{CAR1}-\eqref{CAR2}, we get, for all $F,G\in\hh$,
\ba
\label{SDCAR1}
B^\ast(F)
&=B(J  F),\\
\label{SDCAR2}
\{B^\ast(F),B(G)\}
&=(F,G)_\hh1.
\ea
Equation \eqref{SDCAR2} is sometimes called the selfdual CAR.
\er

The class of so-called 2-point operators is central for the following.

\bd[2-point operator]
\label{def:2pt-op}
An operator $R\in\mL(\hh)$ having the properties
\ba
\label{def:2pt-op-1}
R^\ast
&=R,\\
\label{def:2pt-op-2}
JRJ
&=1-R,\\
\label{def:2pt-op-3}
0
\le 
&\hspace{1.5mm}R
\le 1,
\ea
is called a 2-point operator.
\ed

\br
Since, by definition, a state $\omega$ is a normalized positive linear functional on $\fA$, we have $|\omega(B^\ast(F)B(G))|\le \|F\|_\hh\|G\|_\hh$ for all $F,G\in\hh$ because $\|B(F)\|\le\|F\|_\hh$ for all $F\in\hh$ due to the fact (proven in \cite{Ar87}) that, for all $F\in\hh$, 
\ba
\label{ArakiF}
\|B(F)\|=\frac{1}{\sqrt{2}}\sqrt{\|F\|_\hh^2+\sqrt{\|F\|_\hh^4-|(F,JF)_\hh|^2}},
\ea
\ie, the map $\hh\times\hh\ni [F,G]\mapsto \omega(B^\ast(F)B(G))\in\C$ is a bounded sesquilinear form. Hence, the Riesz representation theorem guarantees the existence of a unique $R\in\mL(\hh)$ such that, for all $F, G\in\hh$, 
\ba
\label{2pt-op}
\omega(B^\ast(F)B(G))
=(F,RG)_\hh.
\ea
Moreover, due to the positivity of $\omega$, we get $R\ge 0$ and, hence, $R^\ast=R$. Since $\omega$ is normalized, \eqref{SDCAR1}-\eqref{SDCAR2} yield $JRJ=1-R$. Finally, since $R\ge 0$, we have $J RJ\ge 0$, \ie, $1-R\ge 0$, and it follows that the operator $R$ which characterizes the 2-point function in \eqref{2pt-op} is a 2-point operator.
\er

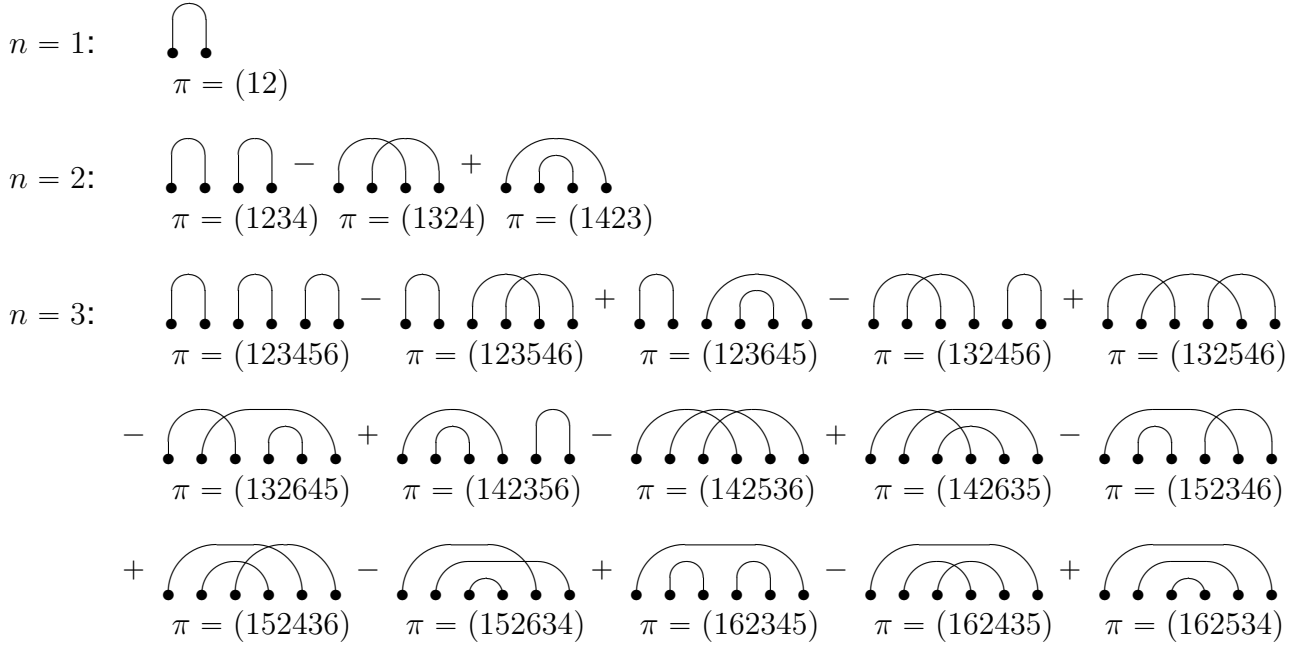
\begin{figure}
\setlength{\unitlength}{8.8mm}
$n=1$:
\begin{picture}(2,2)
\multiput(1,0)(0.5,0){2}{\circle*{0.15}}
\put(1.25,0){\oval(0.5,1.5)[t]}
\put(1,-0.6){$\pi=(12)$}
\end{picture}

$n=2$:
\begin{picture}(22,2)
\multiput(1,0)(0.5,0){4}{\circle*{0.15}}
\put(1.25,0){\oval(0.5,1.5)[t]}
\put(2.25,0){\oval(0.5,1.5)[t]}
\put(1,-0.6){$\pi=(1234)$}
\put(2.8,0.25){$-$}
\multiput(3.5,0)(0.5,0){4}{\circle*{0.15}}
\put(4,0){\oval(1,1.5)[t]}
\put(4.5,0){\oval(1,1.5)[t]}
\put(3.5,-0.6){$\pi=(1324)$}
\put(5.3,0.25){$+$}
\multiput(6,0)(0.5,0){4}{\circle*{0.15}}
\put(6.75,0){\oval(1.5,1.5)[t]}
\put(6.75,0){\oval(0.5,1.0)[t]}
\put(6,-0.6){$\pi=(1423)$}
\end{picture}

$n=3$:
\begin{picture}(22,2)
\multiput(1,0)(0.5,0){6}{\circle*{0.15}}
\put(1.25,0){\oval(0.5,1.5)[t]}
\put(2.25,0){\oval(0.5,1.5)[t]}
\put(3.25,0){\oval(0.5,1.5)[t]}
\put(1,-0.6){$\pi=(123456)$}
\put(3.8,0.25){$-$}
\multiput(4.5,0)(0.5,0){6}{\circle*{0.15}}
\put(4.75,0){\oval(0.5,1.5)[t]}
\put(6,0){\oval(1,1.5)[t]}
\put(6.5,0){\oval(1,1.5)[t]}
\put(4.5,-0.6){$\pi=(123546)$}
\put(7.3,0.25){$+$}
\multiput(8,0)(0.5,0){6}{\circle*{0.15}}
\put(8.25,0){\oval(0.5,1.5)[t]}
\put(9.75,0){\oval(1.5,1.5)[t]}
\put(9.75,0){\oval(0.5,1.0)[t]}
\put(8,-0.6){$\pi=(123645)$}
\put(10.8,0.25){$-$}
\multiput(11.5,0)(0.5,0){6}{\circle*{0.15}}
\put(12,0){\oval(1,1.5)[t]}
\put(12.5,0){\oval(1,1.5)[t]}
\put(13.75,0){\oval(0.5,1.5)[t]}
\put(11.5,-0.6){$\pi=(132456)$}
\put(14.3,0.25){$+$}
\multiput(15,0)(0.5,0){6}{\circle*{0.15}}
\put(15.5,0){\oval(1,1.5)[t]}
\put(16.25,0){\oval(1.5,1.5)[t]}
\put(17,0){\oval(1,1.5)[t]}
\put(15,-0.6){$\pi=(132546)$}
\end{picture}
\phantom{$n=3$:}
\begin{picture}(22,2)
\put(0.3,0.25){$-$}
\multiput(1,0)(0.5,0){6}{\circle*{0.15}}
\put(1.5,0){\oval(1,1.5)[t]}
\put(2.5,0){\oval(2,1.5)[t]}
\put(2.75,0){\oval(0.5,1)[t]}
\put(1.05,-0.6){$\pi=(132645)$}
\put(3.8,0.25){$+$}
\multiput(4.5,0)(0.5,0){6}{\circle*{0.15}}
\put(5.25,0){\oval(1.5,1.5)[t]}
\put(5.25,0){\oval(0.5,1)[t]}
\put(6.75,0){\oval(0.5,1.5)[t]}
\put(4.5,-0.6){$\pi=(142356)$}
\put(7.3,0.25){$-$}
\multiput(8,0)(0.5,0){6}{\circle*{0.15}}
\put(8.75,0){\oval(1.5,1.5)[t]}
\put(9.25,0){\oval(1.5,1.5)[t]}
\put(9.75,0){\oval(1.5,1.5)[t]}
\put(8,-0.6){$\pi=(142536)$}
\put(10.8,0.25){$+$}
\multiput(11.5,0)(0.5,0){6}{\circle*{0.15}}
\put(12.25,0){\oval(1.5,1.5)[t]}
\put(13,0){\oval(2,1.5)[t]}
\put(13,0){\oval(1,1)[t]}
\put(11.55,-0.6){$\pi=(142635)$}
\put(14.3,0.25){$-$}
\multiput(15,0)(0.5,0){6}{\circle*{0.15}}
\put(16,0){\oval(2,1.5)[t]}
\put(15.75,0){\oval(0.5,1)[t]}
\put(17,0){\oval(1,1.5)[t]}
\put(15,-0.6){$\pi=(152346)$}
\end{picture}
\phantom{$n=3$:}
\begin{picture}(22,2)
\put(0.3,0.25){$+$}
\multiput(1,0)(0.5,0){6}{\circle*{0.15}}
\put(2,0){\oval(2,1.5)[t]}
\put(2,0){\oval(1,1)[t]}
\put(2.75,0){\oval(1.5,1.5)[t]}
\put(1.05,-0.6){$\pi=(152436)$}
\put(3.8,0.25){$-$}
\multiput(4.5,0)(0.5,0){6}{\circle*{0.15}}
\put(5.5,0){\oval(2,1.5)[t]}
\put(6,0){\oval(2,1)[t]}
\put(5.75,0){\oval(0.5,0.5)[t]}
\put(4.55,-0.6){$\pi=(152634)$}
\put(7.3,0.25){$+$}
\multiput(8,0)(0.5,0){6}{\circle*{0.15}}
\put(9.25,0){\oval(2.5,1.5)[t]}
\put(8.75,0){\oval(0.5,1)[t]}
\put(9.75,0){\oval(0.5,1)[t]}
\put(8.05,-0.6){$\pi=(162345)$}
\put(10.8,0.25){$-$}
\multiput(11.5,0)(0.5,0){6}{\circle*{0.15}}
\put(12.75,0){\oval(2.5,1.5)[t]}
\put(12.5,0){\oval(1,1)[t]}
\put(13,0){\oval(1,1)[t]}
\put(11.55,-0.6){$\pi=(162435)$}
\put(14.3,0.25){$+$}
\multiput(15,0)(0.5,0){6}{\circle*{0.15}}
\put(16.25,0){\oval(2.5,1.5)[t]}
\put(16.25,0){\oval(1.5,1)[t]}
\put(16.25,0){\oval(0.5,0.5)[t]}
\put(15,-0.6){$\pi=(162534)$}
\end{picture}

\vspace{5mm}
\caption{All the pairings for $n=1$, $n=2$, and $n=3$. The total number of intersections $\#$ in each graph relates to the sign of the permutation $\pi$ as $\sgn(\pi)=(-1)^\#$.}
\label{fig:pairings}
\end{figure}

In the following, $\mE_\mA$ stands for the convex set of all states on any  \Cs algebra $\mA$. We next introduce the class of so-called quasifree states on $\fA$, \ie, the elements of $\mE_\fA$ whose many-point correlation functions factorize in Pfaffian form. Recall that, for all $n\in\N$, the Pfaffian $\pf:\M{2n}\to\C$ is defined, for all $X:=[X_{ij}]_{i,j\in\num{1}{2n}}\in\M{2n}$, by
\ba
\label{pfaff}
\pf(X)
:=\sum_{\pi\in\mP_{2n}}\sgn(\pi) \prod_{i\in\num{1}{n}} X_{\pi(2i-1)\hspace{0.2mm}\pi(2i)},
\ea
where the sum is running over all the $(2n-1)!!=(2n)!/(2^{n}n!)$ pairings of the set $\num{1}{2n}$ (see Figure \ref{fig:pairings}), \ie, 
\ba
\mP_{2n}
:=\{\pi\in\mS_{2n}\,|\, 
&\pi(2i-1)<\pi(2i+1)\mbox{ for all } i\in\num{1}{n-1},\nonumber\\
&\pi(2i-1)<\pi(2i) \mbox{ for all } i\in\num{1}{n}\},
\ea
and $\mS_{2n}$ stands for the group of all permutations of the set $\num{1}{2n}$ and $\sgn(\pi)$ for the sign of the permutation $\pi\in\mS_{2n}$.

\bd[Quasifree state]
\label{def:qfs}
\hspace{0mm}
Let $\omega\in \mE_\fA$ be a state with 2-point operator $R\in\mL(\hh)$. If, for all $n\in\N$ and all $\{F_i\}_{i\in\num{1}{n}}\subseteq\hh$,
\ba
\label{qfs}
\omega\hspace{-0.5mm}\left(\prod\nolimits_{i\in\num{1}{n}} B(F_i)\right)
=\begin{cases}
\pf([(J F_i, RF_j)]_{i,j\in\num{1}{n}}), & n\mbox{ even},\\
\hfill 0, & n\mbox{ odd},\\
\end{cases}
\ea
the state  $\omega$ is called a quasifree state on $\fA$ (induced by the 2-point operator $R$). The subset of $\mE_\fA$ of all quasifree states on $\fA$ is denoted by $\mQ_\fA$. 
\ed

In the following, the elements of the so-called (orthonormal) Kronecker basis $\{\delta_x\}_{x\in\Z}$ of $\h$ are given by $\delta_x(y):=\delta_{xy}$ for all $x,y\in\Z$. Moreover, for all $f\in\h$, the support of $f$ is defined by $\supp(f):=\{x\in\Z\,|\,f(x)\neq 0\}$.

We now define what we call a Fermi family.

\bd[Fermi family]
\label{def:Fs}
Let $\Lm\in\Fin(\Z)$ with $\nu:=\card(\Lm)$. A family $\{Q_i\}_{i\in\num{1}{\nu}}\subseteq\hh$ is called a Fermi family over $\Lm$ if it satisfies the following properties:
\bn[label=(\alph*), ref={\it (\alph*)}]
\setlength{\itemsep}{0mm}
\item 
\label{Fs-a}
$(Q_i,Q_j)_\hh=\delta_{ij}$ for all $i,j\in\num{1}{\nu}$

\item 
\label{Fs-b}
$(Q_i,JQ_j)_\hh=0$ for all $i,j\in\num{1}{\nu}$

\item 
\label{Fs-c}
$\supp((Q_i)_\alpha)\subseteq\Lm$ for all $i\in\num{1}{\nu}$ and all $\alpha\in\num{1}{2}$
\en
\ed

\bx
\label{rem:sFs}
If $\{Q_i\}_{i\in\num{1}{\nu}}\subseteq\hh$ is a Fermi family over $\Lm$, then, for all $i\in\num{1}{\nu}$, all $\alpha\in\num{1}{2}$, and all $x\in\Lm$, there exist $\lm_{i\alpha,x}\in\C$ such that $(Q_i)_\alpha=\sum_{x\in\Lm} \lm_{i\alpha,x}\delta_x$. Hence, Definition \ref{def:Fs} \ref{Fs-a} and \ref{Fs-b} respectively become, for all $i,j\in\num{1}{n}$,
\ba
\sum_{\alpha\in\num{1}{2}}\sum_{x\in\Lm} {\bar \lm_{i\alpha,x}}\lm_{j\alpha,x}
&=\delta_{ij},\\
\sum_{\alpha\in\num{1}{2}}\sum_{x\in\Lm} \lm_{i\alpha,x}\lm_{j3-\alpha,x}
&=0.
\ea
In particular, if $\Lm=\{x_1, \ldots, x_\nu\}\in\Fin(\Z)$ for some $\nu\in\N$ is such that $x_1<\ldots<x_\nu$ if $\nu\ge 2$, the family $\{Q_i\}_{i\in\num{1}{\nu}}\subseteq\hh$, defined, for all $i\in\num{1}{\nu}$, by
\ba
\label{sFs}
Q_i:=\delta_{x_i}\oplus 0,
\ea
is a Fermi family over $\Lm$ which we call the standard Fermi family.
\ex

\br
The conditions from Definition \ref{def:Fs} \ref{Fs-a}-\ref{Fs-b} are equivalent to the fact that, for all $\alpha,\beta\in\num{1}{2}$ and all $i,j\in\num{1}{\nu}$, 
\ba
\label{Fs-d}
(J^\alpha Q_i, J^\beta Q_j)_\hh
=\delta_{\alpha\beta}\delta_{ij}.
\ea
Moreover, using Definition \ref{def:Fs} \ref{Fs-a}-\ref{Fs-b} and \eqref{SDCAR1}-\eqref{SDCAR2}, we get \eqref{dCAR1}-\eqref{dCAR2}, \ie,  for all $i,j\in\num{1}{\nu}$,
\ba
\label{CAR-1}
\{B(Q_i), B(Q_j)\}
&=0,\\
\label{CAR-2}
\{B^\ast(Q_i), B(Q_j)\}
&=\delta_{ij}1,
\ea
and we note (using $J^{\alpha+1}=J^{3-\alpha}$ and $\delta_{3-\alpha\,\beta}=1-\delta_{\alpha\beta}$ for all $\alpha,\beta\in\num{1}{2}$) that \eqref{CAR-1}-\eqref{CAR-2} are equivalent to the fact that, for all $i,j\in\num{1}{\nu}$ and all $\alpha,\beta\in\num{1}{2}$,
\ba
\label{CAR}
\{B(J^\alpha Q_i), B(J^\beta Q_j)\}
=(1-\delta_{\alpha\beta})\delta_{ij}1.
\ea
\er

In the following, if $\Lm\in\Fin(\Z)$, a polynomial in $\fA$ generated by the set $\{a_x\,|\, x\in\Lm\}$ is an arbitrary finite linear combination of arbitrary finite products of elements in that set (see \cite{As23}).

We next specify the local observable algebras as follows.

\bp[Local observable algebras]
\label{prop:LocObs}
Let $\Lm\in\Fin(\Z)$ and set $\fB_\Lm:=\{a_x\,|\,x\in\Lm\}$. Then, the local observable algebra, defined by
\ba
\label{ALm}
\fA_\Lm
:=\hspace{-5mm}\bigcap_{\substack{\fB:\,\textup{\Cs subalgebra of }\fA\\\fB_\Lm\subseteq\fB}}
\hspace{-2mm}\fB,
\ea
is a unital \Cs subalgebras of $\fA$ which equals the set of all polynomials in $\fA$ generated by $\fB_\Lm$. Moreover, the net $(\fA_\Lm)_{\Lm\in\Fin(\Z)}$ is increasing and the closure of the union of all its elements equals $\fA$.
\ep

\bprf
See \cite{As23}.
\eprf

In the following, for any \Cs algebra $\mA$ and any $n\in\N$, we call $\{e_{ab}\}_{a,b\in\num{1}{2}^n}\subseteq\mA$ an $n$-dimensional family of $2\times 2$ matrix units in $\mA$  (see \cite{Gl60} for example) if, for all $a,b,c,d\in\num{1}{2}^n$, 
\ba
\label{MUnt-1}
e_{ab}e_{cd}
&=\delta_{bc} e_{ad},\\
\label{MUnt-2}
e_{ab}^\ast
&=e_{ba},\\
\sum_{a\in\num{1}{2}^n }e_{aa}
\label{MUnt-3}
&=1,
\ea
where, for all $m\in\N$ and all $r:=[r_i]_{i\in\num{1}{m}}, s:=[s_i]_{i\in\num{1}{m}}\in\R^m$, we set $\delta_{rs}:=\prod_{i\in\num{1}{m}}\delta_{r_is_i}$.

\bl[Matrix units]
\label{lem:mu}
Let $\Lm\in\Fin(\Z)$ with $\nu:=\card(\Lm)$ and let $\{Q_i\}_{i\in\num{1}{\nu}}\subseteq\hh$ be a Fermi family over $\Lm$. For all $i\in\num{1}{\nu}$ and all $\alpha,\beta\in\num{1}{2}$, we define $e_{\alpha\beta}^{(i)}\in\fA_\Lm$ by
\ba
\label{eiab}
e_{\alpha\beta}^{(i)}
:=\begin{cases}
B^\ast(Q_i)B(Q_i), & (\alpha,\beta)=(1,1),\\
\hfill S_i B^\ast(Q_i), & (\alpha,\beta)=(1,2),\\
\hfill S_i B(Q_i), & (\alpha,\beta)=(2,1),\\
B(Q_i)B^\ast(Q_i), & (\alpha,\beta)=(2,2),
\end{cases}
\ea
where, for all $i\in\num{1}{\nu}$, the nonlocal multiplicator element $S_i\in\fA_\Lm$ is given, if $\nu\ge 2$, by
\ba
\label{Si}
S_i
:=\begin{cases}
\hfill 1, & i=1,\\
\prod\nolimits_{j\in\num{1}{i-1}}(2B^\ast(Q_j)B(Q_j)-1), & i\in\num{2}{\nu},
\end{cases}
\ea
and by $S_1:=1$ if $\nu=1$.
Moreover, for all $n\in\num{1}{\nu}$ and all $a:=[\alpha_i]_{i\in\num{1}{n}}, b:=[\beta_i]_{i\in\num{1}{n}}\in\num{1}{2}^n$, we define $e_{ab}\in\fA_\Lm$ by
\ba
\label{eab}
e_{ab}
:=\prod_{i\in\num{1}{n}} e_{\alpha_i\beta_i}^{(i)}.
\ea
Then:
\bn[label=(\alph*), ref={\it (\alph*)}]
\setlength{\itemsep}{0mm}
\item 
\label{mu-a}
For all $i\in\num{1}{\nu}$, the family $\{e_{\alpha\beta}^{(i)}\}_{\alpha,\beta\in\num{1}{2}}$ is a $1$-dimensional family of $2\times 2$ matrix units in $\fA_\Lm$, \ie, for all $i\in\num{1}{\nu}$ and all $\alpha,\beta,\gamma,\delta\in\num{1}{2}$, 

\ba
\label{MUi-1}
e^{(i)}_{\alpha\beta} e^{(i)}_{\gamma\delta}
&=\delta_{\beta\gamma} e^{(i)}_{\alpha\delta},\\
\label{MUi-2}
\big(e^{(i)}_{\alpha\beta}\big)^\ast
&=e^{(i)}_{\beta\alpha},\\
\label{MUi-3}
\sum_{\alpha\in\num{1}{2}}e^{(i)}_{\alpha\alpha}
&=1.
\ea
Moreover, for all $i,j\in\num{1}{\nu}$ with $i\neq j$ and all $\alpha,\beta,\gamma,\delta\in\num{1}{2}$, 
\ba
\label{eCom}
e_{\alpha\beta}^{(i)}e_{\gamma\delta}^{(j)}
=e_{\gamma\delta}^{(j)}e_{\alpha\beta}^{(i)}.
\ea

\item 
\label{mu-b}
For all $n\in\num{1}{\nu}$, the family $\{e_{ab}\}_{a,b\in\num{1}{2}^n}$ is a $n$-dimensional family of $2\times 2$ matrix units in $\fA_\Lm$, \ie, for all $a,b,c,d\in\num{1}{2}^n$,
\ba
\label{MUmu-1}
e_{ab} e_{cd}
&=\delta_{bc}e_{ad},\\
\label{MUmu-2}
e_{ab}^\ast
&=e_{ba},\\
\label{MUmu-3}
\sum_{a\in\num{1}{2}^n}e_{aa}
&=1.
\ea

\item 
\label{mu-c}
$\{e_{ab}\}_{a,b\in\num{1}{2}^\nu}$ is a basis of $\fA_\Lm$.
\en
\el

\bprf
See section \ref{sec:Proofs}.
\eprf

\section{Reduced density matrix}
\label{sec:RDM}

In this section, we determine the spectrum of the so-called reduced density matrix which characterizes the local state associated with any quasifree state $\omega\in\mQ_\fA$,  \ie, by definition, the restriction of $\omega$ to the local observable algebra $\fA_\Lm$ over the finite configuration space $\Lm\in\Fin(\Z)$. In order to do so, we make use of the strategy of \cite{ViLaRiKi03}. 

In the following, for any \str algebras $\mA$ and $\mB$, we denote by $\sHom(\mA,\mB)$, $\sIso(\mA,\mB)$, and $\sAut(\mA)$ the set of all \str homomorphisms between $\mA$ and $\mB$, the set of all \str isomorphisms between $\mA$ and $\mB$, and set of all \str automorphisms on $\mA$, respectively. Moreover, we recall from \cite{As23} that, for all $\Lm\in\Fin(\Z)$ with $\nu:=\card(\Lm)$, there exists
\ba
\label{piLm-1}
\pi_\Lm
\in\sIso(\fA_\Lm,\M{2^\nu})
\ea
which satisfies,  for all $a:=[\alpha_i]_{i\in\num{1}{\nu}}, b:=[\beta_i]_{i\in\num{1}{\nu}}\in\num{1}{2}^\nu$,
\ba
\label{piLm-2}
\pi_\Lm(e_{ab})
=E_{\alpha_1\beta_1}\oslash\ldots \oslash E_{\alpha_\nu\beta_\nu}.
\ea
Here, $\{E_{\alpha\beta}\}_{\alpha,\beta\in\num{1}{2}}$ is the $1$-dimensional family of $2\times 2$ matrix units in the \Cs algebra $\M{2}$ defined by $(E_{\alpha\beta})_{ij}:=\delta_{\alpha i}\delta_{\beta j}$ for all $\alpha, \beta, i, j\in\num{1}{2}$ (\ie, the standard basis of $\M{2}$), where we recall that, for all $n\in\N$, the set $\M{n}$ equipped with the usual matrix addition, scalar multiplication, matrix multiplication, the involution which associates to any matrix its conjugate transpose, and with the norm $\|\cdot\|_{\mL(\C^n)}$ (with respect to the usual complex Euclidean scalar product $(\cdot,\cdot)_{\C^n}$ on $\C^n$, see Appendix \ref{app:Toep}) is a unital \Cs algebra with identity $1_n:=\diag[1]_{i\in\num{1}{n}}$ and $\diag[\lm_i]_{i\in\num{1}{n}}\in\M{n}$ stands for the diagonal matrix with diagonal entries $\{\lm_i\}_{i\in\num{1}{n}}\subseteq\C$. Moreover, for all $n,m\in\N$, all $X=[X_{ij}]_{i,j\in\num{1}{n}}\in\M{n}$, and all $Y=[Y_{kl}]_{k,l\in\num{1}{m}}\in\M{m}$, the Kronecker product $X\oslash Y\in\M{nm}$ is defined by  $(X\oslash Y)_{k+(i-1)m, l+(j-1)m}:=X_{ij}Y_{kl}$ for all $i,j\in\num{1}{n}$ and all $k,l\in\num{1}{m}$.

Furthermore, for all $n\in\N$ and all $X\in\M{n}$,  we write $X\ge 0$ if $\tr(X Y^\ast Y)\ge 0$ for all $Y\in\M{n}$ (which is equivalent to $(v,Xv)_{\C^n}\ge 0$ for all $v\in\C^n$) and $\tr(X):=\sum_{i\in\num{1}{n}}X_{ii}$ stands for the usual trace of $X=[X_{ij}]_{i,j\in\num{1}{n}}\in\M{n}$. The Frobenius scalar product on $\M{n}$ is denoted by $(X,Y)_F:=\tr(X^\ast Y)$ for all $X, Y\in\M{n}$ and the corresponding norm by $\|\cdot\|_F$.

Finally, for any sets $A, B$, any function $f:\dom(f)\subseteq A\to B$ with domain $\dom(f)$, and any subset $X\subseteq\dom(f)$, the function $\io_X: X\hookrightarrow\dom(f)$ in the composition $f\circ\io_X$ is the natural injection (inclusion) given by $\io_X(a):=a$ for all $a\in X$ (the arrow with a hook always stands for the corresponding canonical inclusion maps).

\bp[Reduced density matrix]
\label{prop:ls}
Let $\omega\in\mE_\fA$, let $\Lm\in\Fin(\Z)$ with $\nu:=\card(\Lm)$, and set $\omega_\Lm:=\omega\circ\iota_{\fA_\Lm}$. Then, $\omega_\Lm\in\mE_{\fA_\Lm}$ and:
\bn[label=(\alph*), ref={\it (\alph*)}]
\setlength{\itemsep}{0mm}
\item 
\label{ls-a}
There exists a unique $R_\Lm\in\M{2^\nu}$ such that, for all $A\in\fA_\Lm$, 
\ba
\label{omloc}
\omega_\Lm(A)
=\tr(R_\Lm\pi_\Lm(A)).
\ea
The matrix  $R_\Lm$ is called the reduced density matrix (of $\omega$ over $\Lm$).

\item
\label{ls-b}
The reduced density matrix satisfies $R_\Lm\ge 0$ and $\tr(R_\Lm)=1$. It has the expansion
\ba
\label{RLm}
R_\Lm
=\sum_{a,b\in\num{1}{2}^\nu} \omega_\Lm(e_{ba})\, \pi_\Lm(e_{ab}).
\ea
\en
\ep

\bprf
See section \ref{sec:Proofs}.
\eprf

The main assumption which is used in the sequel is the following. 
 
\bass[Diagonal correlations]
\label{ass:2pt}
Let $\omega\in\mE_\fA$ and $\Lm\in\Fin(\Z)$ with $\nu:=\card(\Lm)$. Moreover, let $\{Q_i\}_{i\in\num{1}{\nu}}\subseteq\hh$ be a Fermi family over $\Lm$.
\bn[label=(\alph*), ref={\it (\alph*)}]
\setlength{\itemsep}{0mm}
\item 
\label{2pt-VV}
$\omega_\Lm(B(Q_i)B(Q_j))=0$ for all $i,j\in\num{1}{\nu}$
\item 
\label{2pt-EV}
$\omega_\Lm(B^\ast(Q_i)B(Q_j))=\delta_{ij}\,\omega_\Lm(B^\ast(Q_i)B(Q_i))$ for all $i,j\in\num{1}{\nu}$
\en
\eass

\br
\An{\ref{ass:2pt} \ref{2pt-VV},\ref{2pt-EV}} are equivalent to the fact that, for all $i,j\in\num{1}{\nu}$ and all $\alpha,\beta\in\num{1}{2}$, 
\ba
\label{2pt-VVEV}
\omega_\Lm(B(J^\alpha Q_i)B(J^\beta Q_j))
=(1-\delta_{\alpha\beta})\delta_{ij}\,\omega_\Lm(B(J^\alpha Q_i)B(J^\beta Q_i)).
\ea
If $R\in\mL(\hh)$ is the 2-point operator of $\omega$, \eqref{2pt-VVEV} can be written,  for all $i,j\in\num{1}{\nu}$ and all $\alpha,\beta\in\num{1}{2}$, as 
\ba
(J^{\alpha+1}Q_i, R J^\beta Q_j)_\hh
=(1-\delta_{\alpha\beta})\delta_{ij}(J^{\alpha+1}Q_i, R J^\beta Q_j)_\hh,
\ea
where we used \eqref{SDCAR1} and \eqref{2pt-op}.  In particular, \An{\ref{ass:2pt} \ref{2pt-VV},\ref{2pt-EV}} read, for all $i,j\in\num{1}{\nu}$,
\ba
\label{QRJQ}
(Q_i,RJQ_j)_\hh
&=0,\\
\label{QRQ}
(Q_i,RQ_j)_\hh
&=\delta_{ij}(Q_i,RQ_i)_\hh.
\ea
\er

For the following, recall the notations from Lemma \ref{lem:mu}.

\bl[Factorization]
\label{lem:fact}
Let $\omega\in\mQ_\fA$, let $\Lm\in\Fin(\Z)$ with $\nu:=\card(\Lm)$, and let $\{Q_i\}_{i\in\num{1}{\nu}}\subseteq\hh$ be a Fermi family over $\Lm$. Then, if \An{\ref{ass:2pt} \ref{2pt-VV},\ref{2pt-EV}} hold, we have, for all $n\in\num{1}{\nu}$ and all $a:=[\alpha_i]_{i\in\num{1}{n}}, b\in\num{1}{2}^n$, 
\ba
\label{fact}
\omega_\Lm(e_{ab})
=\delta_{ab} \prod_{i\in\num{1}{n}}\omega_\Lm\big(e_{\alpha_i\alpha_i}^{(i)}\big).
\ea
\el

\bprf
See section \ref{sec:Proofs}.
\eprf

\br
\label{rem:tr}
If a 2-point operator $R\in\mL(\hh)$ is a complex multiple of the identity, it follows from \eqref{def:2pt-op-1}-\eqref{def:2pt-op-2} that $R=1/2$ (and \eqref{def:2pt-op-3} is automatically true). Moreover, if $\omega\in\mQ_\fA$ is induced by $R=1/2$, \eqref{QRJQ}-\eqref{QRQ} are satisfied for any Fermi family and \eqref{fact} holds. Moreover, we have,  for all $i\in\num{1}{\nu}$ and all $\alpha\in\num{1}{2}$,
\ba
\omega_\Lm(e_{\alpha\alpha}^{(i)})
&=\omega_\Lm(B(J^\alpha Q_i) B(J^{3-\alpha} Q_i))\nonumber\\
&=\frac12\hspace{0.5mm}\|J^{3-\alpha} Q_i\|_\hh^2\nonumber\\
&=\frac12.
\ea
Hence, \eqref{fact} yields $\omega_\Lm(e_{ba})=\delta_{ab}/2^\nu$ for all $a,b\in\num{1}{2}^\nu$ and, using \eqref{RLm} and \eqref{MUmu-3}, we get 
\ba
R_\Lm
&=\frac{1}{2^\nu}\hspace{-0.5mm}\sum_{a\in\num{1}{2}^\nu}\pi_\Lm(e_{aa})\nonumber\\
&=\frac{1_{2^\nu}}{2^\nu},
\ea
from which, with \eqref{omloc} (see also \eqref{tr-eaa} and before), for all $A\in\fA_\Lm$,
\ba
\omega_\Lm(A)
=\frac{1}{2^\nu}\,\tr(\pi_\Lm(A)).
\ea
\er

In the following, for any complex Hilbert spaces $\mH_1$ and $\mH_2$, we denote by $\G{U}(\mH_1,\mH_2)$ the set of all unitary operators from $\mH_1$ to $\mH_2$. If the two Hilbert spaces are equal to $\mH$, we simply write $\G{U}(\mH)$ (note that $\G{U}(\mH)$ is a group with respect to the composition of bounded linear operators). 

\bd[Bogoliubov transformations]
\label{def:Bog}
Any $U\in\G{U}(\hh)$ which commutes with $J$, \ie, which satisfies, in $\mL(\hh)$,
\ba
JUJ=U,
\ea
is called a Bogoliubov transformation. The subgroup of $\G{U}(\hh)$ of all Bogoliubov transformations is denoted by $\G{U}_{\!J}(\hh)$. 
\ed

In the following, for all $n\in\N$ and all $X\in\M{n}$, the (multi)set of all eigenvalues of $X$ (repeated according to their multiplicity) is denoted by $\spec(X)$.

The Majorana correlation matrix introduced below has as its entries the 2-point correlation functions of a so-called Majorana family. 

\bd[Majorana family]
\label{def:Ms}
Let $\Lm\in\Fin(\Z)$ with $\nu:=\card(\Lm)$ and let $\{Q_i\}_{i\in\num{1}{\nu}}\subseteq\hh$ be a Fermi family over $\Lm$. The family $\{M_i\}_{i\in\num{1}{2\nu}}\subseteq\hh$ defined, for all $i\in\num{1}{\nu}$, by
\ba
\label{Hodd}
M_{2i-1}
&:=Q_i+JQ_i,\\
\label{Heven}
M_{2i}
&:=\ii(Q_i-JQ_i),
\ea
is called the Majorana family (associated with the Fermi family $\{Q_i\}_{i\in\num{1}{\nu}}$).
\ed

\br
\label{rem:Ms}
Due to \eqref{J} and Definition \ref{def:Fs} \ref{Fs-a}-\ref{Fs-b}, the Majorana family $\{M_i\}_{i\in\num{1}{2\nu}}$ associated with the Fermi family $\{Q_i\}_{i\in\num{1}{\nu}}$ satisfies, for all $i,j\in\num{1}{2\nu}$,
\ba
\label{Ms-1}
JM_i
&=M_i,\\
\label{Ms-2}
(M_i,M_j)_\hh
&=2\delta_{ij}.
\ea
\er

In the following, for all $n\in\N$, any matrix $X\in\M{n}$ is called skew symmetric if $X^T=-X$ and recall that $\Im(X)=(X-\bar X)/(2\ii)$ for all $X\in\M{n}$. Moreover,  
\ba
\G{O}(n)
&:=\{X\in\R^{n\times n}\,|\, X^T X=1_n\},\\
\G{U}(n)
&:=\{X\in\C^{n\times n}\,|\, X^\ast X=1_n\},
\ea
stand for the orthogonal and unitary groups in $n$ dimensions. Finally, we denote by $\oplus$ the usual direct sum of matrices, by $V^\perp$ the orthogonal complement of any subspace $V\subseteq\mH$ of a Hilbert space $\mH$, and by $\spa$ the finite linear span of a set of vectors in $\mH$. 

The spectrum of the reduced density matrix which characterizes the localized state has the following properties.

\bp[Spectrum]
\label{prop:spec}
Let $\omega\in\mQ_\fA$, let $\Lm\in\Fin(\Z)$ with $\nu:=\card(\Lm)$, and let $\{Q_i\}_{i\in\num{1}{\nu}}\subseteq\hh$ be a Fermi family over $\Lm$.
Then:
\bn[label=(\alph*), ref={\it (\alph*)}]
\setlength{\itemsep}{0mm}
\item 
\label{spec-a}
If there exists $U\in\G{U}_{\!J}(\hh)$ such that the family $\{UQ_i\}_{i\in\num{1}{\nu}}\subseteq\hh$ is a Fermi family over $\Lm$ which satisfies \An{\ref{ass:2pt} \ref{2pt-VV},\ref{2pt-EV}}, we have
\ba
\label{specRLm}
\spec(R_\Lm)
=\Big\{\prod\nolimits_{i\in\num{1}{\nu}}\tfrac{1+(-1)^{\alpha_i}\lm_i}{2}\,\Big|\,[\alpha_i]_{i\in\num{1}{\nu}}\in\num{1}{2}^\nu\Big\},
\ea
where, for all $i\in\num{1}{\nu}$, the numbers $\lm_i\in [-1,1]$ are defined by 
\ba
\lm_i
:=1-2\omega_\Lm(B^\ast(UQ_i)B(UQ_i)).
\ea

\item 
\label{spec-b}
There exists $U\in\G{U}_{\!J}(\hh)$ such that the family $\{UQ_i\}_{i\in\num{1}{\nu}}\subseteq\hh$ is a Fermi family over $\Lm$ which satisfies \An{\ref{ass:2pt} \ref{2pt-VV},\ref{2pt-EV}}, \ie, \eqref{specRLm} holds. Moreover, we have 
\ba
\label{spec-b-1}
\spec(\Im(\Omega))
=\big\{\hspace{-1mm}\pm\hspace{-0.5mm}\tfrac{\ii}{2} \lm_i\big\}_{i\in\num{1}{\nu}},
\ea
where $\Omega:=[\Omega_{ij}]_{i,j\in\num{1}{2\nu}}\in\M{2\nu}$, called the Majorana correlation matrix, is defined, for all $i,j\in\num{1}{2\nu}$, by
\ba
\label{Omij}
\Omega_{ij}
:=\frac12\,\omega_\Lm(B(M_i)B(M_j)),
\ea
and $\{M_i\}_{i\in\num{1}{2\nu}}$ is the Majorana family associated with the Fermi family $\{Q_i\}_{i\in\num{1}{\nu}}$.
\en
\ep

\bprf
See section \ref{sec:Proofs}.
\eprf

\br
\label{rem:loc}
For all $U\in\G{U} _{\!J}(\hh)$ and all Fermi familys $\{Q_i\}_{i\in\num{1}{\nu}}\subseteq\hh$ over $\Lm$, the family $\{UQ_i\}_{i\in\num{1}{\nu}}$ satisfies Definition \ref{def:Fs} \ref{Fs-a}-\ref{Fs-b}. Hence, in order for $\{UQ_i\}_{i\in\num{1}{\nu}}\subseteq\hh$ to be a Fermi family over $\Lm$, we only have to check that, for all $i\in\num{1}{\nu}$ and all $\alpha\in\num{1}{2}$,
\ba
\supp((UQ_i)_\alpha)
\subseteq\Lm.
\ea
\er

\section{Von Neumann entropy}
\label{sec:Neu}

In this section, we introduce the von Neumann entropy and express it by means of the eigenvalues of the imaginary part of the Majorana correlation matrix from the preceding section (see again \cite{ViLaRiKi03}). Moreover, we study the functional equation which determines the Shannon entropy uniquely under the conditions of the present setting.

In the following, for any $D\subseteq\R$, we denote by $C(D)$ the set of continuous complex-valued functions on $D$, by  $C_b(D)$ the set of bounded continuous complex-valued functions on $D$, and by $C_0(D)$ the set of continuous complex-valued functions on $D$ with compact support. Moreover, for any open $D\subseteq\R$, we write $C^\infty\hspace{-0.4mm}(D)$ and $C^m\hspace{-0.4mm}(D)$ for any $m\in\N$ for the sets of smooth and $m$ times continuously differentiable complex-valued functions on $D$, respectively.

The following functions play an important role  in the sequel. 

\begin{figure}
\centering
\begin{tikzpicture}
\node (flux) {\includegraphics[width=80mm,height=50mm]{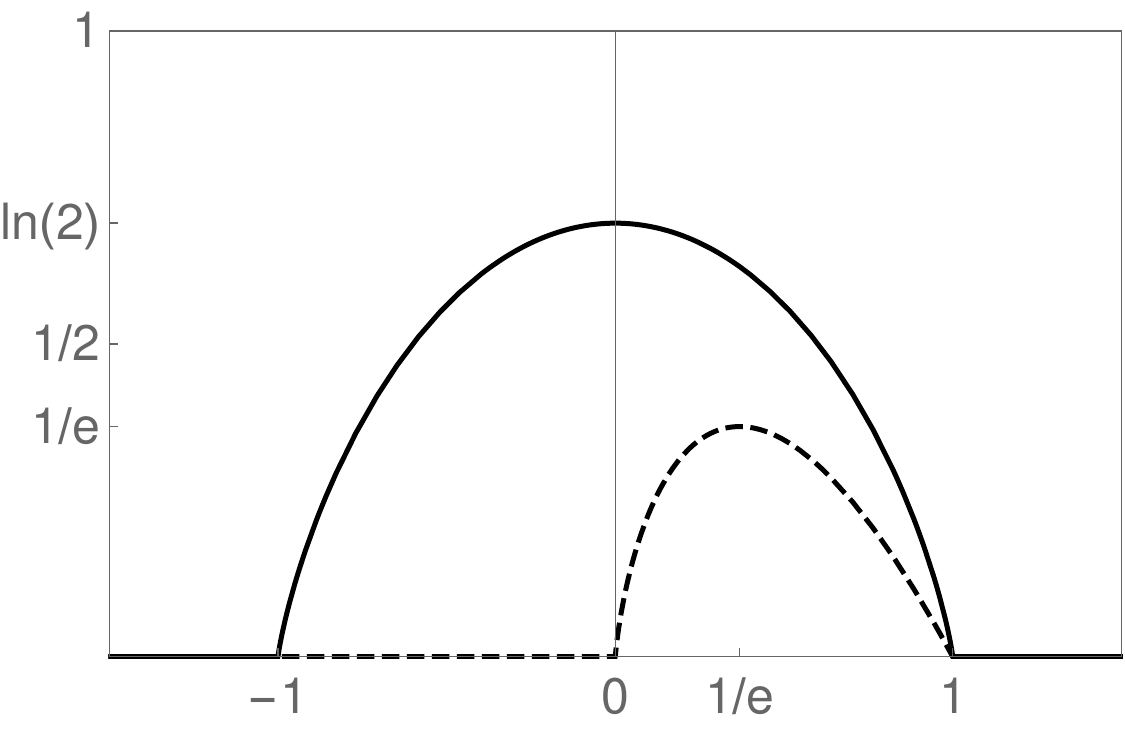}};
\end{tikzpicture}
\vspace{-3.5mm}
\caption{The functions $\wt\ell$ (dashed line) and $\wt\eta$ (continuous line) which are the extensions by zero to the whole of $\R$ of $\ell$ and $\eta$ from \eqref{Shan} and \eqref{eta}.}
\label{fig:Shan}
\end{figure}

\bd[Shannon entropy]
\label{def:Shan}
The function $\ell\in C^\infty\hspace{-0.4mm}(]0,1[)$, called the Shannon entropy, is defined, for all $x\in \,]0,1[$, by
\ba
\label{Shan}
\ell(x)
&:=-x\log(x).
\ea
Moreover, the function $\eta\in C^\infty\hspace{-0.4mm}(]-1,1[)$, sometimes called the binary entropy, is defined,  for all $x\in\,]-1,1[$, by 
\ba
\label{eta}
\eta(x)
:=\ell\bigg(\frac{1+x}{2}\bigg)+\ell\bigg(\frac{1-x}{2}\bigg),
\ea
see Figure \ref{fig:Shan}.
\ed

In the following, we denote by $\mM(\R)$ the $\sigma$-algebra of all Borel sets on $\R$ (see, for example, \cite{As21}), by $\mM_L(\R)\supseteq\mM(\R)$ the $\sigma$-algebra of all Lebesgue measurable sets on $\R$, and by  $\mM_L(]0,1[)$ the so-called trace 
of $\mM_L(\R)$ in $]0,1[\,\in\mM_L(\R)$ (\ie, $\mM_L(]0,1[)=\{M\subseteq\,]0,1[\,|\, M\in \mM_L(\R)\}$). Moreover, for all $n\in\N$, we again write $a:=[\alpha_i]_{i\in\num{1}{n}}\in\num{1}{2}^n$.
 
 The Shannon entropy has the following property.
 
\bp[Functional equation]
\label{prop:Shan}
\bn[label=(\alph*), ref={\it (\alph*)}]
\setlength{\itemsep}{0mm}
\item 
\label{Shan-a}
For all $n\in\N$ and all $\{\lm_i\}_{i\in\num{1}{n}}\subseteq\,]-1,1[$\,, 
\ba
\label{FEq-1}
\sum_{a\in\num{1}{2}^n}\ell\Big(\prod\nolimits_{i\in\num{1}{n}} \tfrac{1+(-1)^{\alpha_i}\lm_i}{2}\Big)
=\sum_{i\in\num{1}{n}} \eta(\lm_i).
\ea

\item 
\label{Shan-b}
Let $f\in\R^{]0,1[}$ be an $(\mM_L(]0,1[),\mM(\R))$-measurable function which satisfies, for all $n\in\N$ and all $\{\lm_i\}_{i\in\num{1}{n}}\subseteq\,]-1,1[$\,, 
\ba
\label{FEq-2}
\sum_{a\in\num{1}{2}^n}f\Big(\prod\nolimits_{i\in\num{1}{n}} \tfrac{1+(-1)^{\alpha_i}\lm_i}{2}\Big)
=\sum_{i\in\num{1}{n}} \sum_{\alpha_i\in\num{1}{2}}f\Big(\tfrac{1+(-1)^{\alpha_i}\lm_i}{2}\Big).
\ea
Then, there exists $C\in\R$ such that, for all $x\in\,]0,1[$\,,
\ba
\label{FEq-4}
f(x)
=Cx\log(x).
\ea
\en
 \ep
 
\bprf
See section \ref{sec:Proofs}.
\eprf
 
In the following, for all $n\in\N$, any matrix $X\in\M{n}$ is called selfadjoint if $X^\ast=X$. Note that, due to Proposition \ref{prop:ls} \ref{ls-b}, if $\Lm\in\Fin(\Z)$ with $\nu:=\card(\Lm)$, the reduced density matrix $R_\Lm\in\M{2^\nu}$ is selfadjoint. Hence, the spectral theorem yields a unique resolution of the identity associated with $R_\Lm$, 
\ba
E
\in\sHom\big(\mB(\R),\M{2^\nu}\big),
\ea
where $\mB(\R)$ stands for the normed unital \str algebra of all bounded Borel functions on $\R$ (see, for example, \cite{As21}). Moreover, we use the notation $f(R_\Lm):=E(f)$ for all $f\in\mB(\R)$. 

Finally, the extension $\wt\ell\in C_0(\R)$ of \eqref{Shan} is defined by $\wt\ell(x):=\ell(x)$ for all $x\in\,]0,1[$ and  $\wt\ell(x):=0$ for all $x\in\R\setminus\,]0,1[$ (see Figure \ref{fig:Shan}).

The von Neumann entropy of the state $\omega_\Lm$ from Proposition \ref{prop:ls} is defined as follows.

\bd[Von Neumann entropy]
\label{def:vNEnt}
Let $\omega\in\mE_\fA$, let $\Lm\in\Fin(\Z)$ with $\nu:=\card(\Lm)$, and let $R_\Lm\in\M{2^\nu}$ be the reduced density matrix of $\omega$ over $\Lm$. The von Neumann entropy $S_\Lm\in[0,\nu\log(2)]$ of $R_\Lm$ is defined by
\ba
\label{vNEnt}
S_\Lm
:=\tr(\wt\ell(R_\Lm)).
\ea
\ed

\br
\label{rem:vNEnt}
Since $\wt\ell\in C_0(\R)\subseteq C_b(\R)\subseteq\mB(\R)$, the right hand side of \eqref{vNEnt} is well-defined. Moreover, there exist $T\in\G{U}({2^\nu})$ and $\{\tau_i\}_{i\in\num{1}{2^\nu}}\subseteq\R$ such $T^\ast R_\Lm T=\diag[\tau_i]_{i\in\num{1}{2^\nu}}$ and, hence, due to the uniqueness of the resolution of the identity, we have $f(R_\Lm)
=T\diag[f(\tau_i)]_{i\in\num{1}{2^\nu}}T^\ast$ for all $f\in\mB(\R)$. Therefore, we get $\spec(R_\Lm)=\{\tau_i\}_{i\in\num{1}{2^\nu}}$ and the properties $R_\Lm\ge 0$ and $\tr(R_\Lm)=1$ imply $\tau_i\in[0,1]$ for all $i\in\num{1}{2^\nu}$. Finally, since \eqref{vNEnt} implies
 \ba
 \label{SLmSum}
 S_\Lm
=\sum_{i\in\num{1}{2^\nu}}\wt\ell(\tau_i),
 \ea
we get $S_\Lm\ge 0$.  Moreover, since $\wt\ell$ is concave, Jensen's inequality yields $\sum_{i\in\num{1}{2^\nu}}\wt\ell(\tau_i)\le 2^\nu\wt\ell(1/2^\nu)=\nu\log(2)$. 
\er

In the following, the extension $\wt\eta\in C_0(\R)$ of \eqref{eta} is defined by $\wt\eta(x):=\eta(x)$ for all $x\in\,]-1,1[$ and  $\wt\eta(x):=0$ for all $x\in\R\setminus\,]-1,1[$ (see Figure \ref{fig:Shan}) and note that
\ba
\label{OdEta}
\Od(\wt\eta)
=0,
\ea
where, for all $\chi\in\C^\R$, we set,  for all $x\in\R$, 
\ba
\Ev(\chi)(x)
&:=\frac{\chi(x)+\chi(-x)}{2},\\
\Od(\chi)(x)
&:=\frac{\chi(x)-\chi(-x)}{2}.
\ea

The von Neumann entropy is determined as follows by the spectrum of the imaginary part of the Majorana correlation matrix $\Omega\in\M{2\nu}$ from \eqref{Omij}.

\bp[Von Neumann entropy]
\label{prop:vNEnt}
Let $\omega\in\mQ_\fA$, let $\Lm\in\Fin(\Z)$ with $\nu:=\card(\Lm)$, and let $\{Q_i\}_{i\in\num{1}{\nu}}\subseteq\hh$ be a Fermi family over $\Lm$.
Then, 
\ba
\label{vNEnt-H}
S_\Lm
=\sum_{i\in\num{1}{\nu}}\wt\eta(\lm_i),
\ea
where $\spec(\Im(\Omega))=\{\pm\ii\lm_i/2\}_{i\in\num{1}{\nu}}$.
\ep

\bprf
See section \ref{sec:Proofs}.
\eprf

\section{Translation invariance}
\label{sec:trans}

In this section, we show that, if the quasifree state under consideration is translation invariant, the Majorana correlation matrix becomes the finite section of a block Toeplitz operator whose symbol can be explicitly computed (see Appendix \ref{app:Toep} for the basic definitions of the block Toeplitz operators used in the following). This fact opens the way for a rigorous analysis of the asymptotic behavior of the von Neumann entropy for the general R/L mover states of Section \ref{sec:R/L}.

\bd[Translation invariance]
\label{def:trans}
\bn[label=(\alph*), ref={\it (\alph*)}]
\setlength{\itemsep}{0mm}
\item 
\label{trans-aa}
The operator $\theta\in\mL(\h)$ defined by $(\theta f)(x):= f(x-1)$ for all $f\in\h$ and all $x\in\Z$ is called the (right) translation on $\h$. 

\item 
\label{trans-bb}
The state $\omega\in\mQ_\fA$ induced by the 2-point operator $R\in\mL(\hh)$ is called translation invariant if $[R,\theta 1_2]=0$.
\en
\ed

In the following, we resort to the usual unitary Fourier transform $\ff:\h\to\fh$ between the 1-particle position Hilbert space $\h$ over $\Z$ and the 1-particle momentum Hilbert space $\fh$ over $\T:=\{z\in\C\,|\,|z|=1\}$ defined by  
\ba
\label{L2}
\fh:=L^2(\T),
\ea
where $L^2(\T):=\{\vi\in\C^\T\,|\, \vi\circ\kappa\in L^2(\Pi)\}$. Here, the (non homeomorphic) bijection $\kappa\in\T^\Pi$ is defined by $\kappa:=\wt\kp\circ\io_\Pi$, where 
\ba
\Pi
:=\,]-\pi,\pi],
\ea
and where the surjection $\wt\kp\in\T^\R$ is given by $\wt\kp(k):=\ei$ for all $k\in\R$, see Figure \ref{fig:kappas} (an arrow with a tail represents an injection and a two-headed arrow a surjection).
Furthermore, $L^2(\Pi)$ denotes the space of all equivalence classes of functions $\vi\in\C^\Pi$ which are $(\mM(\Pi), \mM(\R))$-measurable (\ie, more precisely, for which both the real and imaginary part are $(\mM(\Pi), \mM(\R))$-measurable), where again $\mM(\Pi)$ stands for the trace of $\mM(\R)$ in $\Pi\in\mM(\R)$ (\ie, $\mM(\Pi)=\{M\subseteq\Pi\,|\, M\in \mM(\R)\}$) and for which $|\vi|^2$ is integrable with respect to the (non complete) Borel-Lebesgue measure denoted by $|\cdot|_B$. As usual, the equivalence relation identifies functions which coincide almost everywhere with respect to $|\cdot|_B$ (\ie, which coincide on the complement of a subset of a set of Borel-Lebesgue measure zero). The usual scalar product on $\fh$ is written as $(\vi,\psi)_\fh:=\int_{-\pi}^\pi\rd k/(2\pi)\,\bar\vi(\ei)\psi(\ei)$ for all $\vi, \psi\in\fh$. 
Moreover, $L^\infty(\Pi)$ denotes the space of all equivalence classes of functions $\vi\in\C^\Pi$ which are $(\mM(\Pi), \mM(\R))$-measurable and almost everywhere bounded with respect to $|\cdot|_B$ and we set $L^\infty(\T):=\{\vi\in\C^\T\,|\, \vi\circ\kappa\in L^\infty(\Pi)\}$. Also recall that, for all $f\in\h$, the Fourier transform is defined by the limit (in $\fh$) $\ff f:=\lim_{n\to\infty}\sum_{x\in\num{-n}{n}}f(x)\e_x$, where, for all $x\in\Z$, the plane wave functions $\e_x\in L^\infty(\T)\subseteq\fh$ are given by $\e_x(z):=z^x$ for all $z\in\T$. We write $\wh f:=\ff f\in\fh$ for all $f\in\h$ and $\wh a:=\ff a\ff^\ast\in\mL(\fh)$ for all $a\in\mL(\h)$, where  the adjoint $\ff^\ast$ is given by  $(\ff^\ast\vi)(x)=\Inti\vi(\e^{\ii k})\e_{-x}(\e^{\ii k})$ for all $x\in\Z$, and we also use the notation $\wc\vi:=\ff^\ast\vi\in\h$ for all $\vi\in\fh$ and $\wc\pi:=\ff ^\ast\pi\ff\in\mL(\h)$ for all $\pi\in\mL(\fh)$. 
On the doubled Hilbert spaces, we define, analogously, $\wh F:=\ff^{\oplus 2}F\in\fhh$ for all $F\in\hh$, $\wh A:=\ff^{\oplus 2}A(\ff^\ast)^{\oplus 2}\in\mL(\fhh)$ for all $A\in\mL(\hh)$, $\wc\Phi:=(\ff^\ast)^{\oplus 2}\Phi\in\hh$ for all $\Phi\in\fhh$, and $\wc\Pi:=(\ff^\ast)^{\oplus 2}\Pi\ff^{\oplus 2}\in\mL(\hh)$ for all $\Pi\in\mL(\fhh)$. 

\begin{figure}
\centering
\begin{tikzcd}
\Pi
\arrow[r, tail, two heads, "\textstyle\kp"]
\arrow[d,  hookrightarrow, "\textstyle \io_\Pi"']
	& \T\\
\R
\arrow[ru, two heads, "\textstyle\wt\kp"']
	&
\end{tikzcd}
\caption{The bijection $\kp$ and the surjection $\wt\kp$.}
\label{fig:kappas}
\end{figure}

Finally, for all  $u\in L^\infty(\T)$, the multiplication operator $m[u]\in\mL(\fh)$ on momentum space is (well-) defined (due to H\"older's inequality), for all $\vi\in\fh$, by 
\ba
\label{muvi}
m[u]\vi
:=u\vi.
\ea
Moreover, for all $u:=[u_i]_{i\in\num{1}{3}}\in L^\infty(\T)^3$ , we define $m[u]\in\mL(\fh)^3$ by 
\ba
\label{MultVect}
m[u]
:=[m[u_i]]_{i\in\num{1}{3}}.
\ea

The following assumption is central to the implementation of translation invariance in this section.

\bass[Translation invariance]
\label{ass:trans}
Let $R\in\mL(\hh)$, let $\Lm\in\Fin(\Z)$ with $\nu:=\card(\Lm)\ge 2$, and let $\{Q_i\}_{i\in\num{1}{\nu}}\subseteq\hh$ be a Fermi family over $\Lm$. 
\bn[label=(\alph*), ref={\it (\alph*)}]
\setlength{\itemsep}{0mm}
\item 
\label{trans-a}
$[R,\theta 1_2]=0$

\item
\label{trans-b}
$(\theta 1_2) Q_i=Q_{i+1}$ for all $i\in\num{1}{\nu-1}$
\en
\eass

Under \An{\ref{ass:trans} \ref{trans-a}}, we get the following action on the doubled 1-particle momentum Hilbert space.

\bl[Matrix multiplication operator]
\label{lem:trans}
Let $R\in\mL(\hh)$ fulfill \An{\ref{ass:trans} \ref{trans-a}}. Then, there exist $r_{ij}\in L^\infty(\T)$ for all $i,j\in\num{1}{2}$ such that $\wh R\in\mL(\fhh)$ reads
\ba
\label{mmo}
\wh R
=[m[r_{ij}]]_{i,j\in\num{1}{2}}.
\ea
\el

\bprf
See section \ref{sec:Proofs}.
\eprf

Under \An{\ref{ass:trans} \ref{trans-b}}, we get the following kinematical consequence.

\bl[Standard Fermi family]
\label{lem:StdCtg}
Let $\Lm=\{x_1, \ldots, x_\nu\}\in\Fin(\Z)$ with $\nu\ge 2$ be such that $x_1<\ldots<x_\nu$  and let $\{Q_i\}_{i\in\num{1}{\nu}}\subseteq\hh$ be a Fermi family over $\Lm$ which satisfies \An{\ref{ass:trans} \ref{trans-b}}. Then:
\bn[label=(\alph*), ref={\it (\alph*)}]
\setlength{\itemsep}{0mm}
\item 
\label{StdCtg-a}
$\Lm=\num{x_1}{x_\nu}$

\item
\label{StdCtg-b}
There exist $\lm\in\T$ and $\gamma\in\num{1}{2}$ such that, for all $i\in\num{1}{\nu}$,
\ba
\label{StdFS}
Q_i
=\lm J^\gamma (\delta_{x_i}\oplus 0). 
\ea
If $\lm=1$ and $\gamma=2$, we get the standard Fermi family over $\Lm$ from \eqref{sFs}.
\en
\el

\bprf
See section \ref{sec:Proofs}.
\eprf

In the following, we repeatedly make use of Appendix \ref{app:Toep}. Moreover, we say that $\vi\in\C^\T$ satisfies a property for almost all $z\in\T$ if $\vi\circ\kp$ satisfies the corresponding property almost everywhere on $\Pi$.

Under \An{\ref{ass:trans}}, the Majorana correlation matrix \eqref{Omij} is the finite section of a bounded block Toeplitz operator.  More precisely, we have the following.

\bp[Toeplitz structure]
\label{prop:ToepS}
Let $R\in\mL(\hh)$ satisfy \An{\ref{ass:trans}  \ref{trans-a}}, let $\Lm\in\Fin(\Z)$ with $\nu:=\card(\Lm)\ge 2$, and let $\{Q_i\}_{i\in\num{1}{\nu}}\subseteq\hh$ be a Fermi family over $\Lm$ satisfying \An{\ref{ass:trans}  \ref{trans-b}}. Moreover, let $\wt\Omega\in\M{2\nu}$ be defined, for all $i,j\in\num{1}{2\nu}$, by
\ba
\label{gMcm}
\wt\Omega_{ij}
:=\frac12\,(M_i,RM_j)_\hh.
\ea
Then:
\bn[label=(\alph*), ref={\it (\alph*)}]
\setlength{\itemsep}{0mm}
\item 
\label{ToepS-a}
The operator acting by \eqref{gMcm} on $\C^{2\nu}$ is given by the $\nu$-finite section of the block Toeplitz operator $T[a]\in\mL(\ell^2(\N,\C^2))$ whose block symbol $a\in L^\infty(\T,\M{2})$ is defined, for almost all $z\in\T$, by
\ba
\label{block}
a(z)
:=\frac12\hspace{0.2mm}[((\Gm\wh M_i)(z), r(z) (\Gm\wh M_j)(z))_{\C^2}]_{i,j\in\num{1}{2}},
\ea
where, using  \eqref{mmo}, $r\in L^\infty(\T,\M{2})$ is given by $(r)_{ij}:=r_{ij}\in L^\infty(\T)$ for all $i,j\in\num{1}{2}$, \ie, we have  $m_{\wt\Omega}=T_\nu[a]$ (and ${\wt\Omega}=T_{a,\nu}$).

\item 
\label{ToepS-b}
If, in addition, $R$ satisfies \eqref{def:2pt-op-2}, we have $m_{\Im(\wt\Omega)}=T_\nu[\wt a]$, where the block symbol $\wt a\in L^\infty(\T,\M{2})$ is defined, for almost all $z\in\T$, by
\ba
\label{block2}
\wt a(z)
:= \frac{1}{2\ii}\hspace{0.2mm}[((\Gm\wh M_i)(z), (r(z)-\tfrac121_2) (\Gm\wh M_j)(z))_{\C^2}]_{i,j\in\num{1}{2}}.
\ea
If, in addition, $R$ also satisfies \eqref{def:2pt-op-1}, its entries read, for almost all $z\in\T$, 
\ba
\label{ta11}
(\wt a(z))_{11}
&=\frac{\ii}{2}(1-(\Re(r_{11}(z))+\Re(r_{22}(z))+2\Re(\lm_{\gamma+1}^2 r_{12}(z)))),\\
(\wt a(z))_{12}
&=\frac{(-1)^\gamma}{2}(\Re(r_{11}(z))-\Re(r_{22}(z))-2\ii\Im(\lm_{\gamma+1}^2 r_{12}(z))),\\
(\wt a(z))_{21}
&=\frac{(-1)^{\gamma+1}}{2}(\Re(r_{11}(z))-\Re(r_{22}(z))+2\ii\Im(\lm_{\gamma+1}^2 r_{12}(z))),\\
\label{ta22}
(\wt a(z))_{22}
&=\frac{\ii}{2}(1-(\Re(r_{11}(z))+\Re(r_{22}(z))-2\Re(\lm_{\gamma+1}^2 r_{12}(z)))),
\ea
where $\lm_\gm:=\lm^{(-1)^{\gm}}$ for all $\lm\in\C$ and all $\gm\in\N$.

\item
\label{ToepS-c}
If $\omega\in\mQ_\fA$ is the quasifree state induced by the 2-point operator $R$, the operator acting by the imaginary part of the Majorana correlation matrix on $\C^{2\nu}$ is given by the $\nu$-finite section of the block Toeplitz operator $T[\wt a]\in\mL(\ell^2(\N,\C^2))$ whose block symbol $\wt a\in L^\infty(\T,\M{2})$ is given by \eqref{ta11}-\eqref{ta22}.
\en
\ep

\bprf
See section \ref{sec:Proofs}.
\eprf

\section{R/L mover entropy asymptotics}
\label{sec:R/L}

In this section, we specialize the general setting from the preceding sections to the so-called R/L movers. As  discussed in the Introduction, the definition of the R/L movers is based on the geometric partition of the configuration space into a finite part, called the sample, 
\ba
\label{ZS}
\Z_S
:=\{x\in\Z\,|\, x_L\le x\le x_R\},
\ea
where $x_L, x_R\in\Z$ satisfy $x_L\le x_R$ and $n_S:=x_R-x_L+1\ge 1$, and into two infinite parts to the left and right of the sample which play the role of thermal reservoirs,
\ba
\label{ZL}
\Z_L
&:=\{x\in\Z\,|\, x\le x_L-1\},\\
\label{ZR}
\Z_R
&:=\{x\in\Z\,|\, x\ge x_R+1\}.
\ea
For the convenience of the reader, we recall the definition and give a brief r\'esum\'e of the basic properties of the class of R/L mover states introduced in \cite{As21} (see there for more details).

In the following, if $G, H$ are any groups, we denote by $\Hom(G,H)$ the set of all group homomorphisms from  $G$ to $H$. Moreover, we call the family $\{\tau_t\}_{t\in\R}\subseteq\sAut(\fA)$ a dynamics on $\fA$ if the map $\R\ni t\mapsto\tau_t\in\sAut(\fA)$ is a group homomorphism (with respect to the group laws being the addition of numbers in $\R$ and the composition of \str automorphisms in $\sAut(\fA)$), \ie, a so-called  1-parameter group of \str automorphisms, and if, for any fixed $A\in\fA$, the map $\R\ni t\mapsto\tau_t(A)\in\fA$ is continuous (with respect to the modulus on $\R$ and the \Cs norm on $\fA$). The pair $(\fA,\tau)$ is sometimes called a \Cs dynamical system. 

\bd[Hamiltonian]
\label{def:Ham}
\bn[label=(\alph*),ref={\it (\alph*)}]
\setlength{\itemsep}{0mm}
\item
An operator $H\in\mL(\hh)$ is called a Hamiltonian if
\ba
\label{Ham-1}
H^\ast
&=H,\\
\label{Ham-2}
J H J
&=-H.
\ea

\item
Let $H\in\mL(\hh)$ be a Hamiltonian. The 1-parameter group of \str automorphisms $\tau\in\Hom(\R, \sAut(\fA))$ defined, for all $t\in\R$ and all $F\in\hh$, by
\ba
\label{qfd-H}
\tau^t(B(F))
:=B(\e^{\ii t H}F),
\ea
and suitably extended to the whole of $\fA$, is called the quasifree dynamics (generated by $H$).
\en
\ed

Next, we want to decompose a given Hamiltonian with respect to the geometric partition of the configuration space. To this end, let $\ell^\infty(\Z)$ stand for the usual complex Banach space of bounded complex-valued functions on $\Z$ and, for all  $u\in\ell^\infty(\Z)$, let the multiplication operator $m[u]\in\mL(\h)$ be (well-) defined by $m[u]f:=uf$ for all $f\in\h$ (abusing the notation of \eqref{muvi}). Now, for all $\alpha,\beta\in\{L,S,R\}$, we define
\ba
H_{\alpha\beta}
&:=K_\alpha H K_\beta,
\ea
where $K_\alpha:=m[1_{\Z_\alpha}]1_2$ for all $\alpha\in\{L,S,R\}$. Moreover, we set $H_\alpha:=H_{\alpha\alpha}$ for all $\alpha\in\{L,S,R\}$.

In the following, if $H\in\mL(\hh)$ is a Hamiltonian, we denote by $1_{sc}(H), 1_{ac}(H), 1_{pp}(H)\in\mL(\hh)$ the orthogonal projections onto the singular continuous, the absolutely continuous, and the pure point subspace of $H$, respectively. Moreover, $\mL^1(\hh)$ stands for the 2-sided \str ideal of $\mL(\hh)$ of all trace class operators on $\hh$ and $\slim$ denotes the limit with respect to the strong operator topology on $\mL(\hh)$.

\bass[Hamiltonian]
\label{ass:H}
Let $H\in\mL(\hh)$ be a Hamiltonian. 
\bn[label=(\alph*),ref={\it (\alph*)}]
\setlength{\itemsep}{0mm}
\item
\label{1sc=0}
$1_{sc}(H)=0$

\item
\label{HTheta}
$[H,\theta 1_2]=0$ 

\item
\label{L1}
$H_{LR}\in\mL^1(\hh)$ 
\item
\label{HLR=0}
$H_{LR}=0$

\item
\label{nontrv}
$H\neq z1$ for all $z\in\C$
\en
\eass

\br
\label{rem:nontrv}
If \An{\ref{ass:H} \ref{nontrv}} does not hold, \ie, if there exists $z\in\C$ such that $H=z1$, \eqref{Ham-1}-\eqref{Ham-2} imply that $z=0$. 
\er

\bd[R/L mover generator]
\label{def:Asmpt}
Let $H\in\mL(\hh)$ be a Hamiltonian satisfying \An{\ref{ass:H} \ref{L1}} and let $\beta_L,\beta_R\in\R$ be the inverse reservoir temperatures satisfying
\ba
\label{temp}
0
<\beta_L
\le \beta_R.
\ea
\bn[label=(\alph*),ref={\it (\alph*)}]
\setlength{\itemsep}{0mm}
\item 
For all $\alpha\in\{L,R\}$, the operator $P_\alpha\in\mL(\hh)$, defined by
\ba
P_\alpha
&:=\slim_{t\to\infty} \e^{-\ii t H}K_\alpha\e^{\ii t H}1_{ac}(H),
\ea
is called the ($\alpha$-) asymptotic projection (for $H$).

\item
The operator $\Delta\in\mL(\hh)$, defined by
\ba
\label{R/LGen}
\Delta
:=\sum_{\alpha\in\{L,R\}}\beta_\alpha P_\alpha,
\ea
is called the R/L mover generator (for $H$ and $\beta_L$, $\beta_R$).
\en
\ed

For the following, recall from Section \ref{sec:Neu} that $\mB(\R)$ stands for the normed unital \str algebra of all bounded Borel functions on $\R$.

\bd[Fermi function]
\label{def:Ff}
A function $\rho\in\mB(\R)$ is called a Fermi function if
\ba
\label{def:Ff-1}
\rho
&\ge 0,\\
\label{def:Ff-2}
\Ev(\rho)
&=\frac12.
\ea
\ed

\br
Since $\rho=\Ev(\rho)+\Od(\rho)$ for all  $\rho\in\mB(\R)$ and since $\mB(\R)$ is a \str algebra, $\rho$ is a Fermi function if and only if there exists an odd function $\varrho\in\mB(\R)$ satisfying $|\varrho(x)|\le 1$ for all $x\in\R$ such that $\rho=(1+\varrho)/2$.
\er

In the following, since $\hh$ is a separable Hilbert space, the set $\eig(H)$ of all eigenvalues of any  Hamiltonian $H\in\mL(\hh)$ is a countable subset of $\R$ and, hence, we write $\eig(H)=\{\lambda_i\}_{i\in I}$ with $I\subseteq\N$. 

We next recall the definition of a R/L mover from \cite{As21}.

\bd[R/L mover state]
\label{def:R/L}
Let $H\in\mL(\hh)$ be a Hamiltonian satisfying \An{\ref{ass:H} \ref{1sc=0},\ref{L1}}. Moreover,  
let $R_0\in\mL(\hh)$ be any 2-point operator, called the initial  2-point operator, let $\rho\in\mB(\R)$ be a Fermi function, and let $\beta_L, \beta_R\in\R$ satisfying \eqref{temp} be the inverse reservoir temperatures.
\bn[label=(\alph*),ref={\it (\alph*)}]
\setlength{\itemsep}{0mm}
\item
\label{R/L-1}
If $R\in\mL(\hh)$ has the form $R:=R_{ac}+R_{pp}$, where $R_{ac}, R_{pp}\in\mL(\hh)$ are defined by
\ba
\label{Rac}
R_{ac}
&:=\rho(\Delta H)1_{ac}(H),\\
\label{Rpp}
R_{pp}
&:=\sum_{\lambda\in\eig(H)}
1_\lambda(H) \hspace{0.2mm} R_0\hspace{0.5mm} 1_\lambda(H),
\ea
$R$ is called an R/L mover 2-point operator (for $H$, $R_0$, $\rho$, and $\beta_L, \beta_R$). 

\item
\label{R/L-2}
A state $\omega\in\mE_\fA$ whose 2-point operator is an R/L mover 2-point operator is called  an R/L mover (state).
\en
\ed

\br
\label{rem:uncond}
We know from \cite{As21} that, since $\Delta H$ is selfadjoint, $R_{ac}$ from \eqref{Rac} is (well-) defined by means of the spectral theorem. Moreover, as for $R_{pp}$, the series on the right hand side of \eqref{Rpp}, defined by 
\ba
\slim_{n\to\infty}\sum_{i\in\num{1}{n}}1_{\lambda_i}(H) \hspace{0.2mm} R_0\hspace{0.5mm} 1_{\lambda_i}(H), 
\ea
is unconditionally convergent, \ie, the notation in \eqref{Rpp} is well-motivated. Finally, $R\in\mL(\hh)$ is indeed a 2-point operator.
\er

\bx[NESS]
As discussed in the Introduction, the NESSs serve as the basic motivation for the introduction of the R/L mover states. We briefly sketch the main points in their construction (see \cite{As21} for more details). 
First, let $H\in\mL(\hh)$ be a Hamiltonian, $\rho\in\mB(\R)$ a Fermi function, $\beta_L, \beta_R$ the inverse reservoir temperatures, and $\beta_S\in\R$ with $\beta_S>0$ the inverse sample temperature. Moreover, the so-called initial system consist of a Hamiltonian $H_0\in\mL(\hh)$ given by
\ba
H_0
:=H_L+H_S+H_R,
\ea
of a quasifree dynamics $\tau_0\in\Hom(\R,\sAut(\fA))$ generated by $H_0$  (see \eqref{qfd-H}), and of a state $\omega_0\in\mQ_\fA$ specified by the 2-point operator $R_0\in\mL(\hh)$ given by
\ba
R_0
:=\rho(\Delta_0 H_0),
\ea
where $\Delta_0:=\sum_{\alpha\in\{L,S,R\}}\beta_\alpha K_\alpha\in\mL(\hh)$. 
The NESS $\omega\in\mE_\fA$  (for $H$, $\rho$, and $\beta_L, \beta_S, \beta_R$), whose definition (as a limit point in the weak-$\ast$ topology of the net defined by the ergodic mean between $0$ and $T>0$ of the given initial state time-evolved by the dynamics of interest) stems from \cite{Ru01}, is given, in the setting at hand, for all $A\in\fA$, by
\ba
\label{ness}
\omega(A)
:=\lim_{T\to\infty}\frac1T\int_0^T\rd t\hspace{1.5mm}\omega_0(\tau^t(A)),
\ea
where $\tau\in\Hom(\R,\sAut(\fA))$ is the quasifree dynamics generated by $H$. If $H\in\mL(\hh)$ satisfies \An{\ref{ass:H} \ref{1sc=0},\ref{L1}}, the time-dependent approach to Hilbert space scattering theory on $\hh$ yields that the limit on the right hand side of \eqref{ness} exists and that, for all $n\in\N$, all $\{F_i\}_{i\in\num{1}{2n}}\subseteq\hh$, and for $A=\prod_{i\in\num{1}{2n}} B(F_i)$, 
\ba
\label{limpoly}
\lim_{T\to\infty}\frac1T\int_0^T\rd t\hspace{1mm}
\omega_0(\tau^t(A))
=\lim_{T\to\infty}\frac1T\int_0^T\rd t\hspace{1mm}
\pf(X^{aa}+X^{pp}(t)),
\ea
where the matrix $X^{aa}\in\M{2n}$ and the matrix-valued function $X^{pp}=[X^{pp}_{ij}]_{i,j\in\num{1}{2n}}:\R\to\M{2n}$ with $X^{pp}_{ij}\in AP(\R)$ for all $i,j\in\num{1}{2n}$ are given, for all $i,j\in\num{1}{2n}$ and all $t\in\R$, by
\ba
X^{aa}_{ij}
&=(J F_i, 1_{ac}(H) \rho(\Delta H)1_{ac}(H) F_j),\\
\label{Xpp}
X^{pp}_{ij}(t)
&=(J F_i, 1_{pp}(H) \e^{-\ii t H}T_0 \hspace{0.2mm}\e^{\ii t H}1_{pp}(H)F_j),
\ea
and $AP(\R)$ stands for the complex-valued functions on $\R$ which are almost-periodic (in the sense of H. Bohr). Therefore, the 2-point operator $T\in\mL(\hh)$ of $\omega$ has the form given in Definition \ref{def:R/L} \ref{R/L-1}, \ie, $\omega$ is indeed a R/L mover. Moreover, if $1_{ac}(H)=1$, \eqref{limpoly}-\eqref{Xpp} imply that $\omega\in\mQ_\fA$.
\ex

\bx[Thermal equilibrium state]
\label{ex:KMS}
Let $H$ be a Hamiltonian satisfying \An{\ref{ass:H} \ref{1sc=0},\ref{L1}}, let the inverse reservoir temperatures both be equal to $\beta$, let $\rho$ be a Fermi function, and let the initial  2-point operator be defined by $R_0:=\rho(\beta H)$. Then, $R_{ac}=\rho(\beta H)1_{ac}(H)$, $R_{pp}=\rho(\beta H)1_{pp}(H)$,  and the R/L mover 2-point operator $R$ (for $H$, $R_0$, $\rho$, and $\beta_L, \beta_R$) reads 
\ba
\label{T-KMS}
R
=\rho(\beta H).
\ea
If $\tau$ is the quasifree dynamics generated by $H$ and if $1_{0}(H)=0$, there exists a unique $(\tau,\beta)$-KMS state $\omega\in\mQ_\fA$ whose 2-point operator has the form \eqref{T-KMS} and where the Fermi function $\rho$ is given by the Fermi-Dirac distribution, \ie, for all $x\in\R$, 
\ba
\label{ex:KMS-2}
\rho(x)
:=\frac{1}{1+\e^{-x}},
\ea
see \cite{Ar71}. Note that, in contrast to the so-called gauge-invariant case, there is a minus sign in the exponent of \eqref{ex:KMS-2} (see \cite{As21} for more details).
\ex

\bx[Ground state]
\label{ex:ground}
If $\tau$ is the quasifree dynamics generated by $H$ and if $1_0(H)=0$, there exists a unique $\tau$-ground state $\omega\in\mQ_\fA$, the so-called Fock state, whose 2-point operator has the form \eqref{T-KMS} with $\beta=1$ and 
\ba
\label{rho-ground}
\rho
:=1_{]0,\infty[},
\ea
see \cite{Ar71, ArMa85}.
\ex

In the following, for all $u\in\C^\T$, we denote the zero set of $u$ on $\Pi$ by 
\ba
Z_u
:=(u\circ\kp)^{-1}(\{0\}).
\ea
For all $u:=[u_i]_{i\in\num{1}{3}}, v:=[v_i]_{i\in\num{1}{3}}\in(\R^\T)^3$, we set $uv:=\sum_{i\in\num{1}{3}}u_iv_i$ and $u^2:=uu$, as well as $|u|:=(u^2)^{1/2}$. If $u:=[u_i]_{i\in\num{1}{3}}\in(\R^\T)^3$, we define $\wt u:=[\wt u_i]_{i\in\num{1}{3}}\in(\R^\T)^3$,  for all $i\in\num{1}{3}$, by
\ba
\label{u-tld}
\wt u_i
:=
\begin{cases}
\frac{u_i}{|u|}, & \mbox{on $\kp(Z_{|u|}^c)$},\\
\hfill 0, &  \mbox{on $\kp(Z_{|u|})$},
\end{cases}
\ea
where $T^c:=\T\setminus T$ for all $T\subseteq\T$, $M^c:=\Pi\setminus M$ for all $M\subseteq\Pi$, and, for any sets $A,B$, any $X\subseteq A$, and any $f\in B^A$, we use the notation $f(X):=\{f(a)\,|\,a\in X\}\subseteq B$.  Furthermore, let $TP(\Pi)$ stand for the (ring of) real trigonometric polynomials on $\Pi$ and set $TP(\T):=\{u\in\R^\T\,|\, u\circ\kp\in TP(\Pi)\}$. Recall that if $u\in TP(\T)$, we have $u\neq 0$ if and only if
\ba
\label{Zp}
\card(Z_u)\in\N_0.
\ea
Let $u\in\C^\T$ and let $M\subseteq\Pi$ with $\card(M)\in\N_0$ be the set of points in $\Pi$ on which $u\circ\wt\kp$ is not differentiable (see Figure \ref{fig:kappas}). Then, we define $u':\kp(M^c)\to\C$, for all $z\in\kp(M^c)$, by 
\ba
\label{def:deriv}
u'(z)
:=(u\circ\wt\kp)'(\kp^{-1}(z)),
\ea
where the prime on the right hand side stands for the usual derivative. If $u:=[u_i]_{i\in\num{1}{3}}\in (\R^\T)^3$, we set $u':=[u_i']_{i\in\num{1}{3}}$ (if all the $u_i\circ\wt\kp$ are differentiable on a common open subset of $\R$). Furthermore, whenever the symbol $\pm$ appears several times in the same equation, the latter stands for two equations, one of which corresponds to all the upper signs and the other one to all the lower signs (no ''cross terms'').
Finally, for all $\beta_L, \beta_R\in\R$ satisfying \eqref{temp}, we define $\beta, \delta\ge 0$ by
\ba
\beta
&:=\frac{\beta_R+\beta_L}{2},\\
\delta 
&:=\frac{\beta_R-\beta_L}{2}.
\ea

Next, we make use of \An{\ref{ass:H} \ref{HLR=0}} which means that there is no direct coupling between the two reservoirs, \ie, that the range of the Hamiltonian is bounded by the number $n_S$ of sites of the configuration space $\Z_S$ of the finite sample (see after \eqref{ZS}). This assumption is physically meaningful since the coupling interaction of a real physical sample to a thermal reservoir usually acts by short-range forces across the boundaries of the sample (for a lattice spacing of the order of $10^{-10}m$ and a sample dimension of the order of $10^{-3}m$ [see, for example, \cite{SoGiOtViRe01}], we get $n_S\sim 10^7$).

\bp[R/L mover 2-point operator]
\label{prop:R/L}
Let $H\in\mL(\hh)$ be a Hamiltonian satisfying \An{\ref{ass:H} \ref{HTheta},\ref{HLR=0},\ref{nontrv}}. Moreover, let $R\in\mL(\hh)$ be an R/L mover 2-point operator for $H$, for an initial 2-point operator $R_0\in\mL(\hh)$, for a Fermi function $\rho\in\mB(\R)$, and for inverse reservoir temperatures $\beta_L, \beta_R\in\R$ satisfying \eqref{temp}.
Then:
\bn[label=(\alph*),ref={\it (\alph*)}]
\setlength{\itemsep}{0mm}
\item
\label{R/L-a}
There exist (a smallest) $\mu\in\num{1}{n_S}$ (called the range of $H$) and $c_{\alpha,n}\in\R$ for all $(\alpha,n)\in(\num{0}{2}\times \num{1}{\mu})\cup(\{3\}\times \num{0}{\mu})$ such that $\wh H
=m[u_0]\sigma_0+m[u]\sigma\in\mL(\fhh)$, where,  for all $\alpha\in\num{0}{3}$, the so-called Pauli coefficient functions $u_0\in TP(\T)$ and $u:=[u_i]_{i\in\num{1}{3}}\in TP(\T)^3$ are given, for all $\alpha\in\num{0}{3}$ and all $k\in\Pi$, by
\ba
\label{Ualpha}
u_\alpha(\ei)
=\begin{cases}
\hfill -2\sum_{n\in\num{1}{\mu}}c_{\alpha,n}\sin(nk), & \alpha\in\num{0}{2},\\
c_{3,0}+2\sum_{n\in\num{1}{\mu}} c_{3,n}\cos(nk), & \alpha=3.
\end{cases}
\ea

\item
\label{R/L-b}
Using the following mutually exclusive, exhaustive, and non-empty cases,
\ba
\label{ass-cases}
\mbox{Case}
\begin{cases}
\mbox{1}, & \mbox{$u_0=0$, $u\neq 0$, and $uu'=0$},\\
\mbox{2}, & \mbox{$u_0=0$ and $uu'\neq 0$},\\
\mbox{3}, & \mbox{$u_0\neq 0$ and $u=0$},\\
\mbox{4}, & \mbox{$u_0\neq 0$, $u\neq 0$, and $uu'=0$},\\
\mbox{5}, & \mbox{$u_0\neq 0$, $uu'\neq 0$, and $u_0^2\neq u^2$},\\
\mbox{6}, & \mbox{$u_0\neq 0$ and $u_0^2=u^2$},
\end{cases}
\ea
we have
\ba
[1_{pp}(H), 1_{ac}(H), 1_{sc}(H)]
=\begin{cases}
\hfill [1,0,0], & \mbox{Case 1},\\
\hfill [0,1,0], & \mbox{Cases 2-5},\\
\hfill [1_0(H), 1-1_0(H),0],& \mbox{Case 6},
\end{cases}
\ea
where, in Case 6,  $\dim(\ran(1_0(H)))=\infty$ but $1_0(H)\neq 1$.  In particular, \An{\ref{ass:H} \ref{1sc=0}} is satisfied in all cases.

\item
\label{R/L-c}
In Cases 2-5, we get $R_{pp}=0$ and, thus,
\ba
\label{rhoDH}
R
=\rho(\Delta H).
\ea
Moreover,  we have $\wh R=m[r_0]\sigma_0+m[r]\sigma\in\mL(\fhh)$, where the  Pauli coefficient functions are specified as follows:

Cases 2,4,5\quad
We have $\card(Z_{|u|})\in\N_0$ and, on $\kp(Z_{|u|}^c)$, 
\ba
\label{r0}
r_0
&:=\frac12 (\rho_+ +\rho_-),\\
\label{r}
r
&:=\frac12 (\rho_+ -\rho_-)\wt u,
\ea
where $\rho_\pm\in L^\infty(\T)$ are given by
\ba
\label{rho-pm}
\rho_\pm
&:=\rho\circ((\beta+\delta\hspace{0.2mm}\sign\circ E_\pm') E_\pm),
\ea
and $E_\pm, E'_\pm\in L^\infty(\T)$ by
\ba
\label{E-pm}
E_\pm
&:=u_0\pm|u|,\\
\label{E'-pm}
E_\pm'
&:=u_0'\pm\wt uu'.
\ea

Case 3\quad
We have \eqref{r0}-\eqref{E'-pm} on $\T$, \ie, $E_\pm:=u_0$ and $E_\pm':=u_0'$ and, hence, $r_0:=\rho\circ((\beta+\delta\hspace{0.2mm}\sign\circ u_0')u_0)$ and $r:=0$.
\en
\ep

\bprf 
See, in \cite{As21}, Lemma 55  and (268) at the beginning of the proof of Lemma 58 for \ref{R/L-a}, (264) at the beginning of Lemma 58 and Lemma 58 {\it (b)} for \ref{R/L-b}, and Definition 41, (298)-(300) in the proof of Theorem 61, and Case 3 at the end of the proof of Theorem 61 for \ref{R/L-c}.
\eprf

\br
\label{rem:deriv}
It follows from \eqref{def:deriv} that \eqref{E'-pm} is indeed the derivative (on $\kp(Z_{|u|}^c)$ in Cases 2,4,5 and (on $\T$) in Case 3) of \eqref{E-pm}.
\er

\br
Cases 2-5 are equivalent to the condition $0\notin\eig(\bar V)$, where $\bar V$ stands for the bounded extension of the so-called asymptotic velocity with respect to $\wh H\in\mL(\fhh)$ defined, for all $\Phi\in\dom(P)$, by
\ba
V\Phi
=\lim_{t\to\infty}\frac{1}{t}\hspace{0.5mm} \e^{-\ii t\wh H} P \e^{\ii t\wh H}\Phi,
\ea
and $P$ denotes the position operator on the (doubled)  momentum space (see \cite{As21}). In these cases, the R/L mover generator \eqref{R/LGen} reads
\ba
\wh\Delta
=\beta 1+\delta\hspace{0.2mm} \sign(\bar V),
\ea
where $\sign(\bar V)=m[w_0]\sigma_0+m[w]\sigma$ and $w_0\in L^\infty(\T)$ and $w\in L^\infty(\T)^3$ are given, in Cases 2,4,5, on $\kp(Z^c)$, by 
\ba
w_0
&:=\frac{1}{2}\hspace{0.5mm}(\sign\circ E_+'+\sign\circ E_-'),\\
w
&:=\frac{1}{2}\hspace{0.5mm}(\sign\circ E_+'-\sign\circ E_-')\hspace{0.2mm}\wt u,
\ea
and, in Case 3, on $\T$, by the same expressions, \ie, by $w_0:=\sign\circ u_0'$ and $w:=0$.
\er

\br
Using \eqref{ta11}-\eqref{ta22}, \eqref{mmo}, and Proposition \ref{prop:R/L} \ref{R/L-c}, the symbol $\wt a\in L^\infty(\T,\M{2})$ from Proposition \ref{prop:ToepS} \ref{ToepS-b} can be expressed through the Pauli coefficient functions of the Hamiltonian $\wh H$.  In Cases 2,4,5, it explicitly reads, on $\kp(Z_{|u|}^c)$,
\ba
(\wt a(\cdot))_{11}
&=\frac{\ii}{2} (1-\sigma_+
-(\Re(\lm_{\gm+1}^2)\wt u_1+\Im(\lm_{\gm+1}^2)\wt u_2)\hspace{0.1mm}\sg_-),\\
(\wt a(\cdot))_{12}
&=\frac{(-1)^\gm}{2}
(\wt u_3+\ii (\Re(\lm_{\gm+1}^2)\wt u_2-\Im(\lm_{\gm+1}^2)\wt u_1))\hspace{0.1mm}\sg_-,\\
(\wt a(\cdot))_{21}
&=\frac{(-1)^{\gm+1}}{2}
(\wt u_3-\ii (\Re(\lm_{\gm+1}^2)\wt u_2-\Im(\lm_{\gm+1}^2)\wt u_1))\hspace{0.1mm}\sg_-,\\
(\wt a(\cdot))_{22}
&=\frac{\ii}{2} (1-\sigma_+
+(\Re(\lm_{\gm+1}^2)\wt u_1+\Im(\lm_{\gm+1}^2)\wt u_2)\hspace{0.1mm}\sg_-),
\ea
where the functions $\sg_\pm\in L^\infty(\T)$ are defined by
\ba
\sigma_\pm
:=\rho\circ((\beta+\delta\sign\circ(u_0'+\wt uu'))(u_0+|u|))
\pm \rho\circ((\beta+\delta\sign\circ(u_0'-\wt uu'))(u_0-|u|)).
\ea
In Case 3, we have $(\wt a(\cdot))_{11}=(\wt a(\cdot))_{22}=\ii(1-2\rho\circ((\beta+\delta\sign\circ u_0') u_0))/2$ and $(\wt a(\cdot))_{12}=(\wt a(\cdot))_{21}=0$ on $\T$.
\er

In the following, we write $E:=E_+$ and $E':=E_+'$ for \eqref{E-pm}-\eqref{E'-pm}. 

We now arrive at our desired result. It identifies the limit for infinite string length of the von Neumann entropy density in a general quasifree R/L mover state. 

\bt[R/L mover entropy asymptotics]
\label{thm:R/L-asym}
Let $H\in\mL(\hh)$ be a Hamiltonian satisfying \An{\ref{ass:H} \ref{HTheta},\ref{HLR=0},\ref{nontrv}} and let the Pauli coefficient functions $u_0\in TP(\T)$ and $u\in TP(\T)^3$ of $\wh H\in\mL(\fhh)$ satisfy the conditions of Cases 2-5. Moreover, let $R\in\mL(\hh)$ be an R/L mover 2-point operator for $H$, for an initial 2-point operator $R_0\in\mL(\hh)$, for a Fermi function $\rho\in\mB(\R)$, and for inverse reservoir temperatures $\beta_L, \beta_R\in\R$ satisfying \eqref{temp}, and let $\omega\in\mQ_\fA$ be the R/L mover induced by $R$. Finally, let $\Lm=\{x_1, \ldots, x_\nu\}\in\Fin(\Z)$ with $\nu\ge 2$ be such that $x_1<\ldots<x_\nu$  and let $\{Q_i\}_{i\in\num{1}{\nu}}\subseteq\hh$ be a Fermi family over $\Lm$ which satisfies \An{\ref{ass:trans} \ref{trans-b}}.
Then:
\bn[label=(\alph*), ref={\it (\alph*)}]
\setlength{\itemsep}{0mm}
\item 
\label{R/L-asym-a}
The von Neumann entropy $S_\Lambda$ of the reduced density matrix $R_\Lm\in\M{2^\nu}$ depends only on the length $\nu$ of the string $\Lambda=\num{x_1}{x_\nu}$ and, hence, we denote it by $S_\nu$.

\item 
\label{R/L-asym-b}
For $\nu\to\infty$, we have
\ba
S_\nu
=s_\infty \nu +o(\nu),
\ea
where the asymptotic density $s_\infty\ge 0$ has the form 
\ba
\label{sinfty-3}
s_\infty
:=\sum_{\alpha\in\{L,R\}}\int_{\Pi_\alpha}\frac{\rd k}{2\pi}\hspace{1mm} \wt\eta(2\Od(\rho)(\beta_\alpha E(\ei))),
\ea
and where the sets $\Pi_L, \Pi_R\in\mM(\Pi)$ are defined as follows:

Cases 2,4,5\quad We have 
\ba
\label{PiL}
\Pi_L
&:=\{k\in Z_{|u|}^c\,|\,E'(\ei)<0\},\\
\label{PiR}
\Pi_R
&:=\{k\in Z_{|u|}^c \,|\,E'(\ei)>0\}.
\ea

Case 3\quad We have  \eqref{PiL}-\eqref{PiR} on $\Pi$, \ie, 
\ba
\label{PiL-3}
\Pi_L
&:=\{k\in\Pi\,|\,u_0'(\ei)<0\},\\
\label{PiR-3}
\Pi_R
&:=\{k\in\Pi\,|\,u_0'(\ei)>0\}.
\ea
\en
\et

\br
\label{rem:meas}
Since $u^2\in TP(\T)$ and since $Z_{|u|}=Z_{u^2}$, \eqref{Zp} yields $\card(Z_{|u|})\in\N_0$ in Cases 2,4,5. Hence, since \eqref{def:deriv}, Remark \ref{rem:deriv}, and \eqref{E'-pm} yield that $\vi:=E'\circ\kp\circ\iota_{Z_{|u|}^c}\in C(Z_{|u|}^c)$, the set $\Pi_L=\vi^{-1}(]-\infty,0[)$ is open relative to $Z_{|u|}^c$ and, since $Z_{|u|}^c\in\mM(\Pi)$, we indeed get  $\Pi_L\in\mM(\Pi)$ (and the same holds for $\Pi_R=\vi^{-1}(]0,\infty[)$). In Case 3, we have $\psi:=u_0'\circ\kp\in C(\Pi)$ and, analogously, $\Pi_L=\psi^{-1}(]-\infty,0[)$ and $\Pi_R=\psi^{-1}(]0,\infty[)$ are open relative to $\Pi\in\mM(\Pi)$. 
Finally, in Cases 2,4,5, we set
\ba
\Pi_0
:=\{k\in Z_{|u|}^c \,|\,E'(\ei)=0\},
\ea
and we note that, due to \eqref{E'-pm}, $\card(\Pi_0)\subseteq Z_Q$, where $Q:=u_0'^2u^2-(uu')^2\in TP(\T)$ and $Q\neq 0$ in Cases 2,4,5 (see \cite{As21}). Hence, $\card(\Pi_0)\in\N_0$ and $\Pi_0$ does not contribute to \eqref{sinfty-3}. 
In Case 3, we set $\Pi_0:=\{k\in\Pi \,|\,u_0'(\ei)=0\}$. Since $u_0'\neq 0$ (due to $u_0\neq 0$ and \eqref{Ualpha}) and since $u_0'\in TP(\T)$, we again get $\card(\Pi_0)\in\N_0$ (and  $\card(\Pi_0)\ge 2$ due to \eqref{Ualpha} and the Sturm-Hurwitz theorem).
\er

\br
\label{rem:int}
Let $\al\in\{L,R\}$ be fixed. Since $(\beta_\al E)\circ\wt\kp\in C_b(\R)\subseteq\mB(\R)$, since $\rho\in\mB(\R)$ implies that $2\Od(\rho)\in\mB(\R)$, and since $\wt\eta\in C_0(\R)$, we have $\psi\circ\wt\kp\in\mB(\R)$ (see \cite{As21}), where we set 
\ba
\psi
:=\wt\eta\circ(2\Od(\rho))\circ(\beta_\al E).  
\ea
Hence, we know that $\psi\circ\wt\kp$ is $(\mM(\R), \mM(\R))$-measurable implying that $\psi\circ\kp$ is $(\mM(\Pi), \mM(\R))$-measurable. Since $\psi$ is bounded, we have $\psi\in L^\infty(\T)$ and the integral in \eqref{sinfty-3} is well-defined.
\er

\vspace{5mm}

\bprf
\ref{R/L-asym-a}
Using \eqref{vNEnt-H}, the evenness of \eqref{eta}, $\spec(\Im(\Omega))=\{\pm\ii\lm_i/2\}_{i\in\num{1}{\nu}}$ again from Proposition \ref {prop:vNEnt}, and,  from Proposition \ref{prop:ToepS} \ref{ToepS-c}, 
\ba
2\ii \Im(\Omega)
&=2\ii T_{\wt a, \nu}\nonumber\\
&=T_{b, \nu}, 
\ea
where the symbol  $b\in L^\infty(\T,\M{2})$ is defined by $b:=2\ii\wt a=2a-1_2$, we have 
\ba
S_\Lm
&=\sum_{i\in\num{1}{\nu}}\wt\eta(\lm_i)\nonumber\\
&=\sum_{i\in\num{1}{\nu}}\frac{\wt\eta(\lm_i)+\wt\eta(-\lm_i)}{2}\nonumber\\
&=\frac12\sum_{\lm\hspace{0.2mm}\in\hspace{0.2mm}\spec(T_{b,\nu})}\wt\eta(\lm),
\ea
and, due to \eqref{ta11}-\eqref{ta22}, the symbol $b$ is independent of the position of the string (\ie, without loss of generality, we can choose $x_i=i$ for all $i\in\num{1}{\nu}$ for example).

\ref{R/L-asym-b}
Due to \eqref{2pt-2c}-\eqref{2pt-2e} and since $\bar\lm_\gm=\lm_{\gm+1}$ for all $\lm\in\T$ and all $\gm\in\num{1}{2}$ (see after \eqref{ta22}), \eqref{a11}-\eqref{a22} imply that the symbol $a\in L^\infty(\T,\M{2})$ is selfadjoint, \ie, $(a(z))^\ast=a(z)$ for almost all $z\in\T$, and, thus, so is the symbol $b=2a-1_2$. Moreover, we have $\wt\eta\in C_0(\R)$ and, hence, all the assumptions for the applicability of Szeg\H{o}'s first limit theorem for block symbols are satisfied (see, for example, \cite{BoSi99}). Therefore, using this theorem, we can write
\ba
\label{sinfty}
s_\infty
&:=\lim_{\nu\to\infty} \frac{S_\nu}{\nu}\nonumber\\
&=\frac12\Int s(\ei),
\ea
where the integrand $s\in L^\infty(\T)$ is defined by
\ba
\label{def:symb}
s
:=\tr\circ\wt\eta\circ b,
\ea
and where the definition of $\wt\eta\circ b\in(\M{2})^\T$ is based on the spectral theorem as discussed before Definition \ref{def:vNEnt}. Proceeding as in Remark \ref{rem:vNEnt}, using \eqref{ta11}-\eqref{ta22}, the relation between the standard basis and the Pauli basis given by $r_{11}=r_0+r_3$, $r_{12}=r_1-\ii r_2$, $r_{21}=r_1+\ii r_2$,$r_{22}=r_0-r_3$, and \eqref{r0}-\eqref{r}, we get
\ba
s
&=\sum_{\alpha\in\num{1}{2}}\wt\eta\circ\Big(r_{11}+r_{22}-1+(-1)^\alpha\sqrt{(r_{11}-r_{22})^2+4|r_{12}|^2}\Big)\nonumber\\
&=\sum_{\alpha\in\num{1}{2}}\wt\eta\circ(2(r_0+(-1)^\alpha|r|)-1)\nonumber\\
\label{s(t)}
&=\sum_{\alpha\in\num{1}{2}}\wt\eta\circ(\rho_++\rho_-+(-1)^\alpha|\rho_+-\rho_-|-1).
\ea
Plugging \eqref{rho-pm} into \eqref{s(t)}, separating the contributions with respect to the sign of the modulus term,  and regrouping symmetrically, \eqref{sinfty} becomes
\ba
\label{sinfty-2}
s_\infty
=\frac12\Int \wt\eta(2\rho_+(\ei)-1)
+\frac12\Int \wt\eta(2\rho_-(\ei)-1).
\ea
Since, due to \eqref{Ualpha}, we have, for almost all $z\in\T$,
\ba
\label{sym-1}
E_-(\bar z)
&=-E_+(z),\\
\label{sym-2}
E_-'(\bar z)
&=E_+'(z),
\ea
\eqref{def:Ff-2}, \eqref{OdEta}, and a parity variable transformation in momentum space yields that the second integral on the right hand side of \eqref{sinfty-2} equals the first one and, using again \eqref{def:Ff-2}, we get
\ba
\label{sinfty-6}
s_\infty
=\Int \wt\eta(2\Od(\rho)((\beta+\delta\hspace{0.2mm}\sign(E_+'(\ei)))E_+(\ei))).
\ea
Separating the contributions with respect to the sign of the $\sign\circ E_+'$ term, \eqref{sinfty-6} takes the form \eqref{sinfty-3} with \eqref{PiL}-\eqref{PiR} in Cases 2,4,5 and with \eqref{PiL-3}-\eqref{PiR-3} in Case 3.
\eprf

\br
\label{rem:rhoH}
Since $1_{ac}(H)=1$ in Cases 2-5, if $\beta_L=\beta_R=1$, we get $\Delta=1$ due to \eqref{R/LGen} and the fact that $P_L+P_R=1_{ac}(H)$ (see \cite{As21}). Hence, \eqref{rhoDH} yields 
\ba
R
=\rho(H),
\ea
and \eqref{sinfty-6} leads to the asymptotic density
\ba
s_\infty
=\Int\hspace{0.8mm} \wt\eta(2\Od(\rho)(E(\ei))).
\ea
\er

\br
\label{rem:u0=0}
In applications (see, for example, Example \ref{ex:XYNESS} below or the more general Suzuki models in \cite{As21}), one often encounters the situation, where 
\ba
\label{u0=0}
u_0
=0.
\ea
This special case occurs if and only if $[H,\xi\sigma_3]=0$ (see \cite{As21}), where the parity operator in position space $\xi\in\mL(\h)$ is defined by $(\xi f)(x):=f(-x)$  for all $f\in\h$ and all $x\in\Z$. If \eqref{u0=0} holds, \eqref{sym-1}-\eqref{sym-2} yield,  for almost all $z\in\T$,
\ba
\label{sym-3}
E(\bar z)
&=E(z),\\
\label{sym-4}
E'(\bar z)
&=-E'(z).
\ea
Hence, making a parity variable transformation in momentum space in \eqref{sinfty-6}, writing \eqref{sinfty-6} as half the sum of  \eqref{sinfty-6} plus the parity transformed integral \eqref{sinfty-6}, using that $\beta+\delta\hspace{0.2mm}\sign(E'(\bar z))=\beta-\delta\hspace{0.2mm}\sign(E'(z))$ for almost all $z\in\T$, and regrouping with respect to the sign of the $\sign\circ E'$ term, we get
\ba
\label{sinfty-7}
s_\infty
=\frac12\sum_{\al\in\{L,R\}}\Int\wt\eta(2\Od(\rho)(\beta_\al E(\ei))).
\ea
\er

\br
\label{rem:FD}
Let the Fermi function be given by the Fermi-Dirac distribution \eqref{ex:KMS-2}. Then, since $2\Od(\rho)(x)=\tanh(x/2)$ for all $x\in\R$, \eqref{sinfty-3} becomes
\ba
\label{sinfty-4}
s_\infty
=\sum_{\al\in\{L,R\}}\int_{\Pi_\al}\frac{\rd k}{2\pi}\hspace{0.8mm} \eta(\tanh(\beta_\al E(\ei)/2)).
\ea
Since $|\tanh(\beta_\al E(z)/2)|\le\tanh(\beta_\al\|E\|_{L^\infty(\T)}/2)\le\tanh(\beta_\al\sum_{\beta\in\num{0}{3}}\|u_\beta\|_{L^\infty(\T)}/2)=:a_\al<1$ for all $\al\in\{L,R\}$ and all $z\in\T$ and, since, for all $a\in\,]0,1[$, we have $\eta(x)\ge\eta(a)>0$ for all $x\in[-a,a]\subseteq\,]\!-1,1[$ (see Figure \ref{fig:Shan}), we get the simple lower bound
\ba
\label{sinfty-pos}
s_\infty
&\ge \sum_{\al\in\{L,R\}} \eta(a_\al) |\Pi_\al|_B\nonumber\\
&>0,
\ea
where we used that $\Pi_L, \Pi_R\in\mM(\Pi)$ due to Remark \ref{rem:meas} and that $\sum_{\al\in\{L,R\}}|\Pi_\al|_B=2\pi$.
\er

\begin{figure}
\centering
\begin{tikzpicture}
\node (flux) {\includegraphics[width=80mm,height=50mm]{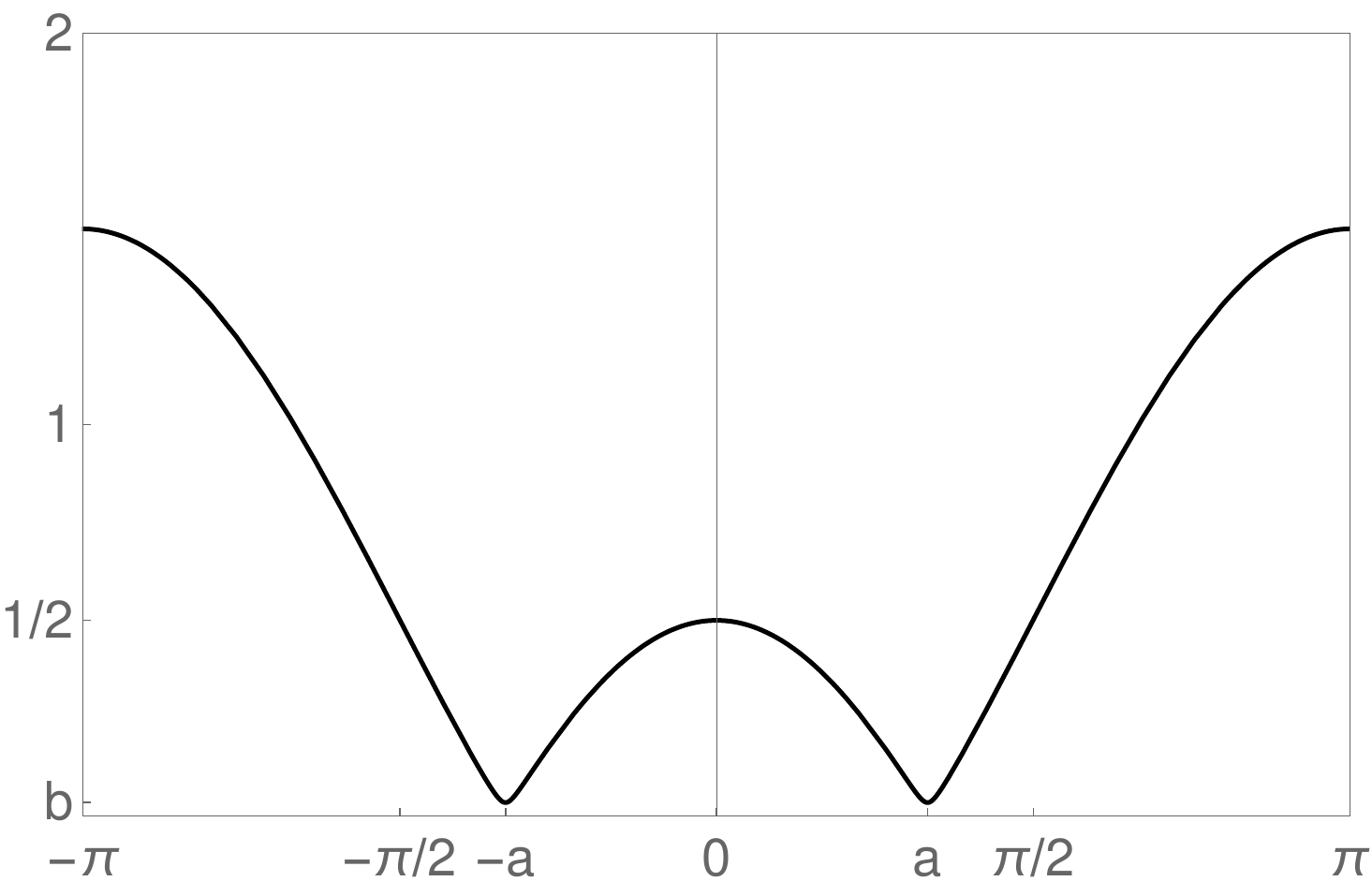}};
\end{tikzpicture}
\caption{The function \eqref{E-XY} for $c_{2,1}=1/50$ and $c_{3,0}=1/2$. We have ${\rm a}=\arccos(625/1248)\approx1.05$ and ${\rm b}=E(\e^{\ii {\rm a}})=\sqrt{1871/39}/200\approx0.03$.}
\label{fig:XY}
\end{figure}

\bx[XY model]
\label{ex:XYNESS}
For the case of the XY model (see Remark \ref{rem:AJW}), the Pauli coefficient functions \eqref{Ualpha} read, for all $k\in\Pi$,
\ba
\label{XY-u0}
u_0(\ei)
&=0,\\
u_1(\ei)
&=0,\\
u_2(\ei)
&=-2 c_{2,1}\sin(k),\\
\label{XY-u3}
u_3(\ei)
&=c_{3,0}+\cos(k),
\ea
where $c_{2,1}, c_{3,0}\in\R$ (in the notation of \eqref{XYDensity} and \eqref{FDensity}, we have $c_{2,1}=-\gamma/2$ and $c_{3,0}=-\lambda$). Hence, \eqref{sinfty-7} and Remark \ref{rem:FD} imply that for the XY NESS discussed in the Introduction, \ie, for the R/L mover whose Hamiltonian is specified by \eqref{XY-u0}-\eqref{XY-u3} and whose Fermi function is given by \eqref{ex:KMS-2}, we have 
\ba
\label{sinfty-5}
s_\infty
=\frac12\sum_{\al\in\{L,R\}}\Int\hspace{0.8mm} \eta(\tanh(\beta_\al E(\ei)/2)),
\ea
where, for all $k\in\Pi$, 
\ba
\label{E-XY}
E(\ei)
&=|u|(\ei)\nonumber\\
&=\sqrt{(c_{3,0}+\cos(k))^2+4c_{2,1}^2\sin^2(k)},
\ea
see Figure \ref{fig:XY}. As in Remark \ref{rem:FD} (and using the notation from there), we can write that $s_\infty\ge \pi \sum_{\al\in\{L,R\}} \eta(a_\al)$, \ie, $s_\infty>0$ (see \cite{As07} and remarks therein). If the inverse reservoir temperatures are both equal to some $\beta$, \eqref{sinfty-5} yields the asymptotic density of the thermal equilibrium state at inverse temperature $\beta$ (see Example \ref{ex:KMS} and Remark \ref{rem:rhoH}).

\ex

In the following, for all $\al\in\{L,R\}$, let  $L^\infty(\Pi_\al)$ denote the space of all equivalence classes of functions $\vi\in\C^{\Pi_\al}$ which are $(\mM(\Pi_\al), \mM(\R))$-measurable and almost everywhere bounded with respect to $|\cdot|_B$. Moreover, for all $\al\in\{L,R\}$ and all inverse reservoir temperatures $\beta_L, \beta_R\in\R$ satisfying \eqref{temp}, we define the function $\vi_\al\in L^\infty(\Pi_\al)$ by
\ba
\label{vi-sg}
 \vi_\al
 :=(\beta_\al E)\circ\kp\circ\io_{\Pi_\al},
\ea
and the sets $S^{\pm}_\al, \Sigma_\al\subseteq\R$ by 
\ba
\label{S-sg}
S^{\pm}_\al
&:=\{\pm\vi_\al(k)\,|\, k\in\Pi_\al\},\\
\label{Sg-sg}
\Sigma_\al
&:=\bigcup_{\sg\in\{\pm\}}S_\al^\sg.
\ea
Moreover, the set $\Sigma\subseteq\R$ is defined by 
\ba
\label{Sg}
\Sigma
:=\bigcup_{\al\in\{L,R\}}\Sigma_\al.
\ea

In Remark \ref{rem:FD} and Example \ref{ex:XYNESS}, the asymptotic density is strictly positive. It's vanishing can be characterized as follows.

\bc[Vanishing asymptotic density]
\label{cor:vad}
Let the assumptions of Theorem \ref{thm:R/L-asym} hold. Then, $s_\infty=0$ if and only if there exists $M\in\mM(\R)$ such that, for almost all $x\in\Sigma$, 
\ba
\label{rhoSgm}
\rho(x)
=1_M(x).
\ea
\ec

\bprf
See section \ref{sec:Proofs}.
\eprf

\br
Since we know that $\spec(\wh H)=\bigcup_{k\in\Pi}\{E_+(\ei), E_-(\ei)\}$  (see \cite{As21}) and since, due to \eqref{sym-1}, we have $\bigcup_{k\in\Pi}\{E_-(\ei)\}=\bigcup_{k\in\Pi}\{E_-(\e^{-\ii k})\}=\bigcup_{k\in\Pi}\{-E_+(\ei)\}$, the spectrum becomes 
\ba
\spec(\wh H)
=\bigcup_{k\in\Pi}\{\pm E(\ei)\}.
\ea
Hence, the set $\spec(\wh H)\setminus(\bigcup_{\al\in\{L,R\}}\beta_\al^{-1}\Sigma_\al)$ consists of finitely many points only.
\er

As an important example, we consider the following class of states.

\bx[Ground state]
Let $\omega\in\mQ_\fA$ be a ground state from Example \ref{ex:ground}. Hence, since $]0,\infty[\,\in\mM(\R)$ and since $\rho(x)=1_{]0,\infty[}(x)$ for all $x\in\R$ , Corollary \ref{cor:vad} yields
\ba
s_\infty
=0.
\ea
\ex

\section{Proofs}
\label{sec:Proofs}

In this section, we collect all the proofs of the foregoing sections carried out in this paper, except for the proof of the main theorem.\\

\noindent{\bf Proof of Lemma \ref{lem:mu}.}\hspace{2mm}
\ref{mu-a}\,
In order to verify that, for all $i\in\num{1}{\nu}$, the family  $\{e_{\alpha\beta}^{(i)}\}_{\alpha,\beta\in\num{1}{2}}$ satisfies \eqref{MUi-1}-\eqref{MUi-3}, we first show that, for all $i,j\in\num{1}{\nu}$ and all $\alpha\in\num{1}{2}$, 
\ba
\label{Si-1}
B(J^\alpha Q_i)S_j
&=\sign(i-j) S_j B(J^\alpha Q_i),\\
\label{Si-2}
[S_i,S_j]
&=0,\\
\label{Si-3}
S_i^\ast
&=S_i,\\
\label{Si-4}
S_i^2
&=1,
\ea
where, for all $x\in\Z$, we define $\sign(x):=1$ if $x\ge 0$ and $\sign(x):=-1$ if $x<0$.  To this end, note that the CAR \eqref{CAR} yield, for all $i,j\in\num{1}{\nu}$ and all $\alpha,\beta\in\num{1}{2}$,
\ba
B(J^\alpha Q_i)(2B(J^\beta Q_j)B(J^{3-\beta} Q_j)-1)
&=2(\delta_{ij}-1) B(J^\beta Q_j)B(J^\alpha Q_i)B(J^{3-\beta} Q_j)\nonumber\\
&\hspace{4mm}+(2(1-\delta_{\alpha\beta})\delta_{ij}-1)B(J^\alpha Q_i),\\
(2B(J^\beta Q_j)B(J^{3-\beta} Q_j)-1)B(J^\alpha Q_i)
&=2(\delta_{ij}-1) B(J^\beta Q_j)B(J^\alpha Q_i)B(J^{3-\beta} Q_j)\nonumber\\
&\hspace{4mm}+(2\delta_{\alpha\beta}\delta_{ij}-1)B(J^\alpha Q_i),
\ea
from which we get, for all $\alpha,\beta\in\num{1}{2}$ and all $i,j\in\num{1}{\nu}$ with $i\neq j$,
 \ba
 \label{comij}
 [B(J^\alpha Q_i), 2B(J^\beta Q_j)B(J^{3-\beta} Q_j)-1]
 =0,
 \ea
 and, for all $\alpha,\beta\in\num{1}{2}$ and all $i,j\in\num{1}{\nu}$ with $i=j$,
\ba
\label{acomii}
\{B(J^\alpha Q_i), 2B(J^\beta Q_i)B(J^{3-\beta} Q_i)-1\}
=0.
\ea
Hence, as for \eqref{Si-1}, for all $\alpha,\beta\in\num{1}{2}$ and all $i,j\in\num{1}{\nu}$ with $i\ge j$, \eqref{Si} and \eqref{comij} yield 
\ba
B(J^\alpha Q_i)S_j
&=S_j B(J^\alpha Q_i)\nonumber\\
&=\sign(i-j) S_j B(J^\alpha Q_i). 
\ea
Moreover, for all $\alpha,\beta\in\num{1}{2}$ and all $i,j\in\num{1}{\nu}$ with $i\le j-1$ (if $\nu\ge 2$), \eqref{Si}, \eqref{comij}, and \eqref{acomii} lead to
\ba
B(J^\alpha Q_i)S_j
&=\Big(\prod\nolimits_{k\in\num{1}{i-1}}\chi_k\Big)B(J^\alpha Q_i)\chi_i\Big(\prod\nolimits_{k\in\num{i+1}{j-1}}\chi_k\Big)\nonumber\\
&=\Big(\prod\nolimits_{k\in\num{1}{i-1}}\chi_k\Big)\Big(\{B(J^\alpha Q_i), \chi_i\}-\chi_i B(J^\alpha Q_i)\Big)\Big(\prod\nolimits_{k\in\num{i+1}{j-1}}\chi_k\Big)\nonumber\\
&=-S_j B(J^\alpha Q_i)\nonumber\\
&=\sign(i-j)S_j B(J^\alpha Q_i),
\ea
where we set $\chi_i:=2B^\ast(Q_i)B(Q_i)-1$ for all $i\in\num{1}{\nu}$ (and we used the convention that a product is equal to $1$ if the index set is empty, see \eqref{Setxy}). Furthermore, since $[2B^\ast(Q_i)B(Q_i)-1, 2B^\ast(Q_j)B(Q_j)-1]=0$ for all $i,j\in\num{1}{\nu}$ due to \eqref{comij}, we get \eqref{Si-2}, \eqref{Si-3}, and $S_i^2=\prod_{j\in\num{1}{i-1}}(2B^\ast(Q_j)B(Q_j)-1)^2$ for all $i\in\num{2}{\nu}$.  Since, due to \eqref{CAR},  we have $(2B^\ast(Q_i)B(Q_i)-1)^2=1$ for all $i\in\num{1}{\nu}$ , we also arrive at \eqref{Si-4}. 

Now, using \eqref{Si-1}, \eqref{Si-4}, and \eqref{CAR}, a direct computation leads to \eqref{MUi-1}. As for \eqref{MUi-2}, we again use  \eqref{Si-1} and also  \eqref{Si-3}. Moreover, \eqref{CAR} yields \eqref{MUi-3}. As for \eqref{eCom}, we again use \eqref{Si-1}, \eqref{Si-2}, \eqref{CAR} and the fact that, for all $i,j\in\num{1}{\nu}$ with $i\neq j$, we have  $(\sign(i-j))^2=1$ and $\sign(i-j) \sign(j-i)=-1$.

\ref{mu-b}\,
Let $n\in\num{1}{\nu}$. Using \eqref{eCom} and \eqref{MUi-1}, we get \eqref{MUmu-1} because, for all $a:=[\alpha_i]_{i\in\num{1}{n}}$, $b:=[\beta_i]_{i\in\num{1}{n}}$, $c:=[\gm_i]_{i\in\num{1}{n}}$, and $d:=[\delta_i]_{i\in\num{1}{n}}\in\num{1}{2}^n$,
\ba
e_{ab} e_{cd}
&=\prod_{i\in\num{1}{n}} e^{(i)}_{\alpha_i\beta_i}e^{(i)}_{\gamma_i\delta_i}\nonumber\\
&=\Big(\prod\nolimits_{i\in\num{1}{n}}\delta_{\beta_i\gamma_i}\Big)\Big(\prod\nolimits_{i\in\num{1}{n}}e^{(i)}_{\alpha_i\delta_i}\Big)\nonumber\\
&=\delta_{bc}e_{ad}.
\ea
Moreover, \eqref{eCom} and \eqref{MUi-2} lead to \eqref{MUmu-2}. As for \eqref{MUmu-3}, \eqref{MUi-3} yields 
\ba
\sum_{a\in\num{1}{2}^n}e_{aa}
&=\prod_{i\in\num{1}{n}}\Big(\sum\nolimits_{\alpha_i\in\num{1}{2}}e^{(i)}_{\alpha_i\alpha_i}\Big)\nonumber\\
&=1.
\ea

\ref{mu-c}\,
The family $\{e_{ab}\}_{a,b\in\num{1}{2}^n}$ spans $\fA_\Lm$ because $\fA_\Lm$ equals the set of all polynomials in $\fA$ generated by $\fB_\Lm$ (see Proposition \ref{prop:LocObs}) and a direct computation yields that it is linearly independent (see \cite{As23} for more details).
\eprf

\vspace{5mm}

\noindent{\bf Proof of Proposition \ref{prop:ls}.}\hspace{2mm}
Note that $\omega_\Lm$ retains the linearity and positivity properties of $\omega$, of course. Moreover, due to \eqref{dCAR2}, we also have $1\in\fA_\Lm$ and, hence, $\omega_\Lm(1)=1$, \ie, $\omega_\Lm\in\mE_{\fA_\Lm}$.

\ref{ls-a}\,
Let us equip $\M{2^\nu}$ with the Frobenius scalar product and recall that the corresponding Frobenius norm is submultiplicative and that the involution (see above) is isometric with respect to the Frobenius norm, \ie, that $\M{2^\nu}$ is a Banach \str algebra (but not a \Cs algebra) with respect to the Frobenius norm. Hence, since we know that, if $\mA$ is a Banach \str algebra and $\mB$ a \Cs algebra, any $\pi\in\sHom(\mA,\mB)$ satisfies $\|\pi(A)\|\le\|A\|$ for all $A\in\mA$, setting $\eta:=\omega_\Lm\circ \pi_\Lm^{-1}$ and using $\pi_\Lm^{-1}\in\sIso(\M{2^\nu},\fA_\Lm)$, we get, for all $X\in\M{2^\nu}$,
\ba
|\eta(X)|
&\le\|\pi_\Lm^{-1}(X)\|\nonumber\\
&\le \|X\|_F,
\ea
\ie, $\eta$ is a continuous linear functional on the Hilbert space $\M{2^\nu}$ (with respect to the Frobenius scalar product). Hence, the Riesz representation theorem guarantees the existence of a unique $S_\Lm\in\M{2^\nu}$ such that $\eta(X)=(S_\Lm, X)_F$ for all $X\in\M{2^\nu}$. Setting $R_\Lm:=S_\Lm^\ast$, we get, for all $A\in\fA_\Lm$,
\ba
\omega_\Lm(A)
&=\omega_\Lm(\pi_\Lm^{-1}(\pi_\Lm(A)))\nonumber\\
&=\eta(\pi_\Lm(A))\nonumber\\
&=(R_\Lm^\ast, \pi_\Lm(A))_F\nonumber\\
&=\tr(R_\Lm\pi_\Lm(A)).
\ea

\ref{ls-b}\,
Since,  for all $X\in\M{2^\nu}$, we have
\ba
\tr(R_\Lm X^\ast X)
&=\omega_\Lm(\pi_\Lm^{-1}(X^\ast X))\nonumber\\
&=\omega_\Lm((\pi_\Lm^{-1}(X))^\ast \pi_\Lm^{-1}(X))\nonumber\\
&\ge 0,
\ea
we get $R_\Lm\ge 0$. Moreover,  due to \eqref{piLm-2}, we also have $\tr(R_\Lm)=\omega_\Lm(\pi_\Lm^{-1}(1_{2^\nu}))=\omega_\Lm(1)=1$. Finally, since we know from Lemma \ref{lem:mu} \ref{mu-c} that $\{e_{ab}\}_{a,b\in\num{1}{2}^\nu}$ is a basis of $\fA_\Lm$, the set $\{\pi_\Lm(e_{ab})\}_{a,b\in\num{1}{2}^\nu}$ is a basis of $\M{2^\nu}$ (see \eqref{piLm-2}). Hence, there exist coefficients $\lm_{ab}\in\C$ for all $a,b\in\num{1}{2}^\nu$ such that $R_\Lm=\sum_{a,b\in\num{1}{2}^\nu}\lm_{ab} \pi_\Lm(e_{ab})$. In order to compute these coefficients, let $n\in\num{1}{\nu}$ be fixed and note that, due to  \eqref{MUmu-1}, we have $e_{ab}=e_{ab}e_{ba}e_{ab}$ for all $a,b\in\num{1}{2}^n$. Hence, using \eqref{piLm-1}, the cyclicity of the trace, and \eqref{MUmu-1}, we get, for all $a,b\in\num{1}{2}^n$,
\ba
\tr(\pi_\Lm(e_{ab}))
&=\tr(\pi_\Lm(e_{ab}e_{ba})\pi_\Lm(e_{ab}))\nonumber\\
&=\tr(\pi_\Lm((e_{ab})^2e_{ba}))\nonumber\\
&=\delta_{ab}\hspace{0.2mm}\tr(\pi_\Lm(e_{aa})).
\ea
Moreover, for fixed $a\in\num{1}{2}^n$,  using \eqref{MUmu-3}, \eqref{MUmu-1}, and again the cyclicity of the trace, we have
\ba
\label{tr-eaa}
\tr(\pi_\Lm(e_{aa}))
&=\tr\Big(\pi_\Lm\Big(1-\sum\nolimits_{b\in\num{1}{2}^n,\hspace{0.5mm}b\neq a}e_{bb}\Big)\Big)\nonumber\\
&=2^\nu-\sum_{\substack{b\in\num{1}{2}^n\\ b\neq a}}\tr(\pi_\Lm(e_{ab}e_{ba}))\nonumber\\
&=2^\nu-(2^n-1)\tr(\pi_\Lm(e_{aa})),
\ea
which implies $\tr(\pi_\Lm(e_{ab}))=2^{\nu-n} \delta_{ab}$ for all $a,b\in\num{1}{2}^n$. Therefore, using \eqref{MUmu-1} and \eqref{MUmu-2}, we get $(\pi_\Lm(e_{ab}), \pi_\Lm(e_{cd}))_F=\delta_{ac}\tr(\pi_\Lm(e_{bd}))
=2^{\nu-n} \delta_{ac}\delta_{bd}$ for all $n\in\num{1}{\nu}$ and all $a,b,c,d\in\num{1}{2}^n$ which, for $n=\nu$ and all $a,b\in\num{1}{2}^\nu$, yields the expansion coefficients
\ba
\omega_\Lm(e_{ab}^\ast)
&=\tr(R_\Lm \pi_\Lm(e_{ab}^\ast))\nonumber\\
&=\sum_{c,d\in\num{1}{2}^\nu}\lm_{cd}\, (\pi_\Lm(e_{ab}), \pi_\Lm(e_{cd}))_F\nonumber\\
\label{Rexp-1}
&=\lm_{ab}.
\ea
\eprf

\vspace{5mm}

\noindent{\bf Proof of Lemma \ref{lem:fact}.}\hspace{2mm}
First, for all $n\in\num{1}{\nu}$ and all $a:=[\alpha_i]_{i\in\num{1}{n}}, b:=[\beta_i]_{i\in\num{1}{n}}\in\num{1}{2}^n$, we set
\ba
N_{ab}
:=\card(\{i\in\num{1}{n}\,|\,\alpha_i\neq\beta_i\}).
\ea
In the following, let $n\in\num{1}{\nu}$ and $a:=[\alpha_i]_{i\in\num{1}{n}}, b:=[\beta_i]_{i\in\num{1}{n}}\in\num{1}{2}^n$ be fixed. We then have the following cases.

{\it Case $N_{ab}=2k-1$ for some $k\in\N$.}\quad
Since, for all $\alpha\in\num{1}{2}$ and all $i\in\num{1}{n}$,
\ba
e_{\alpha\hspace{0.2mm}3-\alpha}^{(i)}
=\begin{cases}
\hfill B(J^\alpha Q_i), & i=1,\\
\big(\prod\nolimits_{j\in\num{1}{i-1}}B(Q_j+JQ_j)B(Q_j-JQ_j)\big) B(J^\alpha Q_i), & i\in\num{2}{n},
\end{cases}
\ea
there exist $m\in\N$ and $\{F_i\}_{i\in\num{1}{2m-1}}\subseteq\hh$ such that $e_{ab}=\prod_{j\in\num{1}{2m-1}}B(F_j)$. Hence, \eqref{qfs} implies that the left hand side of \eqref{fact} vanishes as does its right hand side since $N_{ab}\neq 0$.

{\it Case $N_{ab}=0$.}\quad
Note that \eqref{2pt-VVEV} yields, for all  $\alpha,\beta\in\num{1}{2}$ and all $i,j\in\num{1}{n}$ with $i\neq j$, 
\ba
\label{x-not-y}
\omega_\Lm(B(J^\alpha Q_i)B(J^\beta Q_j))
=0.
\ea
For the special case at hand, we have $a=b$ and, defining the functions $\{G_i\}_{i\in\num{1}{2n}}\subseteq\hh$ by $G_{2i-1}:=J^{\alpha_i}Q_i$ and $G_{2i}:=J^{3-\alpha_i}Q_i$ for all $i\in\num{1}{n}$, \eqref{qfs} yields
\ba
\omega_\Lm(e_{aa})
&=\pf\big([\omega(B(G_i)B(G_j))]_{i,j\in\num{1}{2n}}\big)\nonumber\\
&=\sum_{\pi\in\mP_{2n}}\sgn(\pi)\prod_{i\in\num{1}{n}} \omega_\Lm(B(G_{\pi(2i-1)})B(G_{\pi(2i)}))\nonumber\\
\label{om-BB}
&=\prod_{i\in\num{1}{n}} \omega_\Lm(B(G_{2i-1})B(G_{2i}))\\
\label{om-aa}
&=\prod_{i\in\num{1}{n}} \omega_\Lm(e_{\alpha_i\alpha_i}^{(i)}),
\ea
where, in \eqref{om-BB}, we used \eqref{x-not-y} which implies that only the identity permutation contributes to the sum over $\mP_{2n}$ (see Figure \ref{fig:pairings}).

{\it Case $N_{ab}=2k$ for some $k\in\N$.}\quad
For all $i\in\num{1}{n}$ and all $\alpha,\beta\in\num{1}{2}$, let $f_{\alpha\beta}^{(i)}\in\fA_\Lm$ be defined by
\ba
f_{\alpha\beta}^{(i)}
:=\begin{cases}
B^\ast(Q_i)B(Q_i), & (\alpha,\beta)=(1,1),\\
\hfill B^\ast(Q_i), & (\alpha,\beta)=(1,2),\\
\hfill B(Q_i), & (\alpha,\beta)=(2,1),\\
B(Q_i)B^\ast(Q_i), & (\alpha,\beta)=(2,2),
\end{cases}
\ea
and note that, due to \eqref{eiab} and \eqref{Si-4}, we have $e^{(i)}_{\alpha\beta}=S_i^{\alpha+\beta}f^{(i)}_{\alpha\beta}$ for all $i\in\num{1}{n}$ and all $\alpha,\beta\in\num{1}{2}$. Hence, if $n\ge 2$,  we can write
\ba
e_{ab}
&=\prod_{i\in\num{1}{n}}S_i^{\alpha_i+\beta_i}f^{(i)}_{\alpha_i\beta_i}\nonumber\\
\label{Egm-4}
&=\Big(\prod\nolimits_{i\in\num{1}{n-1}} f^{(i)}_{\alpha_i\beta_i}\Big(\prod\nolimits_{j\in\num{1}{i}} (2B(JQ_j)B(Q_j)-1)^{\alpha_{i+1}+\beta_{i+1}}\Big)\Big) f^{(n)}_{\alpha_{n}\beta_{n}}\\
\label{Egm-5}
&=\Big(\prod\nolimits_{i\in\num{1}{n-1}} f^{(i)}_{\alpha_i\beta_i} (2B(JQ_i)B(Q_i)-1)^{\sum_{j\in\num{i+1}{n}}(\alpha_j+\beta_j)}\Big) f^{(n)}_{\alpha_{n}\beta_{n}},
\ea
where we used \eqref{comij} in \eqref{Egm-4} and \eqref{Egm-5}. Moreover, using Definition \ref{def:Fs}, a direct check yields, as in Lemma \ref{lem:mu} \ref{mu-a}, that, for any fixed $i\in\num{1}{n}$,  the family $\{f_{\alpha\beta}^{(i)}\}_{\alpha,\beta\in\num{1}{2}}$ is a family of $2\times 2$ matrix units in $\fA_\Lm$ (however, in contrast to Lemma \ref{lem:mu} \ref{mu-a}, families at different sites are noncommuting, \ie, we have  $f_{\alpha\beta}^{(i)}f_{\gamma\delta}^{(j)}=(-1)^{(\alpha+\beta)(\gamma+\delta)} f_{\gamma\delta}^{(j)}f_{\alpha\beta}^{(i)}$ for all $i,j\in\num{1}{n}$ with $i\neq j$ and all $\alpha,\beta,\gamma,\delta\in\num{1}{2}$). Hence, setting $\gamma_i:=\sum_{j\in\num{i+1}{n}}(\alpha_j+\beta_j)$ for all $i\in\num{1}{n-1}$, the factors on the right hand side of \eqref{Egm-5} read, for all $i\in\num{1}{n-1}$,
\ba
\label{Egm-6}
f^{(i)}_{\alpha_i\beta_i} (2B(JQ_i)B(Q_i)-1)^{\gamma_i}
&=f^{(i)}_{\alpha_i\beta_i}  (f^{(i)}_{11}-f^{(i)}_{22})^{\gamma_i}\nonumber\\
&=f^{(i)}_{\alpha_i\beta_i}  \Big(\frac{1+(-1)^{\gamma_i}}{2}\,1+\frac{1-(-1)^{\gamma_i}}{2}\,(f^{(i)}_{11}-f^{(i)}_{22})\Big)\nonumber\\
&=\Big(\frac{1+(-1)^{\gamma_i}}{2}+\frac{1-(-1)^{\gamma_i}}{2}\,(\delta_{\beta_i1}-\delta_{\beta_i2})\Big)f^{(i)}_{\alpha_i\beta_i} \nonumber\\
&=(\delta_{\beta_i1}+\delta_{\beta_i2}(-1)^{\gamma_i}) f^{(i)}_{\alpha_i\beta_i}.
\ea
Plugging \eqref{Egm-6} into the right hand side of \eqref{Egm-5} and setting $\lambda:=\prod\nolimits_{i\in\num{1}{n-1}} (\delta_{\beta_i1}+\delta_{\beta_i2}(-1)^{\gamma_i})\in\{-1,1\}$, we thus get $e_{ab}=\lambda\prod_{i\in\num{1}{n}}f^{(i)}_{\alpha_i\beta_i}$ and 
\ba
\label{Egm-7}
\omega_\Lm(e_{ab})
=\lambda\hspace{0.2mm}\omega_\Lm\Big(\prod\nolimits_{i\in\num{1}{n}}f^{(i)}_{\alpha_i\beta_i}\Big).
\ea
Since $N_{ab}=2k$ for some $k\in\N$, there exists $i\in\num{1}{n}$ such that $\alpha_i\neq\beta_i$, \ie,  
\ba
\prod_{i\in\num{1}{n}}f^{(i)}_{\alpha_i\beta_i}
&=f^{(1)}_{\alpha_1\beta_1}\ldots f^{(i)}_{\alpha_i3-\alpha_i}\ldots f^{(n)}_{\alpha_n\beta_n}\nonumber\\
&=f^{(1)}_{\alpha_1\beta_1}\ldots B(J^{\alpha_i}Q_i) \ldots f^{(n)}_{\alpha_n\beta_n}.
\ea
Hence, each product in the Pfaffian \eqref{pfaff} includes a factor of the form $\omega_\Lm(B(J^{\alpha_i}Q_i)B(J^{\beta_j}Q_j))$ or $\omega_\Lm(B(J^{\beta_j}Q_j)B(J^{\alpha_i}Q_i))$ for some $j\in\num{1}{n}$ with $j\neq i$, and any such factor vanishes due to \eqref{x-not-y}. Hence, \eqref{Egm-7} implies $\omega_\Lm(e_{ab})=0$ and, due to $N_{ab}\neq 0$, the right hand side of  \eqref{fact} vanishes, too.
\eprf

\vspace{5mm}

\noindent{\bf Proof of Proposition \ref {prop:spec}.}\hspace{2mm}
\ref{spec-a}\,
Let $Q_i':=UQ_i$ for all $i\in\num{1}{\nu}$ and let us denote by $e^{\prime(i)}_{\alpha\beta}$ and $e'_{ab}$ for all $i\in\num{1}{\nu}$, all $\alpha,\beta\in\num{1}{2}$, and all $a,b\in\num{1}{2}^\nu$, the quantities \eqref{eiab} and \eqref{eab}, respectively, in which the Fermi family $\{Q_i\}_{i\in\num{1}{\nu}}$ has been replaced by the Fermi family $\{Q_i'\}_{i\in\num{1}{\nu}}$. Since $\{Q_i'\}_{i\in\num{1}{\nu}}$ satisfies \An{\ref{ass:2pt} \ref{2pt-VV},\ref{2pt-EV}}, \eqref{RLm} and \eqref{fact} yield
\ba
R_\Lm
=\sum_{a\in\num{1}{2}^\nu} \kp_a\,\pi_\Lm(e'_{aa}),
\ea
where $\kp_a:=\prod_{i\in\num{1}{\nu}}\omega(e^{\prime(i)}_{\alpha_i\alpha_i})$ for all $a:=[\alpha_i]_{i\in\num{1}{\nu}}\in\num{1}{2}^\nu$. Moreover, \eqref{MUmu-1}, \eqref{MUmu-3}, and \eqref{piLm-1} imply that $\pi_\Lm(e'_{aa})\pi_\Lm(e'_{bb})=\delta_{ab} \pi_\Lm(e'_{aa})$ for all $a,b\in\num{1}{2}^\nu$ and $\sum_{a\in\num{1}{2}^\nu}\pi_\Lm(e'_{aa})=1_{2^\nu}$, \ie, the family $\{\pi_\Lm(e'_{aa})\}_{a\in\num{1}{2}^\nu}$ is a complete orthogonal family of (orthogonal) projections in $\M{2^\nu}$ implying that
\ba
\spec(R_\Lm)
=\{\kp_a\,|\, a\in\num{1}{2}^\nu\}.
\ea
Moreover, using \eqref{CAR-2}, we have, for all $i\in\num{1}{\nu}$ and all $\alpha\in\num{1}{2}$, 
\ba
e^{\prime(i)}_{\alpha\alpha}
&=\frac{1-(-1)^\alpha}{2} B^\ast(Q_i')B(Q_i')+\frac{1+(-1)^\alpha}{2} B(Q_i')B^\ast(Q_i')\nonumber\\
&=\frac12\{B^\ast(Q_i'),B(Q_i')\}+\frac{(-1)^\alpha}{2}\,(\{B^\ast(Q_i'),B(Q_i')\}-2B^\ast(Q_i')B(Q_i'))\nonumber\\
&=\frac{1+(-1)^\alpha(1-2B^\ast(Q_i')B(Q_i'))}{2},
\ea
and the normalization of $\omega_\Lm$ yields $\omega_\Lm(e^{\prime(i)}_{\alpha\alpha})=(1+(-1)^\alpha\lm_i)/2$ for all $i\in\num{1}{\nu}$ and all $\alpha\in\num{1}{2}$.  Finally, \eqref{ArakiF} and \eqref{Fs-d} yield $\|B(UQ_i)\|=1$ for all $i\in\num{1}{\nu}$, and we get, for all $i\in\num{1}{\nu}$, 
\ba
0
&\le \omega_\Lm(B^\ast(UQ_i)B(UQ_i))\nonumber\\
&\le \|B^\ast(UQ_i)B(UQ_i)\|\nonumber\\
&=\|B(UQ_i)\|^2\nonumber\\
&=1,
\ea
\ie, $|\lm_i|\le 1$ for all $i\in\num{1}{\nu}$.

\ref{spec-b}\,
Let $\{M_i\}_{i\in\num{1}{2\nu}}$ be the Majorana family from Definition \ref{def:Ms} associated with the Fermi family $\{Q_i\}_{i\in\num{1}{\nu}}$, let $\Omega\in\M{2\nu}$ be given  by \eqref{Omij}, and define $\Xi\in\R^{2\nu\times 2\nu}$ by
\ba
\Xi
:=\Im(\Omega).
\ea
Note that, for all $i,j\in\num{1}{2\nu}$, we have $\bar \Omega_{ij}=\Omega_{ji}$ due to \eqref{SDCAR1} and \eqref{Ms-1} and $\Omega_{ij}=-\Omega_{ji}+\delta_{ij}$ due to \eqref{SDCAR2} and the normalization of $\omega_\Lm$, \ie, we have
\ba
\label{Z-1}
\Omega^\ast
&=\Omega,\\
\label{Z-2}
\Omega^T
&=1_{2\nu}-\Omega,
\ea
and \eqref{Z-1}-\eqref{Z-2} imply that
\ba
\label{ZReIm}
\Omega
=\frac 12 1_{2\nu}+\ii\Xi.
\ea
Hence, due to \eqref{Z-2}, $\Xi\in\R^{2n\times 2n}$ has the additional property 
\ba
\Xi^T
=-\Xi, 
\ea
\ie, $\Xi$ is real and skew symmetric. Therefore, we know (see, for example, \cite{HoJo85}) that there exist $V\in\G{O}(2\nu)$ and $\{\xi_i\}_{i\in\num{1}{\nu}}\subseteq\R$ transforming $\Xi$ into its real canonical form
\ba
\label{Vlm}
V^T \Xi V
&=\bigoplus_{i\in\num{1}{\nu}}
\begin{bmatrix}
0 & \xi_i\\
-\xi_i & 0
\end{bmatrix}.
\ea
Setting $\h_{2\nu}^{\oplus 2}:=\spa(\{M_i\}_{i\in\num{1}{2\nu}})$ and noting that, due to \eqref{Ms-2}, $\{M_i/\sqrt{2}\}_{i\in\num{1}{2\nu}}$ is an orthonormal basis of $\h_{2\nu}^{\oplus 2}$, we define $U\in\mL(\hh)$, for all $i\in\num{1}{2\nu}$, by
\ba
\label{UV}
UM_i
:=\sum_{j\in\num{1}{2\nu}} V_{ji} M_j,
\ea
by $UF:=F$ for all $F\in(\h_{2\nu}^{\oplus 2})^\perp$, and by linear extension (and the Hilbert projection theorem) to the whole of $\hh$. Using \eqref{UV}, \eqref{Ms-2}, and $V\in\G{O}{(2\nu)}$, we get, for all $i,j\in\num{1}{2\nu}$,
\ba
(UM_i, UM_j)_\hh
&=2(V^TV)_{ij}\nonumber\\
&=(M_i, M_j)_\hh,
\ea
and, hence, $(UF,UG)_\hh=(F,G)_\hh$ for all $F,G\in\h_{2\nu}^{\oplus 2}$. Moreover,  we have $(UF,UG)_\hh=(UF,G)_\hh=0=(F,G)_\hh$ for all $F\in\h_{2\nu}^{\oplus 2}$ and all $G\in(\h_{2\nu}^{\oplus 2})^\perp$ because $U\h_{2\nu}^{\oplus 2}\subseteq\h_{2\nu}^{\oplus 2}$ (\ie, $UF\in\h_{2\nu}^{\oplus 2}$ for all $F\in\h_{2\nu}^{\oplus 2}$) and, hence, we get $U\in\G{U}(\hh)$. Since \eqref{Ms-1} implies $J\h_{2\nu}^{\oplus 2}\subseteq\h_{2\nu}^{\oplus 2}$ (from which $J(\h_{2\nu}^{\oplus 2})^\perp\subseteq(\h_{2\nu}^{\oplus 2})^\perp$), we have $JUJM_i=UM_i$ for all $i\in\num{1}{2\nu}$ and $JUJF=J^2F=UF$ for all $F\in(\h_{2\nu}^{\oplus 2})^\perp$, \ie, we arrive at $U\in\G{U}_{\!J}(\hh)$. 
Next, due to Remark \ref{rem:loc}, the family $\{UQ_i\}_{i\in\num{1}{\nu}}\subseteq\hh$ is a Fermi family over $\Lm$ because \eqref{Hodd}-\eqref{Heven} and \eqref{UV} yield, for all $i\in\num{1}{\nu}$, 
\ba
\label{UQi-1}
U Q_i
&=\sum_{j\in\num{1}{2\nu}} \frac{V_{j 2i-1}-\ii V_{j 2i}}{2}\, M_j\\
\label{UQi-2}
&=\sum_{j\in\num{1}{\nu}}\sum_{\alpha\in\num{1}{2}}  c^{(i)}_{j\alpha} J^\alpha Q_j,
\ea
where, for all $i, j\in\num{1}{\nu}$ and all $\alpha\in\num{1}{2}$,
\ba
c^{(i)}_{j\alpha}
&:=\frac{1}{2}\hspace{0.2mm}(V_{2j-1 2i-1}+(-1)^{\alpha} V_{2j 2i}+\ii ((-1)^\alpha V_{2j 2i-1}- V_{2j-1 2i})),
\ea
and because $(JQ_i)_\alpha=\overline{(Q_i)}_{3-\alpha}$ for all $i\in\num{1}{\nu}$ and all $\alpha\in\num{1}{2}$, \ie, we get, for all  $i\in\num{1}{\nu}$ and all $\alpha\in\num{1}{2}$, 
\ba
\supp((UQ_i)_\alpha)
\subseteq\Lm. 
\ea
Finally, we want to check that the Fermi family $\{UQ_i\}_{i\in\num{1}{\nu}}\subseteq\hh$ over $\Lm$ satisfies \An{\ref{ass:2pt} \ref{2pt-VV},\ref{2pt-EV}} in the form \eqref{QRJQ}-\eqref{QRQ}. Using \eqref{UQi-1}, $\Omega_{ij}=(M_i,RM_j)_\hh/2$ for all $i,j\in\num{1}{2\nu}$ from \eqref{Omij}, \eqref{ZReIm}, and the fact that the right hand side of \eqref{Vlm}, denoted by $W\in\M{2\nu}$, satisfies $W_{2i-1 2j-1}=0$, $W_{2i-1 2j}=\delta_{ij}\xi_i$, $W_{2i 2j-1}=-\delta_{ij}\xi_i$, and $W_{2i 2j}=0$ for all $i,j\in\num{1}{\nu}$, we get, for all $i,j\in\num{1}{\nu}$, 
\ba
\label{UQR-1}
(UQ_i, R UQ_j)_\hh
&=\frac12\hspace{0.2mm}((V^T\Omega V)_{2i-1\hspace{0.2mm}2j-1}-\ii (V^T\Omega  V)_{2i-1\hspace{0.2mm} 2j}+\ii (V^T\Omega V)_{2i\hspace{0.2mm} 2j-1} +(V^T\Omega V)_{2i\hspace{0.2mm} 2j})\\
&=\frac12\hspace{0.2mm} \delta_{ij} +\frac{\ii}{2}\hspace{0.2mm} (W_{2i-1\hspace{0.2mm} 2j-1}-\ii W_{2i-1\hspace{0.2mm} 2j}+\ii W_{2i\hspace{0.2mm} 2j-1}+W_{2i\hspace{0.2mm} 2j})\nonumber\\
\label{UQR-3}
&=\delta_{ij} \big(\xi_i+\tfrac12\big),
\ea
\ie, we obtain \eqref{QRQ}. As for \eqref{QRJQ}, since, due to  \eqref{UQi-1}, replacing $UQ_j$ by $JUQ_j$ amounts to changing the sign of the second and the fourth term on the right hand side of \eqref{UQR-1}, we have, for all $i,j\in\num{1}{\nu}$, 
\ba
(UQ_i, R JUQ_j)_\hh
&=\frac{\ii}{2}\hspace{0.2mm} (W_{2i-1\hspace{0.2mm} 2j-1}+\ii W_{2i-1\hspace{0.2mm} 2j}+\ii W_{2i \hspace{0.2mm}2j-1}-W_{2i \hspace{0.2mm}2j})\nonumber\\
&=0,
\ea
\ie, we get \eqref{QRJQ}, too. 
Part \ref{spec-a} of the proposition now yields that $\spec(R_\Lm)$ satisfies \eqref{specRLm} with $\lm_i=1-2(UQ_i,RUQ_i)_\hh=-2\xi_i$ for all $i\in\num{1}{\nu}$. Furthermore, due to the block diagonal structure of $W$ from \eqref{Vlm}, we can write, for all $z\in\C$, 
\ba
\det(\Xi-z1_{2\nu})
&=\det(W-z1_{2\nu})\nonumber\\
&=\prod_{i\in\num{1}{\nu}}(z-\ii\xi_i)(z+\ii\xi_i),
\ea
\ie, we also get \eqref{spec-b-1}.
\eprf

\vspace{5mm}

\noindent{\bf Proof of Proposition \ref{prop:Shan}.}\hspace{2mm}
\ref{Shan-a}\,
Note that $(1+(-1)^\alpha \lambda)/2\in\,]0,1[$ for all $\alpha\in\num{1}{2}$ and all $\lambda\in\,]-1,1[$\,. Hence, the product on the left hand side of \eqref{FEq-1} is indeed in the domain of definition of $\ell$. Moreover, using \eqref{Shan}, the logarithmic functional equation, and the fact that, for all $n\ge 2$ and all $i\in\num{1}{n}$ (for $n=1$, \eqref{FEq-1} is the definition of $\eta$),
\ba
\sum_{a\setminus \alpha_i\in\num{1}{2}^{n-1}}\prod_{j\in\num{1}{n}\setminus\{i\}}\frac{1+(-1)^{\alpha_j}\lambda_j}{2}
&=\prod_{j\in\num{1}{n}\setminus\{i\}}\sum_{\alpha_j\in\num{1}{2}}\frac{1+(-1)^{\alpha_j}\lambda_j}{2}\nonumber\\
&=1, 
\ea
where,  for all $i\in\num{1}{n}$,  we define $a\setminus\alpha_i\in\num{1}{2}^{n-1}$ to be equal to $a$ without its $i$th entry, we get \eqref{FEq-1}.
 
\ref{Shan-b}\,
For all $n\in\N$ and all $i\in\num{1}{n}$, we set $x_i:=(1+\lm_i)/2\in\,]0,1[$ and we call \eqref{FEq-2} the $n$-functional equation. As for the 1-functional equation, the left hand side of \eqref{FEq-2} equals $f(x_1)+f(1-x_1)$ for all $x_1\in\,]0,1[$ and so does the right hand side, \ie, the condition for $n=1$ is void. If $n=2$, \eqref{FEq-2} yields the 2-functional equation, \ie,  for all $x_1,x_2\in\,]0,1[$,
\ba
f(x_1x_2)+f(x_1(1-x_2))+f((1-x_1)x_2)+f((1-x_1)(1-x_2))
&=f(x_1)+f(1-x_1)\nonumber\\
&\hspace{4mm}+f(x_2)+f(1-x_2). 
\ea
Hence, under the stated measurability assumption, we know from \cite{DaJa79} that there exist $A,B,C\in\R$ such that, for all $x\in\,]0,1[$,
\ba
\label{FEq-3}
f(x)
=A+B(4x^3-9x^2+5x)+C x\log(x),
\ea
and that for all $A,B,C\in\R$, the right hand side of \eqref{FEq-3} satisfies the 2-functional equation. Plugging \eqref{FEq-3} into the 3-functional equation, a direct calculation yields $2A-36Bx_1(1-x_1)x_2(1-x_2)x_3(1-x_3)=0$ for all $x_1,x_2,x_3\in\,]0,1[$, \ie, we get $A=B=0$. Finally, since the $n$-functional equations \eqref{FEq-2} are homogeneous in $f$ for all $n\in\N$ and since, due to \ref{Shan-a}, the function $\ell$ satisfies the $n$-functional equations for all $n\in\N$, \eqref{FEq-4} is compatible with all the additional $n$-functional equations for $n\ge 4$.
\eprf

\vspace{5mm}

\noindent{\bf Proof of Proposition \ref{prop:vNEnt}.}\hspace{2mm}
Using Proposition \ref{prop:spec} and \eqref{SLmSum}, we have
\ba
\label{vNEnt-1}
S_\Lm
=\sum_{a\in\num{1}{2}^\nu}\wt\ell\Big(\prod\nolimits_{i\in\num{1}{\nu}}\tfrac{1+(-1)^{\alpha_i}\lm_i}{2}\Big),
\ea
where again $a:=[\alpha_i]_{i\in\num{1}{\nu}}\in\num{1}{2}^\nu$. Now, let 
\ba
N
:=\{i\in\num{1}{\nu}\,|\, \lm_i\in\,]-1,1[\},
\ea
and set $n:=\card(N)\in\num{0}{\nu}$. If $n=\nu$, \eqref{vNEnt-1}, \eqref{FEq-1} for $n=\nu$, and the fact that $\wt\ell$ and $\wt\eta$ are extensions of $\ell$ and $\eta$, respectively, imply \eqref{vNEnt-H}. If $n\in\hspace{1mm}]0,\nu[$, using \eqref{vNEnt-1}, the fact that, for all $\lm\in\{-1,1\}$ and all $r\in\R$, 
\ba
\sum_{\alpha\in\num{1}{2}}\wt\ell\big(\tfrac{(1+(-1)^\alpha\lm)r}{2}\big)
=\wt\ell(0)+\wt\ell(r), 
\ea
the property $\wt\ell(0)=0$, \eqref{FEq-1}, and $\wt\eta(-1)=\wt \eta(1)=0$, we get 
\ba
S_\Lm
&=\sum_{[\alpha_i]_{i\in N}\in\num{1}{2}^{n}}\hspace{1mm}\sum_{[\alpha_j]_{j\in\num{1}{\nu}\setminus N}\in\num{1}{2}^{\nu-n}}\wt\ell\Big(\Big(\prod\nolimits_{j\in\num{1}{\nu}\setminus N}\tfrac{1+(-1)^{\alpha_j}\lm_j}{2}\Big)\Big(\prod\nolimits_{i\in N}\tfrac{1+(-1)^{\alpha_i}\lm_i}{2}\Big)\Big)\nonumber\\
&=\sum_{[\alpha_i]_{i\in N}\in\num{1}{2}^{n}}\ell\Big(\prod\nolimits_{i\in N}\tfrac{1+(-1)^{\alpha_i}\lm_i}{2}\Big)\nonumber\\
&=\sum_{i\in N} \eta(\lm_i)\nonumber\\
&=\sum_{i\in\num{1}{\nu}}\wt\eta(\lm_i),
\ea
where $[\alpha_i]_{i\in N}:=[\alpha_{i_k}]_{k\in\num{1}{n}}\in\num{1}{2}^{n}$ with $i_k\in N$ for all $k\in\num{1}{n}$ and $i_1< i_2<\ldots <i_n$ (and analogously  for $[\alpha_j]_{j\in\num{1}{\nu}\setminus N}\in\num{1}{2}^{\nu-n}$).
For the case $n=0$, since $(1+(-1)^\alpha\lm)/2\in\num{0}{1}$ for all $\alpha\in\num{1}{2}$ and all $\lm\in\{-1,1\}$, the fact that $\wt\ell(0)=\wt\ell(1)=0$ implies that the right hand side of \eqref{vNEnt-1} vanishes. But, since again $\wt\eta(-1)=\wt\eta(1)=0$, the right hand side of \eqref{vNEnt-H} vanishes, too. Finally, the property of the spectrum of $\Im(\Omega)$ stems from Proposition \ref{prop:spec} \ref{spec-b}.
\eprf

\vspace{5mm}

\noindent{\bf Proof of Lemma \ref{lem:trans}.}\hspace{2mm}
Let us write $R=[R_{ij}]_{i,j\in\num{1}{2}}$, where $R_{ij}\in\mL(\h)$ for all $i,j\in\num{1}{2}$. Since $[R,\theta 1_2]=[[R_{ij},\theta]]_{i,j\in\num{1}{2}}$, \An{\ref{ass:trans} \ref{trans-a}}  yields $[R_{ij},\theta]=0$ for all  $i,j\in\num{1}{2}$, \ie, $[\wh R_{ij},\wh \theta\,]=0$ for all  $i,j\in\num{1}{2}$. Moreover,  since 
\ba
\wh \theta
=m[\e_1], 
\ea
we get, for all $i,j\in\num{1}{2}$ and all $\vi\in\fh$,
\ba
\wh R_{ij}\e_1\vi
&=\e_1\wh R_{ij}\vi,\\
\wh R_{ij}\e_{-1}\vi
&=\e_{-1}\wh R_{ij}\vi. 
\ea
Hence, $\wh R_{ij}\e_x=\e_x\wh R_{ij}\e_0$ for all $i,j\in\num{1}{2}$ and all $x\in\Z$ which implies, for all $i,j\in\num{1}{2}$ and all $x,y\in\Z$,
\ba
(\e_x,\wh R_{ij} \e_y)_\fh
=(\e_{x-y}, \wh R_{ij}\e_0)_\fh. 
\ea
Therefore, we know (see, for example, \cite{BoSi06}) that, for all  $i,j\in\num{1}{2}$, there exists $r_{ij}\in L^\infty(\T)$ such that $\wh R_{ij}=m[r_{ij}]$.
\eprf

\vspace{5mm}

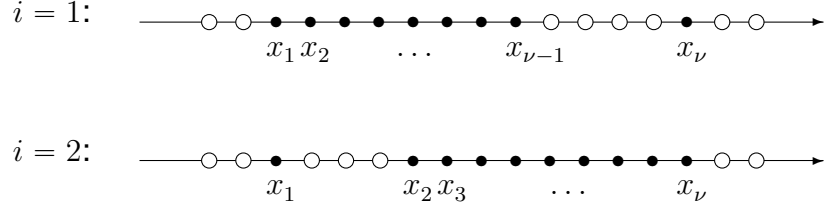
\begin{figure}
\setlength{\unitlength}{9mm}
\begin{center}
$i=1$:\hspace{5mm}
\begin{picture}(10,2)
\put(0,0){\vector(1,0){10}}
\multiput(0.9,0)(0.5,0){2}{\tikzcircle[black, fill=white]{0.1}}
\multiput(2,0)(0.5,0){8}{\circle*{0.16}}
\multiput(5.9,0)(0.5,0){4}{\tikzcircle[black, fill=white]{0.1}}
\put(8,0){\circle*{0.16}}
\multiput(8.4,0)(0.5,0){2}{\tikzcircle[black, fill=white]{0.1}}
\put(1.85,-0.5){$x_1$}
\put(2.35,-0.5){$x_2$}
\put(3.75,-0.5){$\ldots$}
\put(5.35,-0.5){$x_{\nu-1}$}
\put(7.85,-0.5){$x_\nu$}
\end{picture}\\
$i=2$:\hspace{5mm}
\begin{picture}(10,2)
\put(0,0){\vector(1,0){10}}
\multiput(0.9,0)(0.5,0){2}{\tikzcircle[black, fill=white]{0.1}}
\put(2,0){\circle*{0.16}}
\multiput(2.4,0)(0.5,0){3}{\tikzcircle[black, fill=white]{0.1}}
\multiput(4,0)(0.5,0){9}{\circle*{0.16}}
\multiput(8.4,0)(0.5,0){2}{\tikzcircle[black, fill=white]{0.1}}
\put(1.85,-0.5){$x_1$}
\put(3.85,-0.5){$x_2$}
\put(4.35,-0.5){$x_3$}
\put(6,-0.5){$\ldots$}
\put(7.85,-0.5){$x_\nu$}
\end{picture}
\end{center}
\caption{Some of the configurations in the proof of Lemma \ref{lem:StdCtg} \ref{StdCtg-a}.}
\label{fig:ctg}
\end{figure}

\noindent{\bf Proof of Lemma \ref{lem:StdCtg}.}\hspace{2mm}
\ref{StdCtg-a}\,
Let us make the converse hypothesis $\num{x_1}{x_\nu}\setminus\Lm\neq \emptyset$ and set, for all $i\in\num{1}{\nu}$,
\ba
n_i
:=\min(\{n\in\N\,|\,x_i+n\notin\Lm\}). 
\ea
Using that $\card(\{j\in\num{1}{\nu}\,|\,x_j>x_i\})=\nu-i$ for all $i\in\num{1}{\nu}$ and our hypothesis, we have $n_1\le (\nu-1)-1+1=\nu-1$ (there exists at least one $x\in\num{x_1}{x_\nu}$ with $x\notin\Lm$, see Figure \ref{fig:ctg}). 

If $i\in\num{2}{\nu}$, since all $x\in\num{x_1}{x_\nu}$ with $x\notin\Lm$ may lie to the left of $x_i$  (see again Figure \ref{fig:ctg}), we get $n_i\le\nu-i+1\le\nu-1$ for all $i\in\num{2}{\nu}$.  Since \An{\ref{ass:trans} \ref{trans-b}} reads, for all $i\in\num{1}{\nu-1}$, 
\ba
\label{Qi+1}
Q_{i+1}
=(\theta1_2)^iQ_1,
\ea
and since there exist $\lm_{\alpha,j}\in\C$ for all $\alpha\in\num{1}{2}$ and all $j\in\num{1}{\nu}$ such that, for all $\alpha\in\num{1}{2}$,
\ba
\label{Q1alpha}
(Q_1)_\alpha
=\sum_{j\in\num{1}{\nu}}\lm_{\alpha,j}\hspace{0.2mm}\delta_{x_j}, 
\ea
we get $(Q_{n_i+1})_\alpha=\sum_{j\in\num{1}{\nu}}\lm_{\alpha,j}\delta_{x_j+n_i}$ for all $\alpha\in\num{1}{2}$ and all $i\in\num{1}{\nu}$. Given that $Q_{n_i+1}\in\{Q_j\}_{j\in\num{1}{\nu}}$ for all $i\in\num{1}{\nu}$ because $n_i+1\in\num{2}{\nu}$ for all $i\in\num{1}{\nu}$, Definition \ref{def:Fs} \ref{Fs-c} leads to $\lm_{\alpha,i}=0$ for all $\alpha\in\num{1}{2}$ and all $i\in\num{1}{\nu}$, \ie, $Q_1=0$ (which contradicts Definition \ref{def:Fs} \ref{Fs-a}).

\ref{StdCtg-b}\,
Due to \eqref{Qi+1}-\eqref{Q1alpha}, we have $(Q_\nu)_\alpha=\sum_{j\in\num{1}{\nu}}\lm_{\alpha,j}\delta_{x_j+\nu-1}$ for all $\alpha\in\num{1}{2}$. Hence, due to \ref{StdCtg-a}, we get $\lm_{\alpha,j}=0$ for all $\alpha\in\num{1}{2}$ and all $j\in\num{2}{\nu}$, \ie, we find $Q_1=(\lm_1\delta_{x_1})\oplus (\lm_2\delta_{x_1})$, where $\lm_\alpha:=\lm_{\alpha,1}$ for all $\alpha\in\num{1}{2}$. Finally, using Definition \ref{def:Fs} \ref{Fs-a} and \ref{Fs-b}, we get $|\lm_1|^2+|\lm_2|^2=1$ and $\lm_1\lm_2=0$, respectively. 
\eprf

\vspace{5mm}

\noindent{\bf Proof of Proposition \ref{prop:ToepS}.}\hspace{2mm}
\ref{ToepS-a}\,
Using \An{\ref{ass:trans} \ref{trans-b}}, \eqref{Hodd}-\eqref{Heven}, and  $[\theta1_2,J]=0$, we have, for all $i\in\num{1}{\nu-1}$, that $(\theta1_2)M_{2i-1}=Q_{i+1}+(\theta1_2)JQ_i=Q_{i+1}+JQ_{i+1}=M_{2i+1}$ and $(\theta1_2)M_{2i}=\ii(Q_{i+1}-(\theta1_2)JQ_i)=M_{2i+2}$, \ie, for all $i\in\num{1}{2\nu-2}$,
\ba
\label{HTrs}
(\theta1_2)M_i
=M_{i+2}.
\ea
Hence, due to \eqref{HTrs} and \An{\ref{ass:trans} \ref{trans-a}}, we get, for all $i, j\in\num{1}{2\nu-2}$,
\ba
\wt\Omega_{i+2\hspace{0.3mm}j+2}
&=\frac12\hspace{0.5mm}(M_{i+2}, RM_{j+2})_\hh\nonumber\\
&=\frac12\hspace{0.5mm}(M_i, (\theta^\ast1_2)R(\theta1_2)M_j)_\hh\nonumber\\
&=\wt\Omega_{ij}. 
\ea
Next, let us define $b:\num{1}{\nu}^{\times 2}\to\M{2}$, for all $i,j\in\num{1}{\nu}$, by
\ba
b(i,j)
:=\begin{bmatrix}
\wt\Omega_{2i-1\hspace{0.3mm}2j-1} & \wt\Omega_{2i-1\hspace{0.3mm}2j}\\
\wt\Omega_{2i\hspace{0.3mm}2j-1} & \wt\Omega_{2i\hspace{0.3mm}2j}
\end{bmatrix},
\ea
and note that, now, for all $i,j\in\num{1}{\nu-1}$,
\ba
b(i+1,j+1)
=b(i,j). 
\ea
Moreover, if we define the map $\wt b:\num{-(\nu-1)}{\nu-1}\to\M{2}$, for all $x\in\num{-(\nu-1)}{\nu-1}$, by
\ba
\wt b(x)
:=b\big(1+\tfrac{|x|+x}{2},1+\tfrac{|x|-x}{2}\big), 
\ea
we get $b(i,j)=\wt b(i-j)$ for all $i,j\in\num{1}{\nu}$. Hence, the Majorana correlation matrix becomes
\ba
\label{OmBT}
\wt\Omega
&=[\wt\Omega_{ij}]_{i,j\in\num{1}{2\nu}}\nonumber\\
&=[b(i,j)]_{i,j\in\num{1}{\nu}}\nonumber\\
&=[\wt b(i-j)]_{i,j\in\num{1}{\nu}}\nonumber\\
&=\begin{bmatrix}
\wt b(0) & \wt b(-1) & \wt b(-2) & \ldots & \wt b(-(\nu-1))\\
\wt b(1) & \wt b(0) & \wt b(-1) & \ldots & \wt b(-(\nu-2))\\
\wt b(2) & \wt b(1) & \wt b(0) & \ldots & \wt b(-(\nu-3)) \\
\vdots & \vdots & \vdots &\ddots & \vdots\\
\wt b(\nu-1)&  \wt b(\nu-2) &  \wt b(\nu-3) & \ldots & \wt b(0)
\end{bmatrix},
\ea
and, using again \An{\ref{ass:trans} \ref{trans-a}}, the notation $\theta^{-n}:=(\theta^{-1})^n=(\theta^\ast)^n$ for all $n\in\N$ and $\theta^0:=1$, the fact that $\wh\theta=m[\e_1]$, \eqref{mmo}, and $(\Gm [m[c_{ij}]]_{i,j\in\num{1}{2}}\Gm^\ast\Phi)(z)=c(z)\Phi(z)$ for all $c_{ij}\in L^\infty(\T)$ with $i,j\in\num{1}{2}$, all $\Phi\in L^2(\T,\C^2)$, and almost all $z\in\T$, where $c\in L^\infty(\T, \M{2})$ is defined by $(c)_{ij}:=c_{ij}$ for all $i,j\in\num{1}{2}$, we have, for all $x\in\num{-(\nu-1)}{\nu-1}$ and all $i,j\in\num{1}{2}$,
\ba
\label{tbx}
(\wt b(x))_{ij}
&=\frac12\hspace{0.5mm}(M_i, R (\theta1_2)^{-x} M_j)_\hh\nonumber\\
&=\frac12\hspace{0.5mm}(\wh M_i, \wh R (m[\e_{-x}]1_2)\wh M_j)_\fhh\nonumber\\
&=\frac12\hspace{0.5mm}(\Gm\wh M_i, \Gm [m[r_{ij}]]_{i,j\in\num{1}{2}}\Gm^\ast\Gm(m[\e_{-x}]1_2)\Gm^\ast\Gm\wh M_j)_{L^2(\T,\C^2)}\nonumber\\
&=\frac12\Int((\Gm\wh M_i)(\e^{\ii k}), r(\e^{\ii k})(\Gm\wh M_j)(\e^{\ii k}))_{\C^2} \hspace{0.5mm}\e^{-\ii k x}\nonumber\\
&=\Int(a(\e^{\ii k}))_{ij} \hspace{0.5mm}\e^{-\ii k x},
\ea
where we used  \eqref{block}. Since $(\wh M_i)_\alpha\in L^\infty(\T)$ for all $i, \alpha\in\num{1}{2}$ due to Definition \ref{def:Ms} and Definition \ref{def:Fs} \ref{Fs-c}, we get $a\in L^\infty(\T,\M{2})$ (due to H\"older's inequality) and $\wt b(x)=\wc a(x)$ for all $x\in\num{-(\nu-1)}{\nu-1}$. Hence, we know from Toeplitz's theorem (see, for example, \cite{BoSi99}) that $T[a]\in\mL(\ell^2(\N,\C^2))$ and  since, due to \eqref{OmBT} and \eqref{BlockToep},
\ba
\wt\Omega
&=[\wt b(i-j)]_{i,j\in\num{1}{\nu}}\nonumber\\
&=[\wc a(i-j)]_{i,j\in\num{1}{\nu}}\nonumber\\
&=T_{a,\nu}, 
\ea
we arrive at $T_\nu[a]=m_{T_{a,\nu}}=m_{\wt\Omega}$. Note  (see, for example, \cite{BoSi06}) that, for all $a\in L^\infty(\T,\M{2})$,
\ba
\|T[a]\|_{\mL(\ell^2(\N,\C^2))}
=\|a\|_{L^\infty(\T,\M{2})}, 
\ea
and, hence, $a$ is unique in $L^\infty(\T,\M{2})$.

\ref{ToepS-b}\,
Due to \eqref{Hodd}-\eqref{Heven} and \eqref{StdFS}, the Majorana family $\{M_i\}_{i\in\num{1}{2\nu}}$ associated with the Fermi family $\{Q_i\}_{i\in\num{1}{\nu}}$ has the form $(\wh M_{2i-1})_\alpha=\lm_{\gamma+\alpha-1}\e_{x_i}$ and $(\wh M_{2i})_\alpha=(-1)^{\gamma+\alpha-1}\ii\lm_{\gamma+\alpha-1}\e_{x_i}$ for all $i\in\num{1}{\nu}$. Hence, \eqref{block} reads (without using \eqref{def:2pt-op-1}-\eqref{def:2pt-op-3}), for almost all $z\in\T$,
\ba
\label{a11}
(a(z))_{11}
&=\frac{1}{2}(r_{11}(z)+r_{22}(z)+\lm_{\gm+1}^2 r_{12}(z) +\lm_{\gm}^2 r_{21}(z)),\\
(a(z))_{12}
&=\frac{\ii (-1)^\gm}{2}(r_{11}(z)-r_{22}(z)-\lm_{\gm+1}^2 r_{12}(z) +\lm_{\gm}^2 r_{21}(z)),\\
(a(z))_{21}
&=\frac{\ii (-1)^{\gm+1}}{2}(r_{11}(z)-r_{22}(z)+\lm_{\gm+1}^2 r_{12}(z)-\lm_{\gm}^2 r_{21}(z)),\\
\label{a22}
(a(z))_{22}
&=\frac{1}{2}(r_{11}(z)+r_{22}(z)-\lm_{\gm+1}^2 r_{12}(z)-\lm_{\gm}^2 r_{21}(z)).
\ea
Moreover, since $(\wh\zeta\vi)(z)=\overline{\vi(\bar z\hspace{0.4mm})}$ for all $\vi\in\fh$ and almost all $z\in\T$, where $\wh\zeta:=\ff\zeta\ff^\ast\in\bar\mL(\fh)$  (see Definition \ref{def:obs} \ref{obs-c}), \eqref{def:2pt-op-2} rewritten in $\fhh$ is equivalent to 
\ba
\label{2pt-2a}
\wh\zeta r_{11}
&=1-r_{22},\\
\label{2pt-2b}
\wh\zeta r_{12}
&=-r_{21}.
\ea
Plugging \eqref{2pt-2a}-\eqref{2pt-2b} into \eqref{a11}-\eqref{a22}, we get $\overline{a(z)}=1_2-a(\bar z)$ for almost all $z\in\T$ and, hence, \eqref{tbx} implies $\overline{\wt b(x)}=\delta_{x0}1_2-\wt b(x)$ for all $x\in\Z$. Since $\wt\Omega=T_{a,\nu}$ due to \ref{ToepS-a}, \eqref{OmBT} yields 
\ba
\overline{T}_{a,\nu}
&=[\overline{\wt b(i-j)}]_{i,j\in\num{1}{\nu}}\nonumber\\
&=[\delta_{ij}1_2-\wt b(i-j)]_{i,j\in\num{1}{\nu}}\nonumber\\
&=1_{2\nu}-T_{a,\nu},
\ea
and, hence, $\Im(T_{a,\nu})=-\ii(T_{a,\nu}-1_{2\nu}/2)$.  Furthermore, using that $\wt a=-\ii(a-1_2/2)$ and that $T_{\wt a, \nu}=-\ii(T_{a,\nu}-1_{2\nu}/2)$, we get 
\ba
m_{\Im(\wt\Omega)}
&=m_{T_{\wt a, \nu}}\nonumber\\
&=T_\nu[\wt a]. 
\ea
Finally, \eqref{def:2pt-op-1} rewritten in $\fhh$ is equivalent to 
\ba
\label{2pt-2c}
\bar r_{11}
&=r_{11},\\
\label{2pt-2d}
\bar r_{12}
&=r_{21},\\
\label{2pt-2e}
\bar r_{22}
&=r_{22}.
\ea
Plugging \eqref{2pt-2c}-\eqref{2pt-2e} into \eqref{a11}-\eqref{a22} and using $\wt a=-\ii(a-1_2/2)$, we get \eqref{ta11}-\eqref{ta22}.

\ref{ToepS-c}\,
Since the Majorana correlation matrix \eqref{Omij} satisfies $\Omega=\wt\Omega$ due to \eqref{2pt-op}, \eqref{SDCAR1}, and \eqref{Ms-1}, and since the 2-point operator $R$ which induces the quasifree state $\omega$ satisfies \An{\ref{ass:trans} \ref{trans-a}} and \eqref{def:2pt-op-1}-\eqref{def:2pt-op-2} (\eqref{def:2pt-op-3} has not been used in the present proposition), \ref{ToepS-a} and \ref{ToepS-b} yield the assertion.
\eprf

\vspace{5mm}

\noindent{\bf Proof of Corollary \ref{cor:vad}.}\hspace{2mm}
Let us first suppose that $s_\infty=0$.  Since, due to \eqref{sinfty-3}, we have
\ba
\label{s-infty}
s_\infty
=\sum_{\al\in\{L,R\}}\|\wt\eta\circ (2\Od(\rho))\circ\vi_\al\|_{L^\infty(\Pi_\al)}, 
\ea
we get, for all $\al\in\{L,R\}$, that $\wt\eta\circ (2\Od(\rho))\circ\vi_\al(k)=0$ for almost all $k\in\Pi_\al$, \ie, there exists $C_\al'\in\mM(\Pi_\al)$ with $|C_\al'|_B=0$ such that $C_\al:=\{k\in\Pi_\al\,|\, \wt\eta\circ (2\Od(\rho))\circ\vi_\al(k)\neq0\}\subseteq C_\al'$. Moreover, for all $\al\in\{L,R\}$, the sets $A^{\pm}_\al\subseteq\Pi_\al$ and $B^{\pm}_\al\subseteq\R$ are defined by (see  \eqref{vi-sg}-\eqref{S-sg})
\ba
\label{Apm}
A^{\pm}_\al
&:=\{k\in\Pi_\al\,|\, \rho(\pm\vi_\al(k))\notin\{0,1\}\},\\
\label{Bpm}
B^{\pm}_\al
&:=\{x\in S^{\pm}_\al\,|\, \rho(x)\notin\{0,1\}\},
\ea
and, from now on, let $\al\in\{L,R\}$ be fixed. 
Note first that, for all $k\in\Pi_\al$, we have $k\in\Pi_\al\setminus A_\al^-$ if and only if 
$1-\rho(\vi_\al(k))=\rho(-\vi_\al(k))\in \{0,1\}$ 
(due to \eqref{def:Ff-2}), \ie, if and only if $k\in\Pi_\al\setminus A_\al^+$. Hence, we get $A_\al^-=A_\al^+$ and, in the following, we write $A_\al:=A_\al^+$. Moreover, for all $k\in\Pi_\al$, we have  $k\in\Pi_\al\setminus A_\al$ if and only if $\rho(\vi_\al(k))\in \{0,1\}$ if and only if $2\rho(\vi_\al(k))-1\in \{-1,1\}$, \ie, if and only if $|2\Od(\rho)(\vi_\al(k))|=1$ (because $2\Od(\rho)(x)=2\rho(x)-1$ for all $x\in\R$ due to \eqref{def:Ff-2}). Hence, since $|2\Od(\rho)(x)|\le 1$ for all $x\in\R$ (because $0\le \rho(x)\le 1$ for all $x\in\R$ due to \eqref{def:Ff-1}-\eqref{def:Ff-2}) and since $\wt\eta(x)=0$ if and only if $|x|\ge 1$ (see Figure \ref{fig:Shan}), we have $k\in\Pi_\al\setminus A_\al$ if and only if $k\in\Pi_\al\setminus C_\al$, \ie, we get $C_\al=A_\al$. Next, since 
\ba
\label{A-sg}
A_\al
=\Pi_\al\cap(\rho\circ(\beta_\al E)\circ\wt\kp)^{-1}(\R\setminus\{0,1\}),
\ea
since $\rho\in\mB(\R)$ and $(\beta_\al E)\circ\wt\kp\in C_b(\R)\subseteq\mB(\R)$, since $\R\setminus\{0,1\}\in\mM(\R)$, and since $\Pi_\al\in\mM(\R)$ (see Remark \ref{rem:meas}), we get $A_\al\in\mM(\Pi_\al)=\{M\subseteq\Pi_\al\,|\, M\in\mM(\R)\}$. Therefore, due to the fact that $\mM(\Pi_\al)\ni A_\al=C_\al\subseteq C_\al'\in \mM(\Pi_\al)$ with $|C_\al'|_B=0$, we also have 
\ba
\label{A-sg-nll}
|A_\al|_B
=0.
\ea

Next, let us discuss the Cases 2,4,5 and Case 3 separately.

Cases 2,4,5\quad
Recall that, in theses cases, we have $\card(Z_{|u|})\in\N_0$ and $\card(\Pi_0)\in\N_0$ with $\Pi_0\subseteq\Pi\setminus Z_{|u|}$. Hence, there exists $n\in\N_0$ and $\{k_j\}_{j\in\num{0}{n+1}}\subseteq\Pi\cup\{-\pi\}$ satisfying $k_0:=-\pi<k_1<\ldots<k_{n+1}:=\pi$ such that 
\ba
\label{dcmp-1}
\Pi\setminus(\{\pi\}\cup Z_{|u|} \cup \Pi_0)
=\bigcup_{j\in\num{0}{n}} K_j,
\ea
where we set $K_j:=(k_j,k_{j+1})$ for all $j\in\num{0}{n}$. Since $E'\circ\kp\circ\iota_{Z_{|u|}^c}\in C(Z_{|u|}^c)$ (see Remark \ref{rem:meas}), \eqref{dcmp-1} yields that, for all $j\in\num{0}{n}$, either 
$K_j\subseteq\Pi_L$ 
or 
$K_j\subseteq\Pi_R$. 
Hence, due to  \eqref{dcmp-1}, if $\Pi_\al\neq\emptyset$ (note that $\Pi\setminus(Z_{|u|} \cup \Pi_0)=\Pi_L\cup\Pi_R$), there exists $m\in\num{0}{n}$ such that
\ba
\label{dcmp-2}
\Pi_\al\setminus\{\pi\}
=\bigcup_{l\in\num{0}{m}} K_{j_l}. 
\ea
Since $S_\al^\pm=(\pm\vi_\al)(\Pi_\al)$ due to \eqref{S-sg}, \eqref{dcmp-2} yields 
\ba
S_\al^\pm
=\begin{cases}
\{\pm\vi_\al(\pi)\}\cup\big(\bigcup_{l\in\num{0}{m}}(\pm\vi_\al)(K_{j_l})\big), & \pi\in\Pi_\al,\\
\hfill\bigcup_{l\in\num{0}{m}}(\pm\vi_\al)(K_{j_l}), & \pi\notin\Pi_\al.
\end{cases}
\ea
Moreover, since $(\pm\vi_\al)(K_{j_l})\subseteq ((\pm\beta_\al E)\circ\kp)(\bar K_{j_l})$ for all $l\in\num{0}{m}$, where $(\pm\beta_\al E)\circ\kp\in C(\Pi)$ (and $\bar M$ denotes the closure of any $M\subseteq\R$), and since the continuous image of a compact set is a compact set and the continuous image of an interval is an interval, $(\pm\vi_\al)(K_{j_l})$ is a bounded closed, half-open or open interval for all $l\in\num{0}{m}$, \ie, we get $(\pm\vi_\al)(K_{j_l})\in\mM(\R)$ for all $l\in\num{0}{m}$ and, hence, $S_\al^\pm\in\mM(\R)$. 

Now, due to \eqref{Bpm}, we also have $B_\al^\pm=(\pm\vi_\al)(A_\al)$. In order to show that $B_\al^\pm\in\mM(S_\al^\pm)=\{M\subseteq S_\al^\pm\,|\, M\in\mM(\R)\}$ (since $S_\al^\pm\in\mM(\R)$), we want to make use of the Lusin-Souslin theorem (see, for example, \cite{Ke95}) for the sets $A_{\al,l}:=A_\al\cap K_{j_l}\in\mM(\Pi)$ for all $l\in\num{0}{m}$ and (if $\pi\in\Pi_\al$) for $A_{\al, \pi}:=A_\al\cap\{\pi\}\in\mM(\Pi)$. To this end, note that, since $\R$ is a Polish space (\ie, a separable completely metrizable topological space) and since a topological subspace (with respect to the induced topology) of a Polish space is Polish if and only if it is a $G_\delta$ set (\ie, a countable intersection of open sets), $\Pi$ is Polish. Moreover, $\psi_\al:=(\pm\beta_\al E)\circ\kp\in C(\Pi)$ and, for any fixed $
D_\al\in ((\bigcup_{l\in\num{0}{m}}\{A_{\al,l}\})\cup \{A_{\al, \pi}\})$, the function $\psi_\al\circ \iota_{D_\al}$ is injective since $(\pm\beta_\al E)'(z)$ exists for all $z\in\kp(Z_{|u|}^c)$ and (see \eqref{PiL}-\eqref{PiR})
\ba
\label{E'<0}
E'(\kp(\Pi_L))
&<0,\\
\label{E'>0}
E'(\kp(\Pi_R))
&>0.
\ea
Therefore, the Lusin-Souslin theorem yields $\psi_\al(D_\al)\in\mM(\R)$ and, since $A_\al=A_\al\cap\Pi_\al$, \eqref{dcmp-2} leads to $B_\al^\pm=(\pm\vi_\al)(A_\al\cap\Pi_\al)\in\mM(S_\al^\pm)$. 

Next, we want to show that $|B_\al^\pm|_B=0$. To this end, note first that, due to \eqref{A-sg-nll}, for any $\veps>0$, there exists a sequence of compact intervals $I_j\subseteq\R$ for all $j\in\N$ such that $A_\al\subseteq\bigcup_{j\in\N} I_j$ and $\sum_{j\in\N}|I_j|_B\le\veps$. Moreover, since $|E'(z)|\le |u_0'(z)|+|u'(z)|$ for all $z\in\kp(Z_{|u|}^c)$ due to \eqref{E'-pm} (and the Cauchy-Schwarz inequality), there exists $c>0$ such that, for all $z\in\kp(Z_{|u|}^c)$,
\ba
\label{E'-est}
|E'(z)|
\le c,
\ea
where we used that $u_\beta'\in TP(\T)$ for all $\beta\in\num{0}{3}$ and $c:=\sum_{\beta\in\num{0}{3}}\|u_\beta'\|_{L^\infty(\T)}$. Hence, since 
\ba
B_\al^\pm
&=(\pm\vi_\al)(A_\al)\nonumber\\
&\subseteq\bigcup_{j\in\N}\wt\psi_\al(I_j), 
\ea
where $\wt\psi_\al:=(\pm\beta_\al E)\circ\wt\kp\in C(\R)$, and since the continuous image of a compact interval is a compact interval, $B_\al^\pm$ has a sequential covering by compact intervals in $\R$. Moreover, for any fixed $j\in\N$, there exists $r_j\in\N_0$ and $\{x_{j,l}\}_{l\in\num{0}{r_j+1}}\subseteq I_j$ satisfying $x_{j,0}:=\min(I_j)<x_{j,1}<\ldots<x_{j,r_j+1}:=\max(I_j)$ such that 
\ba
\label{dcmp-3}
\mathring{I}_j\cap (\R\setminus \wt Z_{|u|})
=\bigcup_{l\in\num{0}{r_j}} I_{j,l},
\ea
where $\wt Z_{|u|}:=(|u|\circ\wt\kp)^{-1}(\{0\})$ (note that $\card(\wt Z_{|u|}\cap I_j)\in\N_0$ for all $j\in\N$) and where $I_{j,l}:=(x_{j,l},x_{j,l+1})$ for all $l\in\num{0}{r_j}$ (and $\mathring{M}$ denotes the interior of any $M\subseteq\R$). Hence, for all $l\in\num{0}{r_j}$, using the mean value theorem and \eqref{E'-est} for the function $\wt\psi_\al\circ\iota_{\bar I_{j,l}}\in C(\bar I_{j,l})$, we get $|\wt\psi_\al(x)-\wt\psi_\al(y)|\le c|x-y|$ for all $x,y\in \bar I_{j,l}$ and, hence, 
\ba
|\wt\psi_\al(I_j)|_B
&\le c \sum_{l\in\num{0}{r_j}}|I_{j,l}|_B\nonumber\\
&=c |I_j|_B. 
\ea
Therefore, $\sum_{j\in\N}|\wt\psi_\al(I_j)|_B\le c\veps$,  \ie, we arrive at $|B_\al^\pm|_B=0$. 

Case 3\quad
Recall that, in this case, we have $E=u_0$, $E'=u_0'$, $\Pi_L=\{k\in\Pi\,|\,u_0'(\ei)<0\}$, $\Pi_R=\{k\in\Pi\,|\,u_0'(\ei)>0\}$, and $\Pi_0=\{k\in\Pi\,|\,u_0'(\ei)=0\}$.
We again have $S_\al^\pm\in\mM(\R)$ since $\card(\Pi_0)\in\N$ (see Remark \ref{rem:meas}), since \eqref{dcmp-2} holds (due to $u_0'\circ\kp\in C(\Pi)$), since \eqref{S-sg} again yields the decomposition given after \eqref{dcmp-2}, and since $(\pm\beta_\al u_0)\circ\kp\in C(\Pi)$.
As for $B_\al^\pm\in\mM(S_\al^\pm)$, we again proceed as for Cases 2,4,5 by noting that $\psi_\al=(\pm\beta_\al u_0)\circ\kp\in C(\Pi)$ and that \eqref{E'<0}-\eqref{E'>0} become $u_0'(\kp(\Pi_L))<0$ and $u_0'(\kp(\Pi_R))>0$, respectively.
As for $|B_\al^\pm|_B=0$, we note that $u_0$ is differentiable everywhere on $\R$ (so we do not need the decomposition \eqref{dcmp-3}) and that, instead of \eqref{E'-est}, we have $|u_0'(z)|\le c$ for all $z\in\T$, where $c:=\|u_0'\|_{L^\infty(\T)}$. In order to arrive at the conclusion, we again proceed analogously.

Summarizing, we get $S_\al^\pm\in\mM(\R)$, $B_\al^\pm\in\mM(S_\al^\pm)$, and $|B_\al^\pm|_B=0$ for all Cases 2-5. Hence, since $\{x\in\Sigma_\al\,|\, \rho(x)\notin\{0,1\}\}=\bigcup_{\sg\in\{\pm\}} B_\al^\sg$ for all $\al\in\{L,R\}$ and, hence, $\Sigma_0:=\{x\in\Sigma\,|\, \rho(x)\notin\{0,1\}\}=\bigcup_{\al\in\{L,R\}}\bigcup_{\sg\in\{\pm\}} B_\al^\sg\in\mM(\R)$, we arrive at 
\ba
|\Sigma_0|_B
\le \sum_{\al\in\{L,R\}}\sum_{\sg\in\{\pm\}} |B_\al^\sg|_B,
\ea
\ie, $|\Sigma_0|_B=0$. Moreover, since $\rho\in\mB(\R)$, since $\{1\}\in\mM(\R)$, and since $\Sigma\in\mM(\R)$, the set $M\subseteq\R$, defined by
\ba
M
:=\rho^{-1}(\{1\})\cap\Sigma,
\ea
satisfies $M\in\mM(\R)$. Hence, we get \eqref{rhoSgm}.

Finally, the converse statement follows from \eqref{s-infty} by plugging \eqref{rhoSgm} into \eqref{s-infty} and by using the fact that $\wt\eta(\pm 1)=0$.
\eprf

\vspace{5mm}

\noindent{{\bf Acknowledgements}}\,
I would like to thank the anonymous referee for his detailed reading of the manuscript and his constructive suggestions.

\begin{appendix}

\section{Toeplitz operators}
\label{app:Toep}

In this appendix, we provide the basic definitions in position and momentum space of the block Toeplitz operators used in the preceding sections (see, for example, \cite{BoSi99, BoSi06}). 

First, for all $D\subseteq\C$, all $n,m\in\N$, and all $a\in(\nm{n}{m})^D$, we define the entry functions $(a)_{ij}\in \C^D$ by $(a)_{ij}:=(a(\cdot))_{ij}$ for all $i\in\num{1}{n}$ and all $j\in\num{1}{m}$ (see the beginning of Section \ref{sec:infinite}) and, if $m=1$, we set $(a)_i:=(a)_{i1}$ for all $i\in\num{1}{n}$. 

Over position configuration space $\Z$, we set
\ba
\label{l2ZC2}
\ell^2(\Z,\C^2)
&:=\{F\in(\C^2)^\Z\,|\,\mbox{$(F)_i\in\h$ for all $i\in\num{1}{2}$}\},\\
\label{l2N}
\ell^2(\N)
&:=\Big\{f\in\C^\N\,\Big|\, \sum\nolimits_{n\in\N} |f(n)|^2<\infty\Big\},\\
\label{l2NC2}
\ell^2(\N,\C^2)
&:=\{F\in(\C^2)^\N\,|\,\mbox{$(F)_i\in\ell^2(\N)$ for all $i\in\num{1}{2}$}\},\\
\label{tl2}
\wt\ell^2(\N_0)
&:=\{f\in\h\,|\,\mbox{$f(x)=0$ for all $x\in\Z$ with $x\le -1$}\},\\
\label{tl2C2}
\wt\ell^2(\N_0, \C^2)
&:=\{F\in\ell^2(\Z,\C^2)\,|\,\mbox{$(F)_i\in\wt\ell^2(\N_0)$ for all $i\in\num{1}{2}$}\},\\
\ell^2(\Z,\M{2})
&:=\{\alpha\in(\M{2})^\Z\,|\,\mbox{$(\alpha)_{ij}\in\h$ for all $i,j\in\num{1}{2}$}\}.
\ea
Equipped with the scalar products given, for \eqref{l2ZC2}, by $(F,G)_{\ell^2(\Z,\C^2)}:=\sum_{x\in\Z}(F(x), G(x))_{\C^2}$ for all $F,G\in\ell^2(\Z,\C^2)$ (see after \eqref{def:Fin}), for \eqref{l2N}, by $(f,g)_{\ell^2(\N)}:=\sum_{n\in\N}\bar f(n)g(n)$ for all $f,g\in\ell^2(\N)$, for \eqref{l2NC2}, by $(F,G)_{\ell^2(\N,\C^2)}:=\sum_{n\in\N}(F(n), G(n))_{\C^2}$ for all $F,G\in\ell^2(\N,\C^2)$, a direct check yields that \eqref{l2ZC2}-\eqref{l2NC2} become complex Hilbert spaces and \eqref{tl2} and \eqref{tl2C2} closed subspaces of $\h$ and $\ell^2(\Z,\C^2)$, respectively. Here, for all $n\in\N$, we denote the usual complex Euclidean scalar product on $\C^n$ by $(x,y)_{\C^n}:=\sum_{i\in\num{1}{n}}\bar x_i y_i$ for all $x:=[x_i]_{i\in\num{1}{n}}, y:=[y_i]_{i\in\num{1}{n}}\in\C^n$ and the corresponding induced norm by $\|\cdot\|_{\C^n}$.

Let's now turn to momentum configuration space $\T$ and let $\Phi\in(\C^2)^\T$ be fixed and satisfy $\Inti \|\Phi(\e^{\ii k})\|^2_{\C^2}<\infty$. We say that $\Psi\in(\C^2)^\T$ is $L^2(\T,\C^2)$-equivalent to $\Phi$ if  $(\Psi)_i\in [(\Phi)_i]\in \fh$ for all $i\in\num{1}{2}$ (where the square brackets $[\cdot]$ stand for the equivalence class in $\fh$, see the beginning of Section \ref{sec:trans}). Since $L^2(\T,\C^2)$-equivalence also defines an equivalence relation, we again denote by $[\Phi]$ the equivalence class of $\Phi$ and we set 
\ba
\label{L2TC2}
L^2(\T,\C^2)
:=\{[\Phi]\,|\, \mbox{$\Phi\in(\C^2)^\T$ with $[(\Phi)_i]\in \fh$ for all $i\in\num{1}{2}$}\}.
\ea
Omitting, as usual, the square brackets for the equivalence classes in $\fh$ and $L^2(\T,\C^2)$, we write the right hand side of \eqref{L2TC2} as $\{\Phi\in(\C^2)^\T\,|\,\mbox{$(\Phi)_i\in \fh$ for all $i\in\num{1}{2}$}\}$. Equipped with the scalar product 
$(\Phi,\Psi)_{L^2(\T,\C^2)}
:=\Inti (\Phi(\e^{\ii k}), \Psi(\e^{\ii k}))_{\C^2}
=\sum_{i\in\num{1}{2}} ((\Phi)_i, (\Psi)_i)_{\fh}$ 
for all $\Phi,\Psi\in L^2(\T,\C^2)$, \eqref{L2TC2} becomes a Hilbert space. Moreover, let us define (abusing notation analogously)
\ba
\label{H2}
H^2(\T)
&:=\{\vi\in \fh\,|\, \widecheck\vi\in\wt\ell^2(\N_0)\},\\
\label{H22}
H^2(\T,\C^2)
&:=\{\Phi\in L^2(\T,\C^2)\,|\, \mbox{$(\Phi)_i\in H^2(\T)$ for all $i\in\num{1}{2}$}\},\\
\label{LT22}
L^\infty(\T,\M{2})
&:=\{a\in(\M{2})^\T\,|\,\mbox{$(a)_{ij}\in L^\infty(\T)$ for all $i,j\in\num{1}{2}$}\},
\ea
where we recall from Section \ref{sec:trans} that $\wc\vi=\ff^\ast\vi\in\h$ for all $\vi\in\fh$.
Note that the so-called Hardy spaces \eqref{H2} and \eqref{H22} are closed subspaces of $\fh$ and $L^2(\T,\C^2)$, respectively. Moreover, the elements of \eqref{LT22} are called block symbols and \eqref{LT22} is a \Cs algebra with respect to (the pointwise complex scalar multiplication, addition, matrix multiplication, and conjugate transposition, and) the norm defined,  for all $a\in L^\infty(\T,\M{2})$, by
\ba
\label{NrmSymb}
\|a\|_{L^\infty(\T,\M{2})}
:=\esup_{z\in\T}\|a(z)\|_{\mL(\C^2)},
\ea
where $\esup$ stands for the essential supremum and $\|\cdot\|_{\mL(\C^2)}$ is the operator norm on the Hilbert space $\C^2$ (see the beginning of Section \ref{sec:infinite}) equipped with the complex Euclidean scalar product.
In order to be able to define the Toeplitz operators on position and momentum space, we make use of the canonical injection (see the beginning of Section \ref{sec:RDM})
\ba
\iota: H^2(\T)
\hookrightarrow \fh,
\ea
and we denote by $P\in\mL(\fh)$ the usual Riesz projection, \ie, the orthogonal projection from $\fh$ onto $H^2(\T)$. 

Moreover, we define $\Gamma\in\mL(\fhh,L^2(\T,\C^2))$, for all $\vi_1\oplus\vi_2\in \fhh$ and all $i\in\num{1}{2}$, by 
\ba
(\Gamma (\vi_1\oplus \vi_2))_i
:=\vi_i,
\ea
and we note that $\Gamma$ is unitary with inverse $\Gamma^\ast\Phi=(\Phi)_1\oplus (\Phi)_2$ for all $\Phi\in L^2(\T,\C^2)$ (using again the notation from the beginning of Section \ref{sec:infinite}).
Analogously, we define the unitary operator $\Upsilon\in\mL(\hh, \ell^2(\Z,\C^2))$, for all $f_1\oplus f_2\in\hh$ and all $i\in\num{1}{2}$, by
\ba
(\Upsilon( f_1\oplus f_2))_i
:=f_i,
\ea
and we have $\Upsilon^\ast F=(F)_1\oplus (F)_2$ for all $F\in\ell^2(\Z,\C^2)$ and, for all $f_1\oplus f_2\in\hh$ and all $i\in\num{1}{2}$, 
\ba
(\Upsilon (f_1\oplus f_2))_i
=\ff^\ast(\Gamma(\ff1_2) (f_1\oplus f_2))_i.
\ea
 Moreover, we define $\upsilon_0\in\mL(\wt\ell^2(\N_0), \ell^2(\N))$ and $\Upsilon_0\in\mL(\wt\ell^2(\N_0,\C^2), \ell^2(\N,\C^2))$ by $(\upsilon_0 f)(n):=f(n-1)$ for all $f\in \wt\ell^2(\N_0)$ and all $n\in\N$ and by $(\Upsilon_0 F)(n):=F(n-1)$ for all $F\in \wt\ell^2(\N_0,\C^2)$ and all $n\in\N$, respectively (and we again note that $\upsilon_0$ and $\Upsilon_0$ are unitary). Finally, recall the multiplication operator $m[u]\in\mL(\fh)$ for all $u\in L^\infty(\T)$ from \eqref{muvi}.

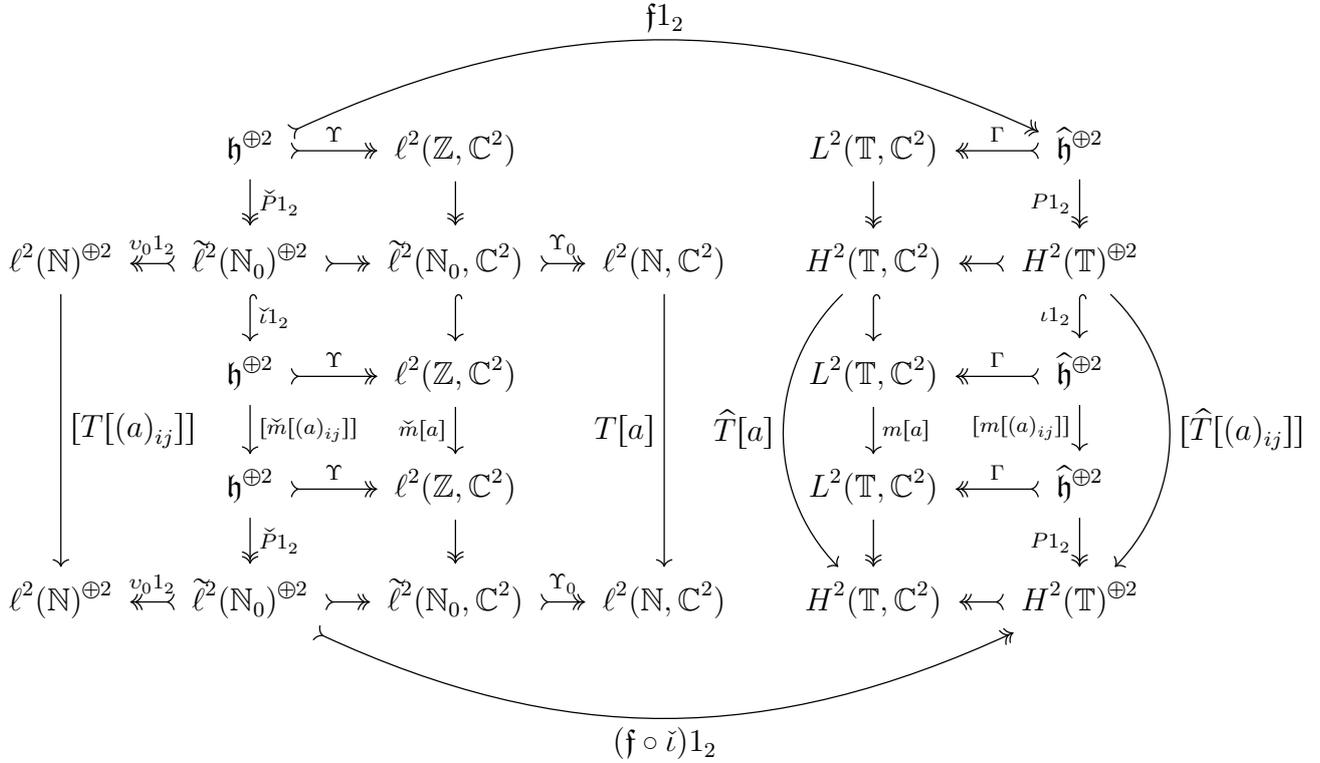
\begin{figure}
\begin{center}
\begin{tikzcd}[row sep=6mm, column sep=6mm]
	& \hh
	\arrow[d, two heads, "\wc P1_2"]
	\arrow[r, tail, two heads, "\Upsilon"]
	\arrow[rrrr, tail, two heads, bend left=25, "\textstyle\ff1_2"]
	& \ell^2(\Z,\C^2)
	\arrow[d, two heads]
	&
	& L^2(\T,\C^2)
	\arrow[d, two heads]
	& \fhh
	\arrow[d, two heads, "P1_2"']
	\arrow[l, tail, two heads, "\Gamma"']
	\\
	\ell^2(\N)^{\oplus 2}
	\arrow[ddd, "\textstyle{[T[(a)_{ij}]]}"]
	& {\wt\ell}^2(\N_0)^{\oplus 2}
	\arrow[d,  hookrightarrow, "\wc\iota1_2"]
	\arrow[l, tail, two heads, "\upsilon_01_2"']
	\arrow[r, tail, two heads]
	& {\wt\ell}^2(\N_0, \C^2)
	\arrow[d,  hookrightarrow]
	\arrow[r, tail, two heads, "\Upsilon_0"]
	& \ell^2(\N, \C^2)
	\arrow[ddd, "\textstyle{T[a]}"']
	& H^2(\T,\C^2)
	\arrow[d,  hookrightarrow]
	\arrow[ddd, bend right=45, "\textstyle{\wh T[a]}"']
	& H^2(\T)^{\oplus 2}
	\arrow[d,  hookrightarrow,"\iota1_2"']
	\arrow[ddd, bend left=45, "\textstyle{[\wh T[(a)_{ij}]]}"]
	\arrow[l, tail, two heads]
	\\
	& \hh
	\arrow[d, "{[\wc m[(a)_{ij}]]}"]
	\arrow[r, tail, two heads, "\Upsilon"]
	& \ell^2(\Z,\C^2)
	\arrow[d, "{\wc m[a]}"']
	&
	& L^2(\T,\C^2)
	\arrow[d, "{m[a]}"]
	& \fhh
	\arrow[d, "{[m[(a)_{ij}]]}"']
	\arrow[l, tail, two heads, "\Gamma"']
	\\
	& \hh
	\arrow[d, two heads, "\widecheck P1_2"]
	\arrow[r, tail, two heads, "\Upsilon"]
	& \ell^2(\Z,\C^2)
	\arrow[d, two heads]
	&
	& L^2(\T,\C^2)
	\arrow[d, two heads]
	& \fhh
	\arrow[d, two heads, "P1_2"']
	\arrow[l, tail, two heads, "\Gamma"']
	\\
	\ell^2(\N)^{\oplus 2}
	& {\wt\ell}^2(\N_0)^{\oplus 2}
	\arrow[l, tail, two heads, "\upsilon_01_2"']
	\arrow[r, tail, two heads]
	\arrow[rrrr, tail, two heads, bend right=25, "\textstyle(\ff\circ\check\iota)1_2"']
	& {\wt\ell}^2(\N_0, \C^2)
	\arrow[r, tail, two heads, "\Upsilon_0"]
	& \ell^2(\N, \C^2)
	& H^2(\T,\C^2)
	& H^2(\T)^{\oplus 2}
	\arrow[l, tail, two heads]
\end{tikzcd}
\caption{The commutative diagram for the action of the Toeplitz operator with symbol $a\in L^\infty(\T,\M{2})$ in position space and in momentum space.}
\label{fig:Toep}
\end{center}
\end{figure}

The block Toeplitz operators on position and momentum space are defined as follows (see Figure \ref{fig:Toep}).

\bdp[Block Toeplitz operators]
\label{prop:Toep}
Let $a$ be a block symbol, \ie, $a\in L^\infty(\T,\M{2})$. Then, on momentum space:
\bn[label=(\alph*), ref={\it (\alph*)}]
\setlength{\itemsep}{0mm}
\item
The block Toeplitz operator $[\wh T[(a)_{ij}]]_{i,j\in\num{1}{2}}\in\mL(H^2(\T)^{\oplus 2})$ is defined by $\wh T[(a)_{ij}]:=Pm[(a)_{ij}]\circ\iota$ for all $i,j\in\num{1}{2}$.

\item
Setting $m[a]:=\Gamma [m[(a)_{ij}]]_{i,j\in\num{1}{2}}\Gamma^\ast\in\mL(L^2(\T,\C^2))$, we have $(m[a]\Phi)(z)=a(z)\Phi(z)$ for all $\Phi\in L^2(\T,\C^2)$ and almost all $z\in\T$. The block Toeplitz operator $\wh T[a]:=\Gamma(P1_2)\Gamma^\ast m[a]\circ\iota\in\mL(H^2(\T,\C^2))$ reads, for all $\Phi\in H^2(\T,\C^2)$ and all $i\in\num{1}{2}$,
\ba
(\wh T[a]\Phi)_i
=P(m[a]\Phi)_i.
\ea
\en
And, on position space:
\bn[label=(\alph*), ref={\it (\alph*)}]\setcounter{enumi}{2}
\setlength{\itemsep}{0mm}
\item 

Defining $[\wc m[(a)_{ij}]]_{i,j\in\num{1}{2}}\in\mL(\hh)$ by $\wc m[(a)_{ij}]:=\ff^\ast m[(a)_{ij}]\ff\in\mL(\h)$ for all $i,j\in\num{1}{2}$, we have $\wc m[(a)_{ij}]f=(a)_{ij}\ast f$ for all $f\in\h$, where $a\ast f:=\sum_{y\in\Z}\wc a(\cdot -y)f(y)$ for all $f\in\h$. The block Toeplitz operator $[T[(a)_{ij}]]_{i,j\in\num{1}{2}}\in\mL(\ell^2(\N)^{\oplus 2})$ is defined by $T[(a)_{ij}]:=\upsilon_0\wc P\wc m[(a)_{ij}]\circ\wc\iota\upsilon_0^\ast$ for all $i,j\in\num{1}{2}$ (where $\wc\iota:\wt\ell^2(\N_0)\hookrightarrow\h$ is the canonical injection), \ie, for all $f\in\ell^2(\N)$ and all $n\in\N$,
\ba
(T[(a)_{ij}]f)(n)
=\sum_{m\in\N}\wc{(a)}_{ij}(n-m) f(m).
\ea

\item
Setting $\wc m[a]:=\Upsilon [\wc m[(a)_{ij}]]_{i,j\in\num{1}{2}}\Upsilon^\ast\in\mL(\ell^2(\Z,\C^2))$, we have $(\wc m[a] F)(x)=\sum_{y\in\Z}\wc a(x-y) F(y)$ for all $F\in\ell^2(\Z,\C^2)$ and all $x\in\Z$, where $\wc a\in\ell^2(\Z,\M{2})$ is defined by $(\wc a)_{ij}:=\wc{(a)}_{ij}$ for all $i,j\in\num{1}{2}$. The block Toeplitz operator $T[a]:=\Upsilon_0\Upsilon(\wc P1_2)\Upsilon^\ast\wc m[a]\Upsilon_0^\ast\in\mL(\ell^2(\N,\C^2))$ reads, for all $F\in \ell^2(\N,\C^2)$ and all $n\in\N$,
\ba
\label{bToep}
(T[a] F)(n)
=\sum_{m\in\N}\wc a(n-m) F(m).
\ea 
\en
\edp

\bprf 
Note that $\{\e_x\}_{x\in\Z}$ and $\{\e_n\}_{n\in\N_0}$ are orthonormal bases of $\fh$ and $H^2(\T)$, respectively. Also note that $P\e_x=1_{\N_0}(x) \e_x$ for all $x\in\Z$ and that the discrete convolution $a\ast f$ is well-defined for all $a\in L^\infty(\T)$ and all $f\in\h$ since $\wc a(x-y)=(\e_x,m[a]\e_y)$ for all $x,y\in\Z$. Using Figure \ref{fig:Toep}, several direct computations lead to all of the assertions.
\eprf

For the following, for all $N\in\N$, we set (as in \eqref{tl2C2})  
\ba
{\wt\ell}^2(\num{1}{N}, \C^2)
:=\{F\in\ell^2(\N,\C^2)\,|\,\mbox{$(F)_i(n)=0$ for all $n\in\N\setminus\num{1}{N}$ and all $i\in\num{1}{2}$}\}.
\ea
For all $N\in\N$, we define the orthogonal projection $P_N\in\mL(\ell^2(\N,\C^2))$ from $\ell^2(\N,\C^2)$ onto ${\wt\ell}^2(\num{1}{N}, \C^2)$, for all $F\in\ell^2(\N,\C^2)$ and all $n\in\N$, by
\ba
(P_NF)(n)
:=1_{\num{1}{N}}(n) F(n).
\ea
Moreover, $\Theta_N\in\mL({\wt\ell}^2(\num{1}{N}, \C^2),\C^{2N})$ is given, for all $F\in {\wt\ell}^2(\num{1}{N}, \C^2)$ and all $n\in\num{1}{N}$, by
\ba
(\Theta_N F)_{2n-1}
&:=(F(n))_1,\\
(\Theta_N F)_{2n}
&:=(F(n))_2,
\ea
and we note that $\Theta_N$ is unitary (with respect to the complex Euclidean scalar product on $\C^{2N}$).
Finally, for all $n\in\N$ and all $X\in\M{n}$, we denote by $m_X\in\mL(\C^n)$ the operator induced by $X$, \ie, for all $v\in\C^n$, 
\ba
m_Xv
:=Xv.
\ea

We next turn to the so-called finite section method which projects the Toeplitz operator \eqref{bToep} onto the subspace ${\wt\ell}^2(\num{1}{N}, \C^2)$ of $\ell^2(\N,\C^2)$ of dimension $2N$. 

\begin{figure}
\begin{center}
\begin{tikzcd}[row sep=6mm, column sep=6mm]
	\ell^2(\N,\C^2)
	\arrow[d, two heads, "{P_N}"]
	&
	\\
	{\wt\ell}^2(\num{1}{N}, \C^2)
	\arrow[d,  hookrightarrow]
	\arrow[r, tail, two heads, "\Theta_N"]
	& \C^{2N}
	\arrow[ddd, "\textstyle{T_N[a]}"]
	\\
	\ell^2(\N,\C^2)
	\arrow[d, "\textstyle{T[a]}"]
	& 
	\\
	\ell^2(\N,\C^2)
	\arrow[d, two heads, "{P_N}"]
	& 
	\\
	{\wt\ell}^2(\num{1}{N}, \C^2)
	\arrow[r, tail, two heads, "\Theta_N"]
	& \C^{2N}
\end{tikzcd}
\caption{The finite section method for $N\in\N$ applied to the block Toeplitz operator $T[a]$ with block symbol $a\in L^\infty(\T,\M{2})$ from \eqref{bToep} (see also Figure \ref{fig:Toep}).}
\label{fig:FinSec}
\end{center}
\end{figure}
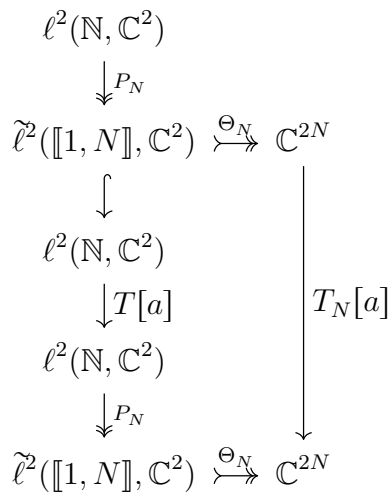

\bp[Finite section method]
For all block symbols $a\in L^\infty(\T,\M{2})$ and all $N\in\N$, the ($N$-) finite section $T_N[a]\in\mL(\C^{2N})$ of the block Toeplitz operator $T[a]\in\mL(\ell^2(\N,\C^2))$ is defined by 
\ba
T_N[a]
:=\Theta_N P_N T[a]\Theta_N^\ast,
\ea
see Figure \ref{fig:FinSec}. For all $N\in\N$, we have $T_N[a]=m_{T_{a,N}}$, where the so-called block Toeplitz matrix $T_{a,N}\in\M{2N}$ is defined by 
\ba
\label{BlockToep}
T_{a,N}
:=\begin{bmatrix}
a_0 & a_{-1} & a_{-2} & \ldots & a_{-(N-1)}\\
a_1 & a_0 & a_{-1} & \ldots & a_{-(N-2)}\\
a_2 & a_1 & a_0 & \ldots & a_{-(N-3)} \\
\vdots & \vdots & \vdots &\ddots & \vdots\\
a_{N-1} & a_{N-2} & a_{N-3} & \ldots & a_0
\end{bmatrix},
\ea
and where we set $a_x:=\wc a(x)\in\M{2}$ for all $x\in\Z$ (\ie, $(a_x)_{ij}=(\wc a(x))_{ij}=(\wc a)_{ij}(x)=\wc{(a)}_{ij}(x)$ for all $x\in\Z$ and all $i,j\in\num{1}{2}$).
\ep

\bprf
Using that $((\Theta_N^\ast v)(n))_i=v_{2n-\delta_{i1}}$ for all $v=[v_j]_{j\in\num{1}{2N}}\in\C^{2N}$ and all $i\in\num{1}{2}$ and that, for all $n,m\in\num{1}{N}$, 
\ba
(T_{a,N})_{2n-1\hspace{0.3mm}2m-1}
&:=(a_{n-m})_{11},\\
(T_{a,N})_{2n-1\hspace{0.3mm}2m}
&:=(a_{n-m})_{12},\\
(T_{a,N})_{2n\hspace{0.3mm}2m-1}
&:=(a_{n-m})_{21},\\
(T_{a,N})_{2n\hspace{0.3mm}2m}
&:=(a_{n-m})_{22},
\ea
a direct computation leads to the assertion.
\eprf
\end{appendix}

\vspace{10mm}

\noindent{\bf Declarations}\quad {\it Competing interests}\,
The author has no competing interests to declare that are relevant to the content of this article. 
{\it Data availability}\,
No data sets were generated or analyzed during the current study.



\begin{thebibliography}{10} 

\bibitem{AmFaOsVe08} Amico L, Fazio R, Osterloh A, and Vedral V 2008
{\it Entanglement in many-body systems}
Rev. Mod. Phys. 80 517-576

\bibitem{Ar68} Araki H 1968
{\it On the diagonalization of a bilinear Hamiltonian by a  Bogoliubov transformation}
Publ. RIMS Kyoto Univ. Ser. A 4 387-412

\bibitem{Ar71} 
Araki H 1971
{\it On quasifree states of CAR and Bogoliubov automorphisms}
Publ. RIMS Kyoto Univ. 6 385-442

\bibitem{Ar84}
Araki H 1984 
{\it On the XY-model on two-sided infinite chain}
Publ. RIMS Kyoto Univ. 20 277-296

\bibitem{Ar87} 
Araki H 1987
{\it Bogoliubov automorphisms and Fock representations of canonical anticommutation relations}
Contemp. Math. 62 23-141

\bibitem{ArMa85}
Araki H and Matsui T 1985
{\it Ground states of the XY model}
Commun. Math. Phys. 101 213-245

\bibitem{As07}
Aschbacher W H 2007
{\it Non-zero entropy density in the XY chain out of equilibrium}
Lett. Math. Phys. 79 1-16

\bibitem{As21} 
Aschbacher W H 2021
{\it Heat flux in general quasifree fermionic right mover/left mover systems}
Rev. Math. Phys. 33 2150018 1-83

\bibitem{As23} 
Aschbacher W H 2023
{\it On Araki's extension of the Jordan-Wigner transformation}
Rev. Math. Phys. 35 2330001 1-55

\bibitem{BoSi99}
B\"ottcher A and Silbermann B 1999
{\it Introduction to large truncated Toeplitz matrices}
(Springer)

\bibitem{BoSi06}
B\"ottcher A and Silbermann B 2006
{\it Analysis of Toeplitz operators}
(Springer)

\bibitem{BrRo8797} Bratteli O and Robinson D W 1987, 1997
{\it Operator algebras and quantum statistical mechanics 1, 2}
(Springer)

\bibitem{CuScPf69}
Culvahouse J W, Schinke D P, and Pfortmiller L G 1969
{\it Spin-spin interaction constants from the hyperfine structure of coupled ions}
Phys. Rev. 177 454-464

\bibitem{DaJa79}
Dar\'oczy Z and J\'arai A 1979
{\it On the measurable solution of a functional equation arising in information theory}
Acta Math. Acad. Sci. Hungar. 34 105-116

\bibitem{EiCrPl10} Eisert J, Cramer M, and Plenio M B 2010
{\it Colloquium: Area laws for the entanglement entropy}
Rev. Mod. Phys. 82 277-306

\bibitem{Em72}
Emch G G 1972
{\it Algebraic methods in statistical mechanics and quantum field theory}
(Reprint, Dover, 2009)

\bibitem{Gl60} Glimm J G 1960
{\it On a certain class of operators algebras}
Trans. Amer. Math. Soc. 95 318-340

\bibitem{GoWi21} Gour G and Wilde M M 2021
{\it Entropy of a quantum channel}
Phys. Rev. Research 3 023096 1-26

\bibitem{He28}
Heisenberg W 1928
{\it Zur Theorie des Ferromagnetismus}
Z. Phys. 49 619-636

\bibitem{HoJo85}
Horn R A and Johnson C R 1985
{\it Matrix analysis}
(Cambridge University Press)

\bibitem{Is25}
Ising E 1925
{\it Beitrag zur Theorie des Ferromagnetismus}
Z. Phys. 31 253-258

\bibitem{ItJiKo05} Its A R, Jin B-Q, and Korepin V E 2005
{\it Entanglement in the XY spin chain}
J. Phys. A: Math. Gen. 38 2975-2990

\bibitem{JiKo04}
Jin B-Q and Korepin V E 2004
{\it Quantum spin chain, Toeplitz determinant and the Fisher-Hartwig
conjecture}
J. Stat. Phys. 116 79-95

\bibitem{JoWi28}
Jordan P and Wigner E 1928
{\it \"Uber das Paulische \"Aquivalenzverbot}
Z. Phys. 47 631-651

\bibitem{Ka62}
Katsura S 1962
{\it Statistical mechanics of the anisotropic linear Heisenberg model}
Phys. Rev. 127 1508-1518

\bibitem{Ke95}
Kechris A S 1995
{\it Classical descriptive set theory}
Graduate Texts in Mathematics 156
(Springer)

\bibitem{Le20}
Lenz W 1920
{\it Beitrag zum Verst\"andnis der magnetischen Erscheinungen in festen K\"orpern}
Physik. Zeitschr. 21 613-615

\bibitem{LiScMa61}
 Lieb E, Schultz T, and Mattis D 1961 
{\it Two soluble models of an antiferromagnetic chain}
Ann. Physics 16 407-466

\bibitem{MiKo04}
Mikeska H-J and Kolezhuk A K 2004
{\it One-dimensional magnetism} in Schollw\"ock U, Richter J, Farnell D J J, and 
Bishop R F (Ed.) 
{\it Quantum Magnetism} Lect. Notes Phys. 645 1-83
(Springer)

\bibitem{Na50}
Nambu Y 1950 
{\it A note on the eigenvalue problem in crystal statistics}
Prog. Theor. Phys. 5 1-13

\bibitem{Ni67} Niemeijer T 1967
{\it Some exact calculations on a chain of spins $1/2$}
Physica 36 377-419

\bibitem{Pr83}
Primas H 1983
{\it Chemistry, quantum mechanics, and reductionism}
(Springer)

\bibitem{Ru01}
Ruelle D 2001
{\it Entropy production in quantum spin systems}
Commun. Math. Phys. 224 3-16

\bibitem{Sa99} Sachdev S 1999
{\it Quantum phase transitions}
(Cambridge University Press)

\bibitem{Se86}
Sewell G 1986
{\it Quantum theory of collective phenomena}
(Reprint, Dover, 2014)

\bibitem{SoGiOtViRe01} Sologubenko A V, Giann\`o K, Ott H R, Vietkine A, and Revcolevschi A 2001
{\it Heat transport by lattice and spin excitations in the spin-chain compounds ${\rm SrCuO_2}$ and ${\rm Sr_2CuO_3}$}
Phys. Rev. B 64 054412 1-11

\bibitem{En82} Van Enter A C D 1982
{\it On a question of Bratteli and Robinson}
Lett. Math. Phys. 6 289-291

\bibitem{ViLaRiKi03}
Vidal G, Latorre J I, Rico E, and Kitaev, A 2003
{\it Entanglement in quantum critical phenomena}
Phys. Rev. Lett. 60 227902 1-4

\bibitem{We78} Wehrl A 1978
{\it General properties of entropy}
Rev. Mod. Phys. 50 221-260
\end{thebibliography}
\end{document}